\documentclass[aps,prd,reprint,superscriptaddress,nofootinbib,longbibliography]{revtex4-2}
\usepackage[T1]{fontenc}
\usepackage[utf8]{inputenc}
\usepackage{amsmath,amssymb,graphicx}
\usepackage{capt-of,placeins}
\usepackage{xcolor}
\def\be{\begin{equation}}
\def\ee{\end{equation}}
\def\ba{\begin{array}}
\def\ea{\end{array}}
\usepackage[normalem]{ulem}
\usepackage{slashed}
\def\sQ3{\widetilde{Q}_3}
\def\sU3{\widetilde{U}_3}
\def\sD3{\widetilde{D}_3}
\usepackage[colorlinks=true,allcolors=blue]{hyperref}
\newcommand{\beq}{\begin{equation}}
\newcommand{\eeq}{\end{equation}}
\newcommand{\bea}{\begin{eqnarray}}
\newcommand{\eea}{\end{eqnarray}}

\begin{document}
\title{Interpreting the High-Recoil LUX-ZEPLIN Event with  \\ Bino-/Singlino-like and Higgsino Dark Matter}

\author{Subhojit Roy}
\email{roysb@ucmail.uc.edu }
\affiliation{High-Energy Physics Division, Argonne National Laboratory, Argonne, IL 60439, USA}
\affiliation{Department of Physics, University of Cincinnati, Cincinnati, Ohio 45221, USA}
\author{Pedro Schwaller}
\email{pedro.schwaller@uni-mainz.de}
\affiliation{PRISMA$^{++}$ Cluster of Excellence $\&$ Mainz Institute for Theoretical Physics, Johannes Gutenberg University, 55099 Mainz, Germany
}
\author{Carlos E. M. Wagner}
\email{cwagner@uchicago.edu}
\affiliation{High-Energy Physics Division, Argonne National Laboratory, Argonne, IL 60439, USA}
\affiliation{Enrico Fermi Institute, Department of Physics, University of Chicago, Chicago, IL 60637, USA}
\affiliation{Kavli Institute for Cosmological Physics, University of Chicago, Chicago, IL 60637, USA}
\affiliation{Leinweber Center for Theoretical Physics, University of Chicago, Chicago, IL 60637, USA}
\affiliation{Perimeter Institute for Theoretical Physics, Waterloo, Ontario N2L 2Y5, Canada}
\date{\today}

\begin{abstract}
The LUX-ZEPLIN (LZ) experiment has recently reported a nuclear-recoil
candidate at $E_R\simeq248$~keV. We study two supersymmetric
interpretations based on distinct neutralino couplings to the $Z$ boson.
In the elastic case, a bino- or singlino-like neutralino scatters through
a diagonal axial $Z$ coupling set by the Higgsino asymmetry
$X=|N_{13}|^2-|N_{14}|^2$ rather than by the total Higgsino fraction.
Since the spin-dependent blind spot does not coincide with the
spin-independent one, Higgs-mediated scattering can be suppressed while
retaining $X\sim10^{-2}$. In the MSSM this favors low $\tan\beta$, in
tension with the observed Higgs mass for a few-TeV squark spectrum, a tension relieved by the
tree-level quartic of the NMSSM. An explicit NMSSM benchmark with
$m_\chi\simeq500$~GeV and $X\simeq0.007$, below current LZ limits,
obtains its relic abundance from a singlet-like pseudoscalar resonance
just above threshold, which also suppresses present-day annihilation.
In the inelastic case, a pseudo-Dirac Higgsino scatters endothermically,
with a splitting of a few hundred keV generated by gauginos at the PeV scale which can be lowered by
opposite-sign bino--wino cancellation, or by weak singlino mixing; the
minimal thermal realization is strongly constrained by solar-neutrino
searches. The two mechanisms predict distinct recoil spectra away from
the candidate.
\end{abstract}

\maketitle

\section{Introduction}
\label{sec:introduction}

The LUX-ZEPLIN (LZ) Collaboration has reported a nuclear-recoil candidate at
$E_R=248\pm23_{\rm stat}\pm23_{\rm sys}~{\rm keV}$ in a
$2.84$ tonne-yr extended-energy search~\cite{LZ:2026axp}. Among the
interaction hypotheses tested by LZ, the maximum local significance is
$3.4\sigma$, reduced to a global significance of $2.6\sigma$ after the
look-elsewhere effect~\cite{LZ:2026axp}. Although this significance is not
sufficient to establish a dark-matter origin, the unusually large recoil
energy makes the event a useful probe of dark-matter scenarios with enhanced
support in the high-energy recoil tail. Any such interpretation must produce
an appreciable rate near $250~{\rm keV}$ while remaining consistent with the
observed recoil population at both lower and higher energies.

For a xenon nucleus, a recoil energy of $E_R\simeq248~{\rm keV}$ corresponds to a momentum transfer
\begin{equation}
q=\sqrt{2m_AE_R}\simeq0.25~{\rm GeV},
\end{equation}
where $m_A$ is the xenon-nucleus mass. At such momentum transfers, the conventional elastic spin-independent (SI) response is significantly suppressed by the loss of nuclear coherence and the associated finite-size nuclear form factor~\cite{Fan:2026kxx,Elahi:2026vlm}. Consequently, an elastic SI interaction normalized to yield an appreciable event rate near $250~{\rm keV}$ would typically predict a much larger recoil rate at lower energies, where nuclear coherence is progressively restored and the form-factor suppression is weaker. Such a spectrum is therefore strongly constrained by the absence of a corresponding excess in the lower-energy recoil data.

Two mechanisms of particular relevance here can instead maintain
appreciable support at high recoil while producing qualitatively different
spectral continuations. In endothermic scattering, the energy required to
access the heavier dark-matter state raises the minimum incident velocity
and strongly suppresses the low-recoil spectrum~\cite{Tucker-Smith:2001myb}. Elastic spin-dependent
(SD) scattering provides a different possibility: the finite-momentum xenon
spin response can retain substantial support in the region where the
coherent SI response is strongly suppressed, while still giving rise to a
correlated population at lower recoil energies~\cite{Fitzpatrick:2012ix,
Anand:2013yka,Elahi:2026vlm}. The contrasting recoil correlations predicted by these two mechanisms,
which have also been compared in Ref.~\cite{DiMauro:2026ldr},
form the central spectral diagnostic of our analysis.

Both structures have close counterparts among the interaction templates
examined by LZ. The Collaboration considered the relativistic operator
basis $\mathcal L_1$--$\mathcal L_{20}$, together with inelastic extensions
of the standard $\mathcal O_1$ and $\mathcal O_4$
interactions~\cite{LZ:2026axp}. Several inelastic $\mathcal O_1$ and
$\mathcal O_4$ templates with splittings of order
$300$--$350~{\rm keV}$ yield local significances above $3\sigma$, with the
largest reaching $3.4\sigma$~\cite{LZ:2026axp}. For elastic scattering, the
$\mathcal O_4=\mathbf S_\chi\!\cdot\!\mathbf S_N$ templates give local
significances of approximately $2.6\sigma$, $2.7\sigma$, and $2.8\sigma$
for $m_\chi=400~{\rm GeV}$, $1~{\rm TeV}$, and $4~{\rm TeV}$,
respectively~\cite{LZ:2026axp}. In the relativistic basis employed by LZ,
the corresponding axial-vector dark-matter current coupled to an
axial-vector quark current is represented by $\mathcal L_{15}$
~\cite{LZ:2026axp,Elahi:2026vlm}. These values characterize the experimental
interaction templates and are not statistical significances assigned to
the supersymmetric realizations developed below.

Supersymmetric neutralino sectors provide natural realizations of both
structures. We take the lightest supersymmetric particle (LSP) to be a
neutralino dark-matter candidate.  A bino-like or singlino-like neutralino framework provides a natural
realization of elastic axial-$Z$ scattering.
In the Minimal Supersymmetric Standard Model (MSSM), bino--Higgsino
admixture generates a diagonal axial $Z$ coupling, while in the
Next-to-Minimal Supersymmetric Standard Model (NMSSM) the neutralino sector
admits additional singlino--Higgsino and bino--singlino--Higgsino
compositions~\cite{Arkani-Hamed:2006wnf, Ellwanger:2009dp}.
Wino decoupling is considered for this study.
The resulting recoil spectrum is dominated by
$\mathcal O_4=\mathbf S_\chi\!\cdot\!\mathbf S_N$ within the leading
one-body point-nucleon treatment adopted here. The Higgsino admixture that
generates the diagonal axial $Z$ coupling also generally induces $CP$-even
Higgs boson exchange and hence dark matter SI scattering. An SD-dominated interpretation
therefore requires suppression of the scalar amplitude without eliminating
the axial $Z$ interaction, which can be achieved at the region close to the well-known SI blind spot conditions~\cite{Arkani-Hamed:2006wnf,Cheung:2012qy,Huang:2014xua,
Badziak:2015exr,Baum:2017enm,Abdallah:2020yag,Chatterjee:2022pxf,
Datta:2022bvg,Ellwanger:2009dp,Arganda:2025fhx,Roy:2024yoh}.
Alternatively, as has been previously emphasized in the literature, a nearly pure Higgsino forms a pseudo-Dirac pair of neutral
Majorana states whose tree-level $Z$ interaction is predominantly
off diagonal. A mass splitting of a few hundred keV then permits
endothermic scattering,
$\chi_1 A \to \chi_2 A,$
through the electroweak $Z$ current
~\cite{Nagata:2014wma,Fan:2026kxx,Freese:2026sga,Rodd:2026tyn, Yin:2026jnn,
Wu:2026nhi,Du:2026guj,Bisal:2026khf,Langhoff:2026ujr,
Cheung:2026byg,Chatterjee:2026scv,Frolovsky:2026tvq}.
Because the minimum incident speed increases with the mass splitting, the
terrestrial recoil rate becomes increasingly sensitive to the extreme
high-velocity tail of the local dark-matter distribution
~\cite{Fan:2026kxx,Rodd:2026tyn}.

In this work, we study these two $Z$-boson mediated interpretations by comparing
their predicted recoil spectra across the extended LZ energy range.
For the elastic interpretation, we determine the diagonal axial coupling
required to produce an order-one yield in the high-recoil region and
quantify the associated spectrum at lower energies. We also examine
how this correlation depends on the finite-momentum xenon spin response.
 For the inelastic branch, we determine how the dark-matter mass, neutral-state
splitting, transition coupling, and high-speed halo population shape the
recoil spectrum, with particular emphasis on the correlated population
above the $248~{\rm keV}$ recoil. This allows us to compare directly the
high-energy continuation of a sub-TeV Higgsino with that of a Higgsino near
the canonical thermal mass scale and to assess how strongly this comparison
depends on the extreme high-velocity tail of the local dark-matter
distribution.

We then connect the recoil properties of the two interpretations to the
underlying supersymmetric neutralino parameters. For the
elastic case, the recoil normalization fixes the diagonal axial $Z$
coupling, or equivalently the Higgsino asymmetry
$|N_{13}|^2-|N_{14}|^2$, rather than the total Higgsino fraction. We study
how this nonzero axial coupling can coexist with suppressed Higgs-mediated
SI scattering in the Minimal Supersymmetric Standard Model (MSSM) and the Next-to-MSSM (NMSSM). We also present a representative
$Z_3$-symmetric NMSSM benchmark that simultaneously realizes the required
Higgsino asymmetry, strong SI scattering suppression, a Standard Model (SM)-like
Higgs state, and a thermal relic abundance close to the observed dark-matter
abundance. The point also has a suppressed present-day annihilation rate and
serves as an illustrative ultraviolet realization rather than the result of a
global fit.
For the inelastic case, instead, 
this means determining how a sub-MeV Majorana splitting of the
pseudo-Dirac Higgsino pair can arise from electroweak gaugino or singlino
mixing, and assessing the corresponding solar-capture constraints.  Our analysis remains at the recoil level and does not reconstruct the full
multidimensional LZ likelihood or perform a global fit of the model.

The paper is organized as follows. Section~\ref{sec:conventions} sets out
the recoil, halo, efficiency, and nuclear-response conventions, and
Sec.~\ref{sec:effective} introduces the two $Z$-boson mediated interactions and
their leading nonrelativistic structures. Sections~\ref{sec:elastic}
and~\ref{sec:inelastic}
 develop the elastic and inelastic recoil analyses,
respectively. Section~\ref{sec:distinguish} brings these results together
by comparing the spectral shapes and the correlated recoil signatures of
the two interpretations. Sections~\ref{sec:uvelastic} and \ref{sec:uvinel} then develop their
supersymmetric realizations, and Sec.~\ref{sec:conclusions} summarizes the
main results.

\section{Recoil framework and conventions}
\label{sec:conventions}

We base our recoil analysis on the LZ extended-energy exposure
$\mathcal E=2.84~{\rm tonne\,yr}$ and the nuclear-recoil candidate with
reconstructed energy~\cite{LZ:2026axp}
\begin{equation}
E_R=248\pm23_{\rm stat}\pm23_{\rm sys}~{\rm keV} \, .
\end{equation}
The LZ likelihood is formulated in the detector-observable space
$\{S1c,\log_{10}(S2c)\}$, where $S1c$ and $S2c$ denote the
position-corrected primary- and secondary-scintillation signals
~\cite{LZ:2026axp}. Here we instead use recoil energy as the physical
variable for calculating and comparing the predicted spectra. The total nuclear-recoil efficiency reported by LZ crosses the $50\%$
level near $E_R=5.4$ and $269.9~{\rm keV}$.  We use these values only as
reference energies and not as hard boundaries in true recoil
energy~\cite{LZ:2026axp}.

For natural xenon, with isotopes labeled by $A$, the differential recoil
rate per unit detector mass is
\begin{equation}
 \frac{dR}{dE_R}(t)
 =
 \frac{\rho_\chi}{m_\chi}
 \sum_A
 \frac{\xi_A}{m_A^{\rm atom}}
 \int d^3v\,
 f_{\rm lab}(\mathbf v,t)\,
 v\,
 \frac{d\sigma_A}{dE_R},
 \label{eq:targetrate}
\end{equation}
where $\rho_\chi$ denotes the local density of the interacting dark-matter
component, $m_\chi$ the incident dark-matter mass, $\xi_A$ the isotope mass
fraction, and $m_A^{\rm atom}$ the atomic mass entering the target number
density. For inelastic scattering we identify
$m_\chi\equiv m_{\chi_1}$. The corresponding nuclear mass $m_A$, distinct
from $m_A^{\rm atom}$, enters the reduced masses, recoil kinematics,
momentum transfer, and differential cross section. The isotope abundances
and nuclear-response conventions are summarized in
Appendices~\ref{app:nuclear} and~\ref{app:operators}.

Our fiducial halo model is a truncated Maxwellian distribution in the
Galactic frame,
$
 \rho_\chi=0.30~{\rm GeV\,cm^{-3}}$,
 $v_0=238~{\rm km\,s^{-1}}$ and 
$v_{\rm esc}=544~{\rm km\,s^{-1}},
$
following Ref.~\cite{Baxter:2021pqo}. The laboratory-frame velocity
distribution accounts for both the motion of the Sun through the Galaxy
and the Earth's orbital motion. Unless stated otherwise, all recoil spectra
are evaluated using the annual average
\begin{equation}
 \frac{d\overline R}{dE_R}
 =
 \frac{1}{T}\int_0^T dt\,
 \frac{dR}{dE_R}(t),
 \label{eq:annualrate}
\end{equation}
where $T$ denotes one year. We use a uniform annual average rather than
weighting the rate by the actual time distribution of the LZ live exposure.
The numerical implementation is described in Appendix~\ref{app:halo}. 
We refer to this choice below as our fiducial Standard Halo Model
(SHM).

The dominant coherent inelastic and elastic SD contributions depend on the
annually averaged mean inverse speed,
\begin{equation}
 \overline{\eta}(v_{\min})
 =
 \frac{1}{T}\int_0^T dt
 \int_{v>v_{\min}}
 \frac{f_{\rm lab}(\mathbf v,t)}{v}\,d^3v,
 \label{eq:eta}
\end{equation}
where $v_{\min}$ is the minimum laboratory-frame speed required to produce
a recoil of energy $E_R$. Additional velocity moments associated with the
subleading operators retained in the elastic calculation are defined in
Appendix~\ref{app:operators}.

For the inelastic interpretation, we assume that the local interacting
dark-matter density is carried by the lighter state $\chi_1$ and retain only
the endothermic process $\chi_1\to\chi_2$. We do not model the present-day
abundance of $\chi_2$ and therefore do not include exothermic
$\chi_2\to\chi_1$ scattering. If $\chi_1$ constitutes only a fraction
$f_\chi$ of the adopted local dark-matter density while retaining the same velocity
distribution, the recoil rate scales linearly with $f_\chi$. At fixed
masses, splitting, nuclear response, and halo model, the coupling required
to reproduce a fixed recoil yield therefore scales as $f_\chi^{-1/2}$.

To separate the underlying recoil spectrum from the effect of the
nuclear-recoil efficiency, we define both raw and efficiency-weighted recoil
yields. For an interval $a<E_R<b$,
\begin{align}
 N^{\rm raw}_{a-b}
 &=
 \mathcal E\int_a^b dE_R\,
 \frac{d\overline R}{dE_R},
 \label{eq:rawcounts}
 \\
 N^{\rm proxy}_{a-b}
 &=
 \mathcal E\int_a^b dE_R\,
 \epsilon_{\rm proxy}(E_R)
 \frac{d\overline R}{dE_R},
 \label{eq:counts}
\end{align}
with
\begin{equation}
 \frac{dN^{\rm proxy}}{dE_R}
 =
 \mathcal E\,\epsilon_{\rm proxy}(E_R)
 \frac{d\overline R}{dE_R}.
 \label{eq:diffproxy}
\end{equation}
Here $\epsilon_{\rm proxy}(E_R)$ denotes our digitization of the final
nuclear-recoil signal efficiency shown in Fig.~S2 of
Ref.~\cite{LZ:2026axp}, which incorporates the trigger and signal
thresholds, single-scatter and analysis selections, and the WIMP-search
ROI. We interpolate the digitized efficiency only over the energy range
shown in Fig.~S2 and do not extrapolate it into the high-energy sideband.
The digitization and interpolation procedure is described in
Appendix~\ref{app:detector}.

The recoil-energy intervals used below are analysis conventions rather than
experimental bins in true recoil energy. For the inelastic calculation, the
benchmark spectra are normalized over $215$--$269.9~{\rm keV}$. For the
elastic calculation, we use $100$--$270~{\rm keV}$ as the high-recoil proxy
interval. Following the approximate recoil-energy mapping of
Ref.~\cite{Rodd:2026tyn}, we use $269.9$--$350~{\rm keV}$ as an
intermediate region and $350$--$590~{\rm keV}$ as the approximate
recoil-energy counterpart of the detector-space high-energy sideband.
These intervals should not be interpreted as official LZ acceptance
windows. In particular, the one-dimensional LZ efficiency does not specify the
nuclear-recoil acceptance in the high-energy sideband. Predictions in
the $269.9$--$350~{\rm keV}$ and $350$--$590~{\rm keV}$ intervals are
therefore quoted as raw recoil yields, without an efficiency correction.

The quantities defined above are recoil-level diagnostics rather than
detector-level event predictions. The LZ analysis constructs signal and
background probability densities directly in
$\{S1c,\log_{10}(S2c)\}$ and performs a simultaneous unbinned
extended-likelihood analysis~\cite{LZ:2026axp}. The one-dimensional
efficiency used here does not specify the full detector response
$P(S1c,S2c|E_R)$, the background probability densities, or the nuisance
parameters entering the likelihood. We therefore use $N^{\rm proxy}$ and
the recoil ratios derived from it only to characterize spectral
correlations; they are neither interpreted as detector-level event counts
nor assigned the official LZ template significances.

The nuclear response is treated separately for the inelastic
$Z$-boson mediated transition and the elastic axial-$Z$ interaction.
For the inelastic transition, the dominant coherent contribution arises
from the vector--vector part of the $Z$-boson mediated interaction. The weak
charge fixes the zero-momentum normalization, while the finite nuclear
size governs the loss of coherence at nonzero momentum transfer. 
We adopt the Helm nuclear form factor, in the Lewin–Smith parametrization~\cite{Lewin:1995rx}, as our fiducial description of the coherent finite-size nuclear response. The Helm model approximates the nuclear density as a uniform sphere convolved with a Gaussian surface profile, thereby incorporating the finite nuclear radius and surface diffuseness and capturing the loss of coherence and associated diffraction structure at nonzero momentum transfer.
As an independent
comparison of the finite-momentum spectral shape, we also use the normalized
elastic ground-state spin-independent structure factors of
Vietze \emph{et al.}~\cite{Vietze:2014vsa}. This comparison probes the
sensitivity of the recoil spectrum to the coherent finite-momentum response;
neither prescription is identified with the nuclear implementation adopted
by LZ. The LZ calculation instead employs \texttt{WimPyDD} with one-body
density matrices developed for DMFormFactor-v6 and the modifications
described in the extended-energy analysis
~\cite{Jeong:2021bpl,LZ:2026axp}.

For elastic axial scattering, the relevant finite-momentum spin response
cannot be represented by a coherent scalar form factor. Our reference
calculation uses the GCN5082\cite{Menendez:2012tm} xenon one-body density matrices evaluated in
the nonrelativistic effective field theory (NREFT) framework, while the
Anand--Fitzpatrick--Haxton (AFH) response functions provide an independent
nuclear-structure comparison~\cite{Anand:2013yka}. The calculation retains
the correlated proton and neutron amplitudes, their finite-momentum
interference, and all nonzero operators generated by the leading one-body
point-nucleon $Z$ current. The isotope treatment, nuclear-response inputs,
and normalization conventions are detailed in
Appendix~\ref{app:nuclear}.

\section{Effective $Z$ interactions}
\label{sec:effective}

The two recoil mechanisms considered here arise naturally from the
structure of Majorana neutral currents. For a Majorana fermion, the
diagonal vector current vanishes identically, whereas an off-diagonal
vector current between two distinct Majorana states and a diagonal
axial-vector current can both be nonzero. We therefore consider
\begin{align}
 \mathcal L_{\rm inel}
 &=
 i\,\frac{g}{2c_W}\,\kappa_V Z_\mu
 \bar\chi_2\gamma^\mu\chi_1,
 \label{eq:inelL}
 \\
 \mathcal L_{\rm el}
 &=
 \frac12 g_A^\chi Z_\mu
 \bar\chi\gamma^\mu\gamma^5\chi,
 \qquad
 g_A^\chi=-\frac{g}{2c_W}\kappa_A ,
 \label{eq:elL}
\end{align}
where $g$ is the $SU(2)_L$ gauge coupling,
$s_W\equiv\sin\theta_W$, $c_W\equiv\cos\theta_W$, and
$\chi_1$ ($\chi_2$) denotes the lighter (heavier) state of the inelastic
pair. The transition vector bilinear is anti-Hermitian, and the explicit
factor of $i$ in Eq.~\eqref{eq:inelL} therefore ensures a Hermitian
interaction. The factor of $1/2$ in Eq.~\eqref{eq:elL} is the conventional
normalization for a bilinear of identical Majorana fields.
The inelastic interaction is
specified by $(m_\chi,\delta,\kappa_V)$, with
\begin{equation}
 m_\chi\equiv m_{\chi_1},
 \qquad
 \delta\equiv m_{\chi_2}-m_{\chi_1}>0,
\end{equation}
whereas the elastic interaction is specified by $(m_\chi,\kappa_A)$.
With the normalization of Eq.~\eqref{eq:inelL}, $\kappa_V=1$ corresponds
to the exact tree-level pure-Higgsino transition strength.

For momentum transfers satisfying $q^2\ll m_Z^2$, the $Z$ gauge boson can
be integrated out. The resulting effective couplings to a nucleon
$N=p,n$ are
\begin{align}
 f_N&=\frac{G_F\kappa_V}{\sqrt2}\,w_N,
 &
 B_N&=\frac{G_F\kappa_V}{\sqrt2}\,D_N,
 \notag\\
 b_N&=-\frac{G_F\kappa_A}{\sqrt2}\,w_N,
 &
 C_N&=-\frac{G_F\kappa_A}{\sqrt2}\,D_N ,
 \label{eq:matching}
\end{align}
where the vector weak charges are
\begin{equation}
 w_p=-(1-4s_W^2),\qquad w_n=1,
 \label{eq:wN}
\end{equation}
and the axial combination is
\begin{equation}
 D_N\equiv\Delta u^N-\Delta d^N-\Delta s^N .
 \label{eq:DN}
\end{equation}
Here $\Delta q^N$ denotes the contribution of quark flavor
$q=u,d,s$ to the spin of the nucleon $N$, defined through the axial-current
matrix element
\begin{equation}
 \langle N|\bar q\gamma^\mu\gamma^5q|N\rangle
 =
 2\Delta q^N s_N^\mu ,
\end{equation}
with $s_N^\mu$ the nucleon spin four-vector. We use the central
$\overline{\rm MS}$ values at $2~{\rm GeV}$ from
Ref.~\cite{Park:2025rxi}, with the neutron charges obtained by
$u\leftrightarrow d$. Their uncertainties are not propagated in the
present recoil calculation.

With these conventions, matching the weak currents onto nucleon currents gives the effective interactions~\cite{Anand:2013yka,Barello:2014uda,Bishara:2017pfq},
\begin{align}
 \mathcal L_{\rm inel}^{N}
 &=
 i\bar\chi_2\gamma_\mu\chi_1
 \sum_{N=p,n}
 \bar N\gamma^\mu(f_N+B_N\gamma^5)N,
 \notag\\
 \mathcal L_{\rm el}^{N}
 &=
 \frac12\bar\chi\gamma_\mu\gamma^5\chi
 \sum_{N=p,n}
 \bar N\gamma^\mu(b_N+C_N\gamma^5)N .
 \label{eq:nucleoncontact}
\end{align}
For a xenon isotope $A$, the coherent vector coupling is proportional to
the weak charge
\begin{equation}
 Q_W^{(A)}
 =
 N_A-(1-4s_W^2)Z_A,
 \label{eq:weakcharge}
\end{equation}
where $N_A$ and $Z_A$ are the neutron and proton numbers, respectively.
The associated finite-size form factor is normalized as $F_A(0)=1$.

We introduce the conventional free-nucleon cross-section normalizations~\cite{Essig:2007az,Nagata:2014aoa,Anand:2013yka},
\begin{align}
 \sigma_n^{\rm inel}
 &=
 \kappa_V^2\frac{G_F^2\mu_n^2}{2\pi}(\hbar c)^2,
 \label{eq:inel_sigma}
 \\
 \sigma_{\rm SD}^{N}
 &=
 \frac{3G_F^2\mu_N^2D_N^2\kappa_A^2}{2\pi}(\hbar c)^2,
 \qquad N=p,n,
 \label{eq:sigma}
\end{align}
where,
\begin{equation}
 \mu_N=\frac{m_\chi m_N}{m_\chi+m_N}
\end{equation}
is the dark-matter--nucleon reduced mass.

For the inelastic benchmark spectra, we adopt the reference normalization~\cite{Essig:2007az, Rodd:2026tyn}
\begin{equation}
 \sigma_n^{\rm ref}
 =
 7.3\times10^{-39}\ {\rm cm}^2 .
 \label{eq:sigmaref}
\end{equation}
We define the corresponding physical transition coupling by
\begin{equation}
 \kappa_V^{\rm ref}(m_\chi)
 =
 \left[
 \frac{2\pi\,\sigma_n^{\rm ref}}
 {G_F^2\mu_n^2(\hbar c)^2}
 \right]^{1/2}.
 \label{eq:kapparef}
\end{equation}
Over the mass range considered here,
$\kappa_V^{\rm ref}\simeq1$. We nevertheless distinguish this reference
coupling from the exact pure-Higgsino value $\kappa_V=1$: the former is
defined by the fixed cross-section normalization in
Eq.~\eqref{eq:sigmaref}, whereas the latter denotes the physical
tree-level Higgsino transition strength.

The quantity $\sigma_n^{\rm inel}$ is a convenient free-nucleon
normalization and should not be interpreted as a physical zero-threshold
inelastic cross section, since the endothermic threshold is set by the
nuclear kinematics. Likewise, $\sigma_{\rm SD}^{N}$ denotes the conventional
spin-dependent free-nucleon normalization. The xenon calculation retains
the correlated proton and neutron amplitudes rather than treating their
contributions independently.

The leading nonrelativistic operators follow directly from the current
structure.
For the inelastic interaction, the vector--vector term maps onto
$\mathcal O_1$ and the corresponding coherent spin-independent
nuclear-density response, conventionally denoted the $M$ response in
the NREFT notation. The accompanying
vector--axial term generates $\mathcal O_7$, $\mathcal O_9$, and a
splitting-dependent $\mathcal O_{10}$ contribution. A separate NREFT
evaluation shows that these transition-spin contributions are negligible
for the benchmarks considered here. The central inelastic calculation
therefore retains only the coherent $\mathcal O_1$ response, proportional
to $
 (Q_W^{(A)} F_A(q))^2$.

For elastic scattering, the axial--axial interaction maps primarily onto
$ \mathcal O_4=\mathbf S_\chi\!\cdot\!\mathbf S_N$,
while the accompanying axial--vector current generates
$\mathcal O_8$ and $\mathcal O_9$. Within the leading one-body
point-nucleon treatment adopted here, the complete elastic recoil spectrum
is $\mathcal O_4$ dominated, with the additional operators modifying the
reference rate only at the percent level. These contributions and their
interference are nevertheless retained. The operator definitions, momentum
conventions, matching coefficients, velocity moments, and nuclear-response
contractions are collected in Appendix~\ref{app:operators}.

Finite nuclear size is incorporated through the coherent form factors for
the inelastic transition and through the finite-momentum NREFT responses
for the elastic interaction. At the nucleon level, however, the weak
currents are treated at leading point-nucleon order. A more complete
finite-momentum treatment would include momentum-dependent vector and
axial nucleon form factors, weak-magnetism and induced-pseudoscalar terms,
and consistent one- and two-body current corrections~\cite{Bishara:2017pfq,Klos:2013rwa}. These effects are discussed in
Appendix~\ref{app:finiteq}. The percent-level non-$\mathcal O_4$
contribution quoted above therefore refers only to the additional operators
generated within the adopted point-nucleon treatment and should not be
interpreted as an estimate of the full hadronic or nuclear-response
uncertainty.

\section{Elastic axial $Z$ scattering}
\label{sec:elastic}

In the elastic interpretation, the high-recoil signal is generated by the
diagonal axial $Z$ coupling of a Majorana dark-matter particle. Unlike the
endothermic scenario of Sec.~\ref{sec:inelastic}, there is no inelastic
threshold; the recoil spectrum is instead controlled by the strength of the
axial interaction and the finite-momentum xenon spin response. We first
determine the coupling required to produce an order-one high-recoil yield
and then quantify the correlated spectrum predicted at lower recoil
energies.

\begin{table}[t]
\begin{ruledtabular}
\begin{tabular}{ccccc}
$m_\chi$ [GeV] & $\kappa_A$  &
$\sigma_{\rm SD}^n \times 10^{-42}\,{\rm cm}^2$ & $R_{\rm low}$ \\
\hline
300  & 0.012  & 4.0 & 6.7 \\
500  & 0.013  & 5.2 & 5.3 \\
1000 & 0.017  & 8.8 & 4.6 \\
2000 & 0.023  & 16.3 & 4.3 \\
\end{tabular}
\end{ruledtabular}
\caption{Elastic axial-$Z$ benchmarks evaluated with the GCN5082
reference response in the fiducial SHM and normalized to
$N^{\rm proxy}_{100-270}=1$. Here,
$R_{\rm low}=N^{\rm proxy}_{5.4-100}/N^{\rm proxy}_{100-270}$
denotes the correlated low-to-high recoil ratio, and
$\sigma_{\rm SD}^n$ is the corresponding conventional free-neutron
spin-dependent cross section. The corresponding AFH results are given
in Appendix~\ref{app:nuclear}.}
\label{tab:elastic}
\end{table}

\subsection{Benchmark normalization and nuclear-response dependence}
\label{benchmarkelastic}

Within the leading one-body point-nucleon $Z$ current introduced in
Sec.~\ref{sec:effective}, the coupling $\kappa_A$ fixes the dominant
axial--axial interaction together with the accompanying axial--vector
terms. Since our analysis remains at the recoil level, we define the
elastic benchmark by
\begin{equation}
 N^{\rm proxy}_{100-270}=1,
 \label{eq:elasticmatch}
\end{equation}
where $N^{\rm proxy}$ is defined in Eq.~\eqref{eq:counts}. The
$100$--$270~{\rm keV}$ interval is an analysis convention used to
characterize the high-recoil yield and is not an experimental bin in true recoil energy. 
We stress that this is only a recoil-level reference normalization.
The corresponding $|X|\simeq0.013$ lies slightly above the approximate
observed $L_{15}^{v}$ upper limit, $|X|\lesssim0.012$, near this mass,
as discussed in Sec.~\ref{sec:uvelastic}.  The public $L_{15}^{v}$
release does not provide a best-fit coupling, and
$N^{\rm proxy}_{100-270}$ is not a detector-level fitted signal yield.

Using GCN5082 as the reference nuclear response, the
$m_\chi=500~{\rm GeV}$ benchmark requires
\begin{equation}
 \kappa_A\simeq0.013,~~
 |g_A^\chi|\simeq4.8\times10^{-3},~~
 \sigma_{\rm SD}^n\simeq5.21\times10^{-42}\ {\rm cm}^2 .
 \label{eq:elastic500}
\end{equation}
Here $\sigma_{\rm SD}^n$ is the conventional free-neutron
spin-dependent normalization defined in Eq.~\eqref{eq:sigma}. The xenon
rate itself retains the correlated proton and neutron amplitudes, their
finite-momentum interference, and all operators generated by the adopted
one-body current.

Applying the same normalization condition with the AFH response gives
\begin{equation}
 \kappa_A\simeq0.02,
 \qquad
 \sigma_{\rm SD}^n\simeq1.36\times10^{-41}\ {\rm cm}^2 .
\end{equation}
The coupling inferred from a fixed xenon recoil yield therefore depends
appreciably on the nuclear-response prescription. This difference reflects
the mapping between the underlying axial interaction and the
finite-momentum xenon response. We adopt GCN5082 as the reference response
and use AFH as an independent nuclear-structure comparison; the
corresponding AFH results are collected in
Appendix~\ref{app:nuclear}. 

The value of $\sigma_{\rm SD}^n$ in Eq.~\eqref{eq:elastic500} should not
be compared directly with the reported neutron-only SD exclusion curve by
taking a ratio of cross sections. Such a comparison can involve different
proton--neutron coupling assignments, nuclear responses, current
treatments, halo assumptions, and detector likelihoods. A quantitative
exclusion test requires the reference and the experimental limit to be
evaluated within a common nuclear and detector framework. We therefore
treat $\sigma_{\rm SD}^n$ as a conventional normalization and focus on
recoil correlations calculated consistently within a specified response
model.
\subsection{Correlated low-energy recoil yield}
Once the high-recoil normalization in Eq.~\eqref{eq:elasticmatch} is
fixed, the elastic spectrum predicts a correlated yield at lower recoil
energies. We quantify this by
\begin{equation}
 R_{\rm low}
 =
 \frac{N^{\rm proxy}_{5.4-100}}
      {N^{\rm proxy}_{100-270}} .
 \label{eq:Rlow}
\end{equation}
Because the denominator is unity by construction, $R_{\rm low}$ directly
gives the efficiency-weighted recoil yield in $5.4$--$100~{\rm keV}$ per
unit proxy recoil in $100$--$270~{\rm keV}$.

For the $m_\chi=500~{\rm GeV}$ GCN5082 benchmark,
$ R_{\rm low}\simeq5.3$.
Thus an elastic axial interaction normalized to one high-recoil proxy
recoil predicts approximately $5.3$ correlated proxy recoils below
$100~{\rm keV}$. Over the mass range shown in
Table~\ref{tab:elastic}, the ratio increases toward lower $m_\chi$.
Using the AFH response instead gives
$R_{\rm low}\simeq7.2$, showing that the precise low-to-high recoil
correlation also depends on the xenon nuclear response. These quantities
are recoil-level diagnostics rather than detector-level event counts, as
discussed in Sec.~\ref{sec:conventions}.

Table~\ref{tab:elastic} shows that the low-energy yield remains several
times larger than the high-recoil normalization throughout the mass range
considered. Although its precise value depends on $m_\chi$ and on the
adopted nuclear response, a substantial lower-energy continuation is a
characteristic prediction of the elastic axial interpretation.

\subsection{$\mathcal O_4$ dominance and comparison with LZ templates}

The diagonal axial $Z$ interaction is approximately isovector, with proton
and neutron axial coefficients of opposite sign and comparable magnitude.
Its leading nonrelativistic contribution is
\begin{equation}
 \mathcal O_4=\mathbf S_\chi\!\cdot\!\mathbf S_N \, \,.
\end{equation}
The accompanying axial--vector current also generates
$\mathcal O_8$ and $\mathcal O_9$, whose finite-momentum contributions and
interference are retained.

For the $m_\chi=500~{\rm GeV}$ GCN5082 benchmark, we compare the complete
point-current spectrum with a pure isovector $\mathcal O_4$ spectrum.
After normalizing both to unit raw yield over
$5.4$--$269.9~{\rm keV}$, their recoil shapes differ by less than $1\%$
across the numerical grid. The elastic spectrum is therefore strongly
$\mathcal O_4$ dominated within the leading one-body point-nucleon
treatment adopted here. This statement concerns only the additional
operators generated within that treatment; finite-momentum nucleon form
factors, induced-pseudoscalar and weak-magnetism terms, and two-body
currents are not included and are discussed separately in
Appendix~\ref{app:finiteq}.

This close shape correspondence allows a comparison with the
$\mathcal O_4$ interaction templates analyzed by LZ. For zero splitting,
LZ reports local significances of approximately $2.6\sigma$,
$2.7\sigma$, and $2.8\sigma$ for both isoscalar and isovector
$\mathcal O_4$ templates at $m_\chi=400$, $1000$, and
$4000~{\rm GeV}$, respectively~\cite{LZ:2026axp}. These values refer to
the experimental templates evaluated within the LZ likelihood and should
not be assigned directly to the physical axial-$Z$ interaction considered
here.

An independent elastic interpretation likewise finds an
$\mathcal O_4$-dominated recoil spectrum from a diagonal $Z$ coupling
in singlet--doublet Majorana dark matter~\cite{Elahi:2026vlm},
providing a complementary realization of the same leading recoil
structure.
The elastic recoil analysis therefore fixes the axial coupling required
in the high-recoil region and predicts a correlated lower-energy spectrum.
Embedding this interaction in a supersymmetric neutralino sector introduces
an additional requirement: the Higgsino admixture responsible for the
diagonal $Z$ coupling generally also induces Higgs-mediated SI scattering.
A viable realization must suppress this scalar contribution without
eliminating the axial interaction. We address this model-building
requirement in Sec.~\ref{sec:uvelastic}.

\section{Inelastic $Z$ mediated scattering}
\label{sec:inelastic}

\subsection{Kinematics and reference normalization}

For the endothermic process
$\chi_1 A\rightarrow\chi_2 A$, with
$m_\chi\equiv m_{\chi_1}$ and
$\delta\equiv m_{\chi_2}-m_{\chi_1}>0$, energy conservation gives
~\cite{Barello:2014uda}
\begin{equation}
 v_{\min}(E_R)
 =
 \frac{m_AE_R/\mu_A+\delta}
 {\sqrt{2m_AE_R}},
 \qquad
 \mu_A=
 \frac{m_\chi m_A}{m_\chi+m_A},
 \label{eq:vmin}
\end{equation}
where $m_A$ is the nuclear mass of isotope $A$ and $\mu_A$ is the
dark-matter--nucleus reduced mass. The distinction between $m_A$ and the
atomic masses entering the target number density is specified in
Appendix~\ref{app:nuclear}. The minimum incident speed is itself minimized
at
\begin{equation}
 E_R^*
 =
 \frac{\mu_A}{m_A}\,\delta,
 \qquad
 v_{\min}^*
 =
 \sqrt{\frac{2\delta}{\mu_A}}.
 \label{eq:vminstar}
\end{equation}
Compared with elastic scattering, a positive mass splitting raises the
incident speed required to produce a given recoil. The
$\delta/\sqrt{E_R}$ term in Eq.~\eqref{eq:vmin} therefore strongly
suppresses the low-energy spectrum, while Eq.~\eqref{eq:vminstar}
identifies the recoil energy at which the kinematic requirement is least
restrictive. For splittings of a few hundred keV, xenon scattering probes
the extreme high-speed tail of the local dark-matter distribution
~\cite{Fan:2026kxx,Rodd:2026tyn}.

For the reference spectra, we use
$\sigma_n^{\rm ref}$ defined in Eq.~\eqref{eq:sigmaref}. Over the mass range considered
here, this corresponds to $\kappa_V^{\rm ref}\simeq1$ and hence to an
electroweak-strength Higgsino-like transition. We consider two reference
masses,
\begin{equation}
 m_\chi=500~{\rm GeV},
 \qquad
 m_\chi=1.1~{\rm TeV},
\end{equation}
and for each mass determine the splitting by imposing
\begin{equation}
 N^{\rm raw}_{215-269.9}=1.
 \label{eq:inelmatch}
\end{equation}
Thus the interaction strength is held fixed at the reference value while
Eq.~\eqref{eq:inelmatch} is solved for $\delta$. No independent fit of the
transition coupling to the LZ event is performed.

\subsection{Coupling--splitting relation and recoil spectrum}

At fixed $m_\chi$, $\delta$, halo model, and nuclear response, the coherent
transition rate scales as $\kappa_V^2$. If
$\kappa_V^{\rm ref}(m_\chi)$ denotes the physical coupling associated with
$\sigma_n^{\rm ref}$, the coupling required to satisfy
Eq.~\eqref{eq:inelmatch} is
\begin{equation}
 \kappa_V^{\rm req}(m_\chi,\delta)
 =
 \frac{\kappa_V^{\rm ref}(m_\chi)}
 {\sqrt{
 N^{\rm raw,ref}_{215-269.9}(m_\chi,\delta)
 }},
 \label{eq:kappareq}
\end{equation}
where $N^{\rm raw,ref}_{215-269.9}$ is evaluated at the reference
interaction strength. Since $\kappa_V^{\rm ref}\simeq1$ for the masses
considered here, Eq.~\eqref{eq:kappareq} directly tracks the loss of
kinematic support in the normalization interval. As
$N^{\rm raw,ref}_{215-269.9}\rightarrow0$, the required coupling diverges.

\begin{figure}[t]
\includegraphics[width=\columnwidth]{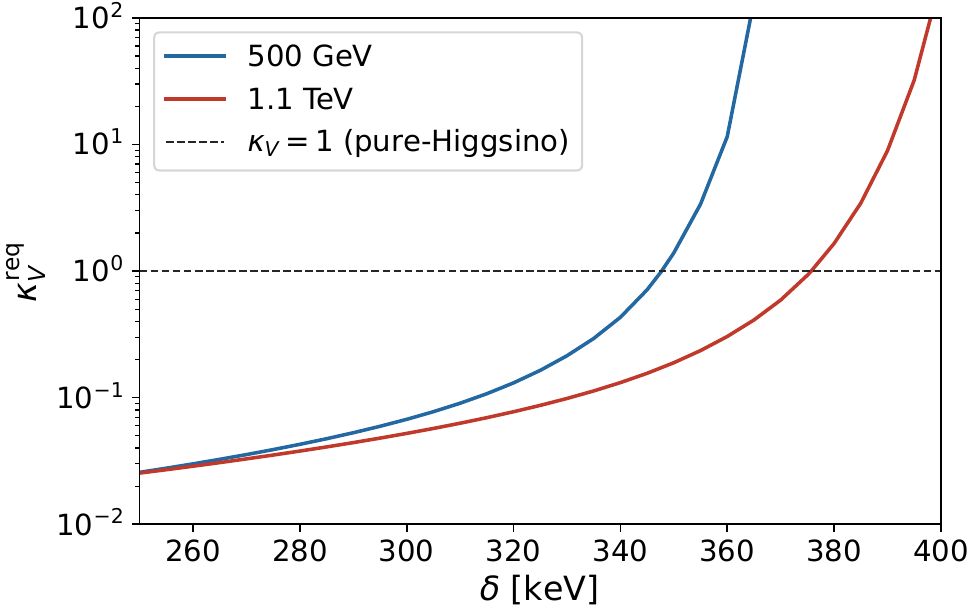}
\caption{The required transition coupling $\kappa_V^{\rm req}$ for one raw
recoil in the $215$--$269.9~{\rm keV}$ normalization interval, evaluated
in the fiducial annually averaged SHM with the Helm coherent response.
The curves correspond to the two reference masses,
$m_\chi=500~{\rm GeV}$ (blue) and $1.1~{\rm TeV}$ (red). The horizontal dashed black line marks
the exact pure-Higgsino electroweak strength, $\kappa_V=1$, defined in
Eq.~\eqref{eq:inelL}. 
}
\label{fig:coupling}
\end{figure}

Figure~\ref{fig:coupling} shows the required transition coupling for the
two reference masses in the fiducial annually averaged SHM. Increasing
$\delta$ progressively removes the high-speed halo phase space capable of
producing recoils in the normalization interval, so a larger transition
coupling is required to maintain Eq.~\eqref{eq:inelmatch}. At fixed
$m_\chi$, $\delta$, halo velocity distribution, and nuclear response, the
rate scales as $\rho_\chi\kappa_V^2$. The same rate-matching condition
therefore implies
\begin{equation}
 \kappa_V^{\rm req}\propto\rho_\chi^{-1/2}.
\end{equation}
The dependence on the coherent nuclear response is discussed in
Appendix~\ref{app:nuclear}.

The recoil spectrum is controlled by the interplay of endothermic
kinematics and the finite-size nuclear response. Schematically,
\begin{equation}
 \frac{d\overline R}{dE_R}
 \propto
 \sum_A
 \left[Q_W^{(A)}\right]^2
 F_A^2(q_A)\,
 \overline\eta\!\left(v_{\min}^{(A)}(E_R)\right),
 \label{eq:inel_schematic}
\end{equation}
where $q_A=\sqrt{2m_AE_R}$. Isotope abundances, target-counting factors,
and the remaining isotope-dependent normalizations are suppressed in
Eq.~\eqref{eq:inel_schematic}. The mean inverse speed determines the
available halo phase space, whereas $F_A(q_A)$ controls the
finite-momentum diffraction structure.

The Helm spectra display a pronounced suppression near
$E_R\simeq270$--$280~{\rm keV}$. This feature is a nuclear-diffraction
minimum rather than an inelastic kinematic threshold. The individual xenon
isotopes have slightly different nuclear radii and therefore different
form-factor zeros, so the natural-xenon sum develops a deep minimum rather
than an exact common zero. Its location is set mainly by the nuclear
response and consequently changes little between the two reference masses.

The dark-matter mass and splitting instead determine how this diffraction
structure is populated. Varying $m_\chi$ and $\delta$ changes
$v_{\min}(E_R)$, the available high-speed phase space, the relative weights
of neighboring diffraction lobes, and the kinematic endpoint. Above the
diffraction minimum, the Helm response rises into the next lobe, allowing
the spectrum to recover when the required incident speeds remain
accessible. At lower recoil energies, the same diffraction structure is
largely obscured by the endothermic suppression.
\subsection{High-recoil correlations}
Solving Eq.~\eqref{eq:inelmatch} at the two reference masses gives the
matched splittings listed in Table~\ref{tab:inelastic}. Although the two
values of $\delta$ differ by only $28~{\rm keV}$, the corresponding
high-recoil yields differ sharply. In the fiducial SHM, the
$500~{\rm GeV}$ solution predicts about $94\%$ fewer raw recoils in
$350$--$590~{\rm keV}$ than the $1.1~{\rm TeV}$ solution.

\begin{table}[t]
\begin{ruledtabular}
\begin{tabular}{rrrr}
$m_\chi$ [GeV] &
$\delta$ [keV] &
$N^{\rm raw}_{269.9-350}$ &
$N^{\rm raw}_{350-590}$ \\
500  & 347.7 & 0.94 & 0.32 \\
1100 & 375.7 & 3.28 & 5.44 \\
\end{tabular}
\end{ruledtabular}
\caption{Helm-fiducial inelastic reference points in the baseline annually
averaged SHM. For each $m_\chi$, the splitting $\delta$ is determined by
imposing $N^{\rm raw}_{215-269.9}=1$ at
$\sigma_n^{\rm ref}=7.3\times10^{-39}\,{\rm cm}^2$. The last two columns
give the correlated raw recoil yields above the normalization interval.
These recoil-energy intervals are analysis conventions rather than
official LZ true-recoil bins. The independently matched Vietze comparison
is given in Appendix~\ref{app:nuclear}.}
\label{tab:inelastic}
\end{table}

As discussed in Sec.~\ref{sec:conventions}, the
$350$--$590~{\rm keV}$ interval is only an approximate recoil-energy
mapping of the detector-space high-energy sideband and is not an official
true-recoil acceptance window~\cite{Rodd:2026tyn}. The values in
Table~\ref{tab:inelastic} are therefore raw recoil yields, with no
high-sideband efficiency or acceptance correction applied.

The high-recoil prediction depends on the coherent nuclear response because
this energy range samples the xenon diffraction structure. Repeating the
same matching procedure with the Vietze response changes both the matched
splittings and the integrated yields above the normalization interval.
Nevertheless, the same mass ordering is obtained in the fiducial SHM:
the $500~{\rm GeV}$ solution predicts about $90\%$ fewer raw recoils in
$350$--$590~{\rm keV}$ than the $1.1~{\rm TeV}$ solution. The
corresponding values are listed in Table~\ref{tab:responses_comp}. The strong
mass dependence is therefore present in both coherent-response
prescriptions, although its magnitude remains form-factor dependent.

This mass dependence is primarily kinematic. The isotope-dependent reduced
masses and the matched splittings determine the minimum speeds required to
populate the higher-recoil intervals, while Eq.~\eqref{eq:inelmatch} fixes
only the yield in $215$--$269.9~{\rm keV}$. The spectrum above the
normalization interval therefore carries information that is not fixed by
the recoil matching.

\subsection{Sensitivity to the high-velocity tail}

The high-recoil ordering found above is specific to the baseline annually
averaged SHM and is not halo independent. Endothermic scattering with a
few-hundred-keV splitting probes the extreme upper tail of the
laboratory-frame speed distribution, where a small redistribution of
phase-space weight can produce a large change in
$\overline\eta(v_{\min})$~\cite{Fan:2026kxx,Rodd:2026tyn}.

To assess this dependence without committing to a specific alternative
halo model, we consider normalized deformations of the detector-frame
speed distribution while retaining the same annual laboratory-motion
prescription. For the deformation families studied here, additional
support centered below roughly $800~{\rm km\,s^{-1}}$ produces relatively
small changes. By contrast, support introduced in the approximate
$800$--$900~{\rm km\,s^{-1}}$ region and above has a much larger effect.
These scales characterize the present reference points and deformation
prescription and should not be interpreted as universal velocity
thresholds.

Redistributing phase-space weight into this extreme-speed region changes
both the splitting selected by Eq.~\eqref{eq:inelmatch} and the predicted
yields in $269.9$--$350$ and $350$--$590~{\rm keV}$. Within the
deformations considered here, even the relative ordering of the
$500~{\rm GeV}$ and $1.1~{\rm TeV}$ high-recoil yields can reverse.
The mass hierarchy obtained with the Helm and Vietze responses should
therefore be regarded as a baseline-SHM prediction rather than as a
halo-independent consequence of inelastic scattering.

The Large Magellanic Cloud (LMC) provides a physically motivated source of
additional high-speed dark-matter phase space
~\cite{Smith-Orlik:2023kyl,Graham:2024syw}. The generic speed
deformations used here are not an LMC phase-space model and should not be
interpreted as an LMC uncertainty band. A quantitative Milky-Way--LMC
prediction for these reference points would require a specified
simulation-derived Galactic-frame phase-space distribution, including its
normalization, directional structure, and time dependence. The numerical
implementation and halo-deformation study are described in
Appendix~\ref{app:halo}.

\begin{figure*}[t]
\centering
\includegraphics[width=0.48\linewidth]{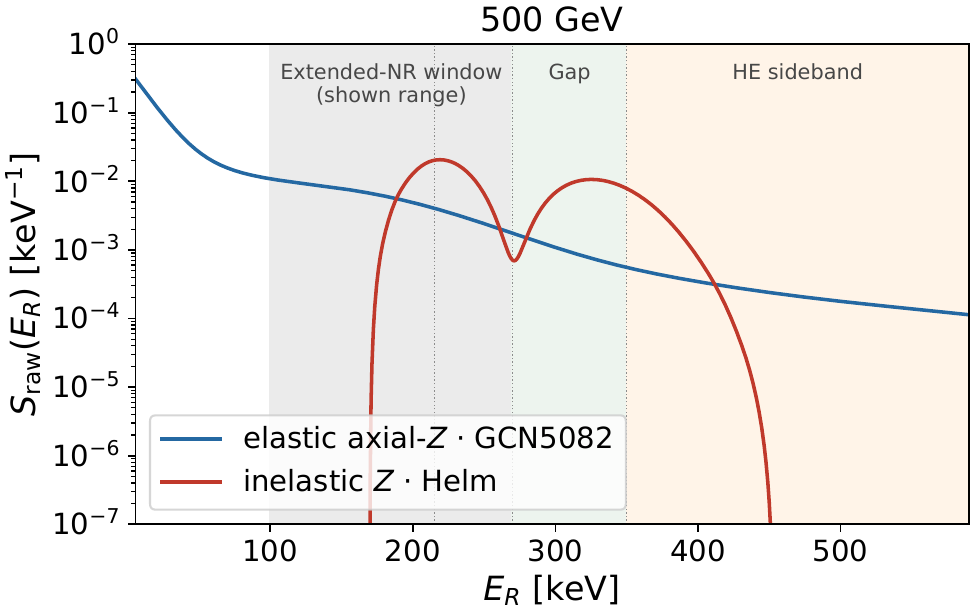} \,\,\,
\includegraphics[width=0.48\linewidth]{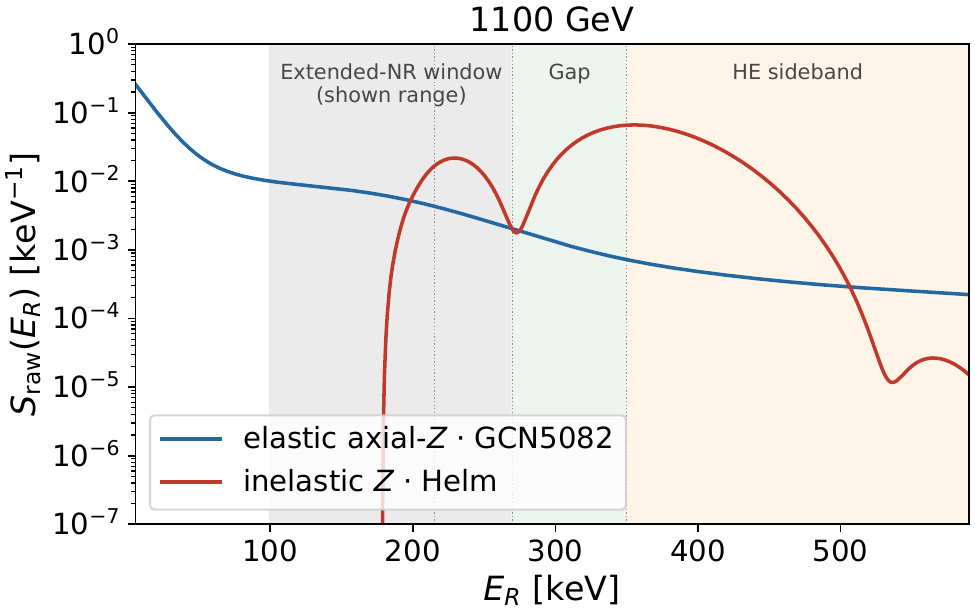} \,\,\,
\caption{Shape-normalized recoil spectra for elastic axial-$Z$ and
inelastic $Z$-boson mediated scattering at $m_\chi=500~{\rm GeV}$ (left) and
$1.1~{\rm TeV}$ (right), evaluated in the baseline annually averaged
SHM. Each curve shows $S_{\rm raw}(E_R)$ and is normalized to unit raw
yield over $100$--$269.9~{\rm keV}$. The elastic spectra use the GCN5082
nuclear response, while the inelastic spectra use the Helm coherent nuclear form factor with the splittings listed in Table~\ref{tab:inelastic}. The shaded
regions indicate the $100$--$269.9~{\rm keV}$ normalization interval,
the $269.9$--$350~{\rm keV}$ intermediate region, and the
$350$--$590~{\rm keV}$ interval used as the approximate recoil-energy
mapping of the detector-space high-energy sideband
~\cite{Rodd:2026tyn}. These intervals are analysis conventions rather
than official LZ true-recoil bins. No detector efficiency or
high-sideband nuclear-recoil acceptance is considered.}
\label{fig:shape}
\end{figure*}
\section{Spectral discriminants}
\label{sec:distinguish}

The inelastic and elastic benchmarks are defined with different
normalization prescriptions. Eq.~\eqref{eq:inelmatch} fixes one raw
recoil in $215$--$269.9~{\rm keV}$ for the inelastic case, whereas
Eq.~\eqref{eq:elasticmatch} fixes one efficiency-weighted proxy recoil in
$100$--$270~{\rm keV}$ for the elastic case. Their absolute recoil yields
therefore cannot be compared directly. To remove this normalization
dependence and compare only the spectral shapes, we define
\begin{equation}
 S_{\rm raw}(E_R)
 =
 \frac{d\overline R/dE_R}
 {\displaystyle
  \int_{100~{\rm keV}}^{269.9~{\rm keV}}
  dE'_R\,d\overline R/dE'_R},
 \label{eq:shape}
\end{equation}
so that
\begin{equation}
 \int_{100~{\rm keV}}^{269.9~{\rm keV}}
 dE_R\,S_{\rm raw}(E_R)=1.
\end{equation}
No detector-efficiency weighting enters this normalization.

Figure~\ref{fig:shape} combines the complementary recoil correlations
established in Secs.~\ref{sec:elastic} and~\ref{sec:inelastic}. At lower
recoil energies, elastic axial-$Z$ scattering retains substantial support,
consistent with the correlated yield quantified by $R_{\rm low}$.
Endothermic scattering instead strongly suppresses the low-energy spectrum
because the transition to the heavier state raises the minimum incident
speed. The lower-energy recoil spectrum therefore provides a direct
distinction between the two interpretations.

A different source of discrimination emerges at higher recoil energies.
For the inelastic interaction, the spectrum is shaped jointly by
endothermic kinematics and the coherent xenon form factor. The pronounced
suppression near $270$--$280~{\rm keV}$ is the nuclear-diffraction minimum
discussed in Sec.~\ref{sec:inelastic}.
Above this minimum, the next diffraction lobe can restore recoil support
provided that the halo contains particles with speeds above the required
$v_{\min}$. Since $v_{\min}(E_R)$ depends on both $m_\chi$ and the mass
splitting $\delta$, the $500~{\rm GeV}$ and $1.1~{\rm TeV}$ inelastic
benchmarks retain markedly different high-recoil shapes even after their
overall normalizations are removed. This behavior reflects the
mass-dependent high-recoil yields quantified in
Table~\ref{tab:inelastic}, whose magnitude is further sensitive to the
extreme high-velocity tail of the halo.

The two sides of the $248~{\rm keV}$ recoil thus carry complementary
information. Elastic axial scattering predicts a sizeable correlated
spectrum at lower energies, whereas endothermic scattering suppresses this
continuation and shifts the discriminating information toward the
high-recoil tail. In the latter case, the spectral shape is especially
sensitive to the dark-matter mass, the mass splitting, and the high-speed
halo population. These differences persist after the overall interaction
normalization is removed and therefore constitute genuine shape-level
predictions. A quantitative assessment of their experimental discrimination requires
the full multidimensional LZ detector response, background model, and
likelihood framework.
%
%
\section{Supersymmetric Realization I  : Elastic Bino-like and Singlino-like dark matter}
\label{sec:uvelastic}

The recoil results translate into  distinct requirements on the
supersymmetric neutralino sector.  The elastic interpretation requires a single Majorana neutralino
with a small but nonzero diagonal axial $Z$ coupling.  For the elastic
branch, the recoil normalization  fixes the asymmetry between the two
neutral-Higgsino components of the LSP.  We determine how this
requirement can coexist with suppressed Higgs-mediated SI direct detection scattering rate in
bino--Higgsino and singlino--Higgsino realizations, and give an explicit
$Z_3$-NMSSM benchmark.

In the MSSM, bino and wino mixing break the approximate Dirac symmetry of
the neutral-Higgsino pair.  For real parameters and the basis
\begin{equation}
 \psi^T=
 (\widetilde B,\widetilde W^0,
  \widetilde H_d^0,\widetilde H_u^0),
\end{equation}
the neutralino mass term is
$-\frac12\psi^T M_4\psi+\mathrm{h.c.}$, with~\cite{Martin:1997ns}
\begin{equation}
\resizebox{.98\columnwidth}{!}{$
 M_4=
 \begin{pmatrix}
 M_1&0&-m_Zs_Wc_\beta&m_Zs_Ws_\beta\\
 0&M_2&m_Zc_Wc_\beta&-m_Zc_Ws_\beta\\
 -m_Zs_Wc_\beta&m_Zc_Wc_\beta&0&-\mu\\
 m_Zs_Ws_\beta&-m_Zc_Ws_\beta&-\mu&0
 \end{pmatrix}$},
 \label{eq:uvmatrix}
\end{equation}
where $M_1$ and $M_2$ are the signed bino and wino soft masses,
$s_\beta=\sin\beta$, and $c_\beta=\cos\beta$.  We use
$v=174\,\mathrm{GeV}$, with
$v_u=v\sin\beta$ and $v_d=v\cos\beta$.

In the $Z_3$-invariant NMSSM, the MSSM field content is extended by a
gauge-singlet chiral superfield $\widehat S$.  The corresponding
superpotential is~\cite{Ellwanger:2009dp}
\beq
{\cal W}
=
{\cal W}_{\rm MSSM}\big|_{\mu=0}
+
\lambda \widehat S\,\widehat H_u\cdot\widehat H_d
+
\frac{\kappa}{3}\widehat S^3 ,
\label{eqn:superpot}
\eeq
where ${\cal W}_{\rm MSSM}\big|_{\mu=0}$ denotes the MSSM superpotential
without the supersymmetric Higgsino mass term (the so-called $\mu$-term).  Here,
$\widehat H_u$ and $\widehat H_d$ are the MSSM Higgs-doublet superfields.
The dimensionless couplings $\lambda$ and $\kappa$ govern the
singlet--doublet interaction and the singlet self-interaction, respectively.
The singlet superfield $\widehat S$ also contains a fermionic component,
the singlino $\widetilde S$, which enlarges the neutralino sector of the
MSSM. 
When the scalar component of $\widehat S$ acquires a vacuum expectation value
$s$, the $\lambda$ term generates an effective Higgsino mass parameter,
while the singlino obtains the signed mass parameter
\begin{equation}
 \mu_{\rm eff}=\lambda s,
 \qquad
 m_{\widetilde S}=2\kappa s .
 \label{eq:z3masses}
\end{equation}
Thus, the Higgsino mass scale is generated dynamically through the singlet
vacuum expectation value, providing the usual NMSSM solution to the
$\mu$ problem~\cite{Kim:1983dt}. The corresponding soft supersymmetry-breaking Lagrangian is
\begin{equation}
\begin{aligned}
-\mathcal{L}_{\rm soft}
={}&-\mathcal{L}^{\rm soft}_{\rm MSSM}\big|_{B\mu=0}
+m_S^2|S|^2
\\
&+\left(
\lambda A_\lambda S H_u\!\cdot\!H_d
+\frac{\kappa}{3}A_\kappa S^3
+{\rm h.c.}
\right) \, ,
\end{aligned}
\label{eqn:lagrangian}
\end{equation}
where
$-\mathcal{L}^{\rm soft}_{\rm MSSM}\big|_{B\mu=0}$ denotes the MSSM
soft-breaking Lagrangian with the Higgs bilinear soft term removed.
The parameter $m_S^2$ is the soft mass squared of the singlet scalar,
while $A_\lambda$ and $A_\kappa$ are the dimensionful trilinear soft
parameters.
In the NMSSM, the additional singlet scalars and singlino degrees of freedom modify
both the Higgs and neutralino sectors and are central to the phenomenology
considered here.  We therefore summarize the relevant features of these
sectors in the following subsections.

The elastic recoil result maps directly onto the neutralino mixing matrix.
For
\begin{equation}
 \chi \equiv \chi_1 =
 N_{11}\widetilde B
 +N_{12}\widetilde W^0
 +N_{13}\widetilde H_d^0
 +N_{14}\widetilde H_u^0
 +N_{15}\widetilde S ,
 \label{eq:neutralinoComposition}
\end{equation}
the diagonal $Z$ coupling is controlled by the difference of the two
neutral-Higgsino probabilities,
\begin{equation}
 X\equiv |N_{13}|^2-|N_{14}|^2 .
 \label{eq:Xdef}
\end{equation}
With the convention of Sec.~\ref{sec:effective},
\begin{equation}
 \mathcal L_Z
 =
 -\frac{gX}{4c_W}
 Z_\mu\bar\chi\gamma^\mu\gamma^5\chi,
 \qquad
 \kappa_A=|X| .
 \label{eq:uvaxial}
\end{equation}
The $500~{\rm GeV}$ GCN5082 unit-yield reference of
Sec.~\ref{sec:elastic} therefore corresponds to
\begin{equation}
 |X| \simeq 1.3\times10^{-2}.
 \label{eq:Xtarget}
\end{equation}
This value defines the unit-high-recoil-yield reference used to
characterize the elastic spectrum; it is not the normalization adopted
for the explicit ultraviolet benchmark.  A more direct comparison with
the LZ analysis can be made using the relativistic axial--axial
interaction $L_{15}$ considered by the Collaboration
~\cite{LZ:2026axp}.  In the LZ convention, the relativistic operator
coefficients are normalized using the reference electroweak scale
$v_{\rm LZ}=246.2~{\rm GeV}$.  With the conventions of
Sec.~\ref{sec:effective} and Appendix~\ref{app:operators}, the physical
axial-$Z$ interaction maps onto
\begin{equation}
 d_{15}^{s}v_{\rm LZ}^{\,2} \simeq -0.0275\,X,
 \qquad
 d_{15}^{v}v_{\rm LZ}^{\,2} \simeq -0.6104\,X ,
 \label{eq:L15_mapping}
\end{equation}
where $d_{15}^{s,v}$ denote the isoscalar and isovector coefficients,
respectively.  The interaction is therefore dominantly isovector, with
\begin{equation}
 \left|\frac{d_{15}^{s}}{d_{15}^{v}}\right|\simeq0.045 .
\end{equation}

The corresponding numerical limits are provided in the public LZ
HEPData release~\cite{LZ:2026axp}.  The release tabulates observed
upper limits on the pure-isovector $L_{15}^{v}$ interaction at discrete
dark-matter masses, including $400$ and $1000~{\rm GeV}$ but not the
mass of our benchmark.  Interpolating between these neighboring entries
gives, near $m_\chi\simeq505~{\rm GeV}$, an observed upper limit
corresponding approximately to
\begin{equation}
 |X| \lesssim 0.012 .
 \label{eq:L15_Xlimit}
\end{equation}
Different simple interpolations between the two tabulated masses give
$|X|\simeq0.0116$--$0.0119$ for the upper endpoint, which we summarize
by the value in Eq.~\eqref{eq:L15_Xlimit}.%
These limits are provided for one isospin component at a time,
whereas the physical $Z$ interaction contains the correlated
isoscalar component in Eq.~\eqref{eq:L15_mapping}.  The comparison in
Eq.~\eqref{eq:L15_Xlimit} should therefore be regarded as an
operator-level approximation rather than as a dedicated recast of the
full axial-$Z$ interaction.  Moreover, the public release does not
provide a best-fit $L_{15}^{v}$ coupling or a profile-likelihood scan,
so it does not determine a statistically preferred value of $X$.

For the explicit realization below, we adopt
$|X|\simeq0.007$ as a representative normalization comfortably below
the operator-specific observed upper limit in Eq.~\eqref{eq:L15_Xlimit}.
For numerical illustrations of the normalization dependence, we vary
the coupling by $\pm0.002$ around this representative value,
\begin{equation}
 0.005\leq |X|\leq0.009 .
 \label{eq:Xillustrative}
\end{equation}
The upper end is also close to the face-value neutron-only SD
normalization guide discussed below. This interval is illustrative and
is not an experimentally determined confidence interval.

The same Higgsino mixing that generates the diagonal axial-$Z$
interaction generally also induces $CP$-even Higgs exchange and hence
spin-independent dark-matter scattering.  The relevant neutralino
parameter space must therefore retain a nonzero axial coupling while
suppressing the scalar amplitude.  This distinguishes the SI blind
spots of interest here from SD blind spots, for which the required
axial interaction itself vanishes, $X\rightarrow0$.

The recoil requirement alone does not determine the thermal abundance.
A nearly pure bino typically annihilates inefficiently in a standard
thermal history unless a resonance, coannihilation channel, or another
depletion mechanism is available.  Bino--Higgsino mixing can instead
increase annihilation and coannihilation rates, moving the system
toward the well-tempered neutralino regime
~\cite{Arkani-Hamed:2006wnf}.  More generally, however, the relic
density depends on the complete neutralino, chargino, and Higgs spectra
and need not be determined by the Higgsino asymmetry required by the
elastic recoil interpretation.

\subsection{MSSM bino--Higgsino realization}
\label{sec:mssmbino}

Consider the MSSM with the wino sufficiently heavy that the relevant light
neutralino system is bino--Higgsino.  Away from exact bino--Higgsino
degeneracy, the Higgsino components ($N_{13}$ and $N_{14}$) of a bino-like LSP satisfy
~\cite{Djouadi:2001kba,Pierce:2013rda}
\begin{align}
 \frac{N_{13}}{N_{11}}
 &\simeq
 \frac{m_Zs_W
 \left(m_\chi\cos\beta+\mu\sin\beta\right)}
 {\mu^2-m_\chi^2},
 \notag\\
 \frac{N_{14}}{N_{11}}
 &\simeq
 -\frac{m_Zs_W
 \left(m_\chi\sin\beta+\mu\cos\beta\right)}
 {\mu^2-m_\chi^2}.
 \label{eq:mssmHiggsinoComponents}
\end{align}
Their difference gives
\begin{equation}
 X
 \simeq
 -|N_{11}|^2
 \frac{m_Z^2s_W^2\cos2\beta}
 {\mu^2-m_\chi^2}.
 \label{eq:uvbino}
\end{equation}
Here, $m_\chi$ denotes the signed neutralino mass eigenvalue associated with the LSP.
Eq.~\eqref{eq:uvbino} shows that the recoil target does not determine
a unique total Higgsino fraction or bino--Higgsino mass separation. 
Thus the separation required for a given axial coupling depends explicitly
on $\tan\beta$.  At moderate or large $\tan\beta$,
$|\cos2\beta|\rightarrow1$, whereas at
$\tan\beta\rightarrow1$ the diagonal axial coupling vanishes.  The
corresponding SD blind spot is
$ \tan\beta=1$,
which is incompatible with the recoil target
~\cite{Cheung:2012qy}.

The axial-$Z$ requirement by itself does not force low $\tan\beta$.
For example, fixing the physical LSP mass near
$m_\chi=500~{\rm GeV}$ and taking $\tan\beta=10$ with a decoupled
wino, exact neutralino diagonalization admits a solution near
\begin{equation}
 M_1\simeq503~{\rm GeV},
 \qquad
 \mu\simeq-720~{\rm GeV},
 \label{eq:mssmElasticTarget}
\end{equation}
for which the LSP is about $98\%$ bino, with a total Higgsino
probability of about $1.7\%$ and $|X|\simeq0.007$. Thus, for a
$500~{\rm GeV}$ neutralino, the representative axial coupling does not
by itself require a highly compressed bino--Higgsino spectrum. At
larger $\tan\beta$, the same value of $|X|$ can be obtained with a
substantially larger separation between the bino and Higgsino mass
parameters.

A stronger restriction arises when SI suppression is imposed simultaneously.
In the limit where the SM-like Higgs boson dominates, with heavy squarks and
the nonstandard $CP$-even Higgs states effectively decoupled, the coupling of a
bino--Higgsino LSP pair to the SM-like Higgs vanishes approximately for
\begin{equation}
 \frac{m_\chi}{\mu}
 \simeq
 -\sin2\beta \, .
 \label{eq:MSSMlightHblind}
\end{equation}
The signed neutralino eigenvalue enters this relation.  Importantly, this
SI blind spot does not coincide with the SD scattering blind spot, $X=0$: Higgs-mediated SI scattering
can be strongly suppressed while retaining a nonzero diagonal axial
$Z$ coupling~\cite{Huang:2014xua}.

Combining Eqs.~\eqref{eq:uvbino} and
\eqref{eq:MSSMlightHblind} gives
\begin{equation}
 |X|
 \simeq
 |N_{11}|^2
 \frac{m_Z^2s_W^2}{m_\chi^2}
 \frac{\sin^2 2\beta}{|\cos2\beta|}.
 \label{eq:mssmCombinedBlind}
\end{equation}
For $m_\chi\simeq500~{\rm GeV}$ and a predominantly bino LSP,
Eq.~\eqref{eq:mssmCombinedBlind} with $|X|\simeq0.007$ selects,
in the simplified light-Higgs blind-spot limit, approximately
\begin{equation}
 \tan\beta \simeq 2.1\text{--}2.2,
 \qquad
 \mu \simeq -(645\text{--}660)~{\rm GeV},
 \label{eq:mssm_target}
\end{equation}
with $M_1$ near the physical LSP mass. Thus the more compressed spectrum
arises only after the axial requirement is combined with the simple
SM-like-Higgs blind-spot condition; it is not a generic consequence of the
axial $Z$ coupling alone.

If the heavy MSSM-like $CP$-even state $H$ is not decoupled, destructive
interference between light- and heavy-Higgs mediated diagrams opens additional
SI blind-spot regions.  For approximately equal up- and down-type scalar
nucleon weights, the cancellation condition can be written as~\cite{Huang:2014xua}
\begin{equation}
 \frac{m_\chi}{\mu}
 +\sin2\beta
 -\frac{m_h^2}{2m_H^2}
 \cos2\beta
 \left(
 \tan\beta-\frac{1}{\tan\beta}
 \right)
 \simeq0 .
 \label{eq:MSSMgeneralBlind}
\end{equation}
At moderate or large $\tan\beta$, where
$\cos2\beta\simeq-1$ and $1/\tan\beta$ is small,
\begin{equation}
 \frac{m_\chi}{\mu}
 +\sin2\beta
 +\frac{m_h^2}{2m_H^2}\tan\beta
 \simeq0 .
 \label{eq:MSSMlargeTanbBlind}
\end{equation}
Larger-$\tan\beta$ solutions are in principle possible once the complete
$CP$-even Higgs boson amplitude is included.  The low-$\tan\beta$ branch above is
specific to the simpler limit in which the SM-like-Higgs coupling with the pair of dark matter itself
lies close to the blind spot criteria.

Low $\tan\beta$ also has a direct consequence for the observed Higgs boson mass.  In the
MSSM decoupling limit,
\begin{equation}
 m_h^2\lesssim m_Z^2\cos^22\beta+\Delta_{\tilde t},
 \label{eq:mssmHiggsMassLowTanb}
\end{equation}
where $\Delta_{\tilde t}$ are radiative corrections, mainly associated with the stop sector~\cite{Slavich:2020zjv}.  For
$\tan\beta\simeq2.1$--$2.2$, the tree-level contribution remains relatively small, so a
near-$125~{\rm GeV}$ SM-like Higgs requires a large radiative correction, associated with stops at the $\mathcal{O}(10-100)$~TeV  scale depending on the stop mixing parameter $X_t$~\cite{Giudice:2004tc,Draper:2013oza,PardoVega:2015eno}.
This is not a formal obstruction to the MSSM, since sufficiently heavy
and/or strongly mixed stops can supply the required uplift, but it motivates
the realization of the NMSSM discussed below.

In the large-$\tan\beta$ regime, Eq.~(\ref{eq:uvbino}) implies that,
for $m_\chi=500~{\rm GeV}$ and
$|X|=0.005$--$0.009$,
\[
 \left|\frac{m_\chi}{\mu}\right|
 \simeq 0.63\text{--}0.73 ,
\]
with the ratio increasing for larger $m_\chi$ or larger $|X|$.
The heavy-Higgs blind-spot condition therefore requires approximately
\begin{equation}
\sin 2\beta
+\frac{m_h^2}{2m_H^2}\tan\beta
\gtrsim 0.6 \,.
\label{eq:blindsp}
\end{equation}
Null results from searches for additional heavy neutral Higgs bosons at
the LHC place significant constraints on the MSSM heavy-Higgs
sector~\cite{ATLAS:2020zms,CMS:2022goy}. In particular, for
$\tan\beta\sim 10$, these searches imply a lower bound on the heavy-Higgs
mass of approximately $m_H > 1~{\rm TeV}$, in the decoupling regime
where $m_H\simeq m_A$. This bound suppresses the heavy-Higgs contribution
to Eq.~(\ref{eq:blindsp}), leaving the left-hand side smaller than the
required value by more than a factor of two. Increasing $\tan\beta$
to values of the order $60$, for which the corresponding bound approaches
$m_H\sim 2~{\rm TeV}$, does not significantly improve the situation,
since the enhancement from larger $\tan\beta$ is offset
by the stronger suppression from the heavier Higgs mass. At somewhat
smaller values, $\tan\beta\simeq 7$, the weaker bound
$m_H\sim 600~{\rm GeV}$ increases the heavy-Higgs contribution, but
a factor of about 1.5 mismatch still remains.
In summary, there is some tension between the blind-spot solution at
larger values of $\tan\beta$, the current bounds on the heavy Higgs
boson mass, and the realization of a scenario with sufficiently large
values of $|X| \simeq 0.005$--$0.009$ and
$m_\chi \gtrsim 500~{\rm GeV}$. A realization of the MSSM scenario at
large $\tan\beta$ will therefore require lower values of $m_\chi$,
for which the quality of the LZ signal is somewhat reduced. We plan
to explore this issue in a subsequent study.

The relic-density requirement remains logically separate from the
direct-detection conditions.  A few-percent Higgsino admixture can increase
electroweak annihilation and make nearby Higgsino-like neutralinos and the
chargino relevant for coannihilation, moving the spectrum toward the
well-tempered bino--Higgsino regime
~\cite{Arkani-Hamed:2006wnf}.  For a predominantly bino LSP with
$m_\chi\simeq500~{\rm GeV}$ and $|X|\simeq0.007$, however, the diagonal
$Z\chi\chi$ coupling is small and $m_\chi$ lies far from the
$Z$ resonance.  $s$-channel $Z$ exchange is therefore insufficient by
itself to obtain the observed thermal abundance.

One possible additional depletion mechanism is the MSSM pseudoscalar ($m_A$)
funnel, $m_A\simeq2m_\chi$.
For $m_\chi\simeq500~{\rm GeV}$ this places $m_A$ near
$1~{\rm TeV}$.  Whether the funnel reproduces the observed abundance
depends on the $A\chi\chi$ coupling, pole offset, width, and thermally
averaged annihilation rate.  Since $m_H\simeq m_A$ in this regime,
heavy-Higgs exchange must simultaneously be included in the SI amplitude.
For sufficiently heavy squarks, destructive $h$--$H$ interference can
satisfy Eq.~\eqref{eq:MSSMgeneralBlind}
~\cite{Huang:2014xua}.  The pseudoscalar funnel, the generalized SI blind
spot, and the required value of $X$ must therefore be realized within the
same MSSM spectrum.


\subsection{$Z_3$-NMSSM  realization}
\label{sec:nmssmelastic}

The $Z_3$-invariant NMSSM modifies the low-$\tan\beta$ Higgs-mass
constraint because the SM-like Higgs receives the additional tree-level
contribution~\cite{Ellwanger:2009dp}
\begin{equation}
 m_h^2 \supset \lambda^2v^2\sin^22\beta,
 \label{eq:nmssmTreeHiggsUplift}
\end{equation}
in the $v\simeq174~{\rm GeV}$ convention used here.  This term is largest at low $\tan\beta$, precisely in the region where
the $|X|\simeq0.007$ axial requirement combined with the simple
bino--Higgsino SI blind spot selects $\tan\beta$ near $2.1$--$2.2$. The NMSSM can therefore retain the
neutralino structure required for the axial $Z$ coupling while reducing
the stop-sector radiative correction needed for a near-$125~{\rm GeV}$
SM-like Higgs.

The singlet sector also modifies both the SI amplitude and the mechanisms
available for obtaining the relic abundance.  We first summarize the
bino-like dark matter analytic limit relevant to the numerical benchmark and then
briefly comment on the alternative of the singlino-like dark matter

\paragraph{Bino-like LSP.}

For a bino-like LSP, a nearby singlino modifies the Higgsino components and thereby affects both the coupling of the LSP pair to the SM-like Higgs and the axial quantity $X$.  With a decoupled wino and an aligned SM-like Higgs, the
tree-level SI blind-spot condition can be written as
~\cite{Abdallah:2020yag}
\begin{align}
 0\simeq{}&
 m_\chi+\mu_{\rm eff}\sin2\beta
 +\frac{2\lambda^2v^2}
 {m_{\widetilde S}-m_\chi}
 \notag\\
 &+
 \frac{\lambda^4v^4
 \left(\mu_{\rm eff}\sin2\beta-m_\chi\right)}
 {\left(m_{\widetilde S}-m_\chi\right)^2
  \left(\mu_{\rm eff}^2-m_\chi^2\right)} .
 \label{eq:NMSSMbinoBlind}
\end{align}
The MSSM bino--Higgsino relation is recovered when the singlino decouples.
If the final term is parametrically subleading, this reduces to
\begin{equation}
 m_\chi
 +
 \frac{2\lambda^2v^2}
 {m_{\widetilde S}-m_\chi}
 \simeq
 -\mu_{\rm eff}\sin2\beta .
 \label{eq:NMSSMbinoBlindApprox}
\end{equation}
The singlino contribution therefore opens SI cancellations not available in
the pure MSSM bino--Higgsino limit.  The relative signs and mass separation
of the bino-like LSP and singlino can compensate the Higgsino contribution
while leaving $X$ nonzero.

The NMSSM also admits additional SD blind spots.  Besides the
MSSM-like $\tan\beta=1$ solution, cancellations among bino, Higgsino, and
singlino components can drive $X\rightarrow0$~\cite{Abdallah:2020yag}.  Such regions suppress the axial interaction
required by the elastic recoil interpretation and are therefore not relevant to the
present recoil interpretation.

The singlet sector also enlarges the possible relic-density mechanisms.
Besides bino--Higgsino well tempering and neutralino coannihilation,
singlet-like scalar or pseudoscalar states can mediate resonant
annihilation.  In particular,
$ m_{a_s}\simeq2m_\chi$ defines a singlet-pseudoscalar funnel. Unlike the MSSM $A$-funnel, the singlet resonance need not track the heavy-doublet Higgs scale. The mass of the singlet-like pseudoscalar $a_s$ can instead be adjusted through the trilinear soft parameter $A_\kappa$, while the heavy-doublet Higgs states remain decoupled. The
Higgsino admixture can therefore generate the required axial $Z$ coupling
while a separate singlet-sector process controls the thermal abundance.

\paragraph{Singlino-like  LSP.}
A singlino-like LSP provides a second elastic realization.
Controlled Higgsino mixing can generate the required
$|X|\sim10^{-2}$, while SI suppression can arise from the
singlino--Higgsino blind spot, bino-assisted cancellations, or
interference involving a singlet-like $CP$-even scalar.  In the
$Z_3$-NMSSM, however, the singlino mass is directly correlated with the
singlet-scalar spectrum.  In the limit of small singlet--doublet mixing
and neglecting loop and doublet-dependent contributions, this correlation
takes the schematic form
$m_{\widetilde S}^2\simeq m_{h_s}^2+m_{a_s}^2/3$.
Thus, bringing the singlino to the LSP mass scale generally also brings
the singlet-scalar sector into the phenomenologically relevant region.
In particular, a relatively light $CP$-even singlet with even a small
doublet admixture can regenerate a sizeable SI scattering amplitude.

The simultaneous realization of the required Higgsino asymmetry,
suppressed SI scattering, and the correct thermal abundance is therefore
more correlated in the singlino-like branch than in the bino-like
realization that will be discussed below.  The latter offers greater parametric
freedom, since the bino mass fixes the LSP scale while the singlet sector
can separately participate in the SI cancellation and in resonant
annihilation.  We therefore present an explicit bino-like benchmark
in the following, while the singlino mixing relations, blind spots, and scalar-mass
correlations are collected in Appendix~\ref{app:singlino}.


\subsection{Representative $Z_3$-NMSSM benchmark.}

Motivated by the additional flexibility of the bino-like branch, we now
present an explicit realization of the elastic scenario.
Table~\ref{tab:nmssm_benchmark} gives a representative
$Z_3$-NMSSM point obtained with \texttt{NMSSMTools~6.2.2}
~\cite{Ellwanger:2004xm,Ellwanger:2005dv,Das:2011dg} interfaced to
\texttt{micrOMEGAs~6.2.2}
~\cite{Belanger:2006is,Belanger:2008sj, Alguero:2023zol}.
The LSP is strongly bino dominated, with a percent-level Higgsino component
generating the required axial $Z$ coupling, while the $CP$-even and singlet
Higgs sectors control the SI amplitude and thermal abundance, respectively.
The benchmark is illustrative and is not obtained from a global likelihood or
a dedicated collider recast.

\begin{table*}[t]
\begin{ruledtabular}
\begin{tabular}{l r l r}
\multicolumn{2}{c}{Input parameters} &
\multicolumn{2}{c}{Spectrum and dark-matter observables}\\
\hline
$\lambda$ & $0.5$ & $m_{\chi_1^0}$ & $505.3$\\
$\kappa$ & $0.5$ & $m_{\chi_2^0},\,m_{\chi_3^0}$ & $653.9,\ 656.6$\\
$\tan\beta$ & $2.195$ & $m_{\chi_4^0},\,m_{\chi_5^0}$ & $1309,\ 2521$\\
$\mu_{\rm eff}$ & $-638$ & $m_{\chi_1^\pm},\,m_{\chi_2^\pm}$ & $653,\ 2521$\\
$M_1$ & $509$ & $m_{h_1},\,m_{h_2},\,m_{h_3}$ & $124,\ 1090,\ 1171$\\
$M_2$ & $2500$ & $m_{a_1},\,m_{a_2}$ & $1015,\ 1165$\\
$M_3$ & $3000$ & $m_{H^\pm}$ & $1151$\\
$A_\lambda$ & $-118$ & $m_{\tilde t_1},\,m_{\tilde t_2}$ & $2476,\ 3047$\\
$A_\kappa$ & $565$ & $|N_{11}|^2$ & $0.99$\\
$A_t$ & $5000$ & $|N_{13}|^2+|N_{14}|^2$ & $0.011$\\
$A_b,\,A_\tau,\,A_\mu$ & $2500,\ 2500,\ 2500$
& $|N_{15}|^2$ & $2.57\times10^{-5}$\\
$M_{Q_{1,2,3}}$ & $3000$
& $X=|N_{13}|^2-|N_{14}|^2$ & $0.00693$\\
$M_{U_{1,2,3}},\,M_{D_{1,2,3}}$ & $2500,\ 2500$
& $\sigma_{\rm SI}^p$ & $7.9\times10^{-54}\ {\rm cm}^2$\\
$M_{L_{1,2,3}},\,M_{E_{1,2,3}}$ & $2500,\ 2500$
& $\sigma_{\rm SI}^n$ & $3.5\times10^{-52}\ {\rm cm}^2$\\
& & $\sigma_{\rm SD}^p$ & $1.9\times10^{-42}\ {\rm cm}^2$\\
& & $\sigma_{\rm SD}^n$ & $1.5\times10^{-42}\ {\rm cm}^2$\\
& & $\Omega_\chi h^2$ & $0.123$\\
\end{tabular}
\end{ruledtabular}
\caption{Representative $Z_3$-NMSSM benchmark for the elastic
axial-$Z$ interpretation of the observed LZ event with bino-like dark
matter, consistent with the relevant theoretical and experimental
constraints. Shown are the pertinent input
parameters defining the benchmark, together with the resulting masses,
neutralino mixings, and dark-matter observables. Dimensionful input
parameters and masses are given in GeV. The bino-like LSP realizes the
required Higgsino asymmetry, $X \simeq 0.007$.}
\label{tab:nmssm_benchmark}
\end{table*}

The lightest neutralino is predominantly bino, with a bino fraction of $98.9\%$ and a total Higgsino admixture of $1.0\%$.
 The resulting Higgsino asymmetry is 
\begin{equation}
 X=|N_{13}|^2-|N_{14}|^2 \simeq0.007 .
 \label{eq:nmssm_benchmark_X}
\end{equation}
Thus, the benchmark realizes the representative
$|X|\simeq0.007$ normalization introduced above. Since the
free-nucleon SD quantities obtained using \texttt{micrOMEGAs} employ
its internal nucleon inputs, we use $X$ as the common model-level
quantity when comparing the benchmark with the recoil calculation of Sec.~\ref{sec:elastic}.

The benchmark also illustrates the effect of the NMSSM Higgs quartic at low
$\tan\beta$. With $\lambda=0.5$ and $\tan\beta=2.195$, the physical stop
masses are $2.48$ and $3.05~{\rm TeV}$, while the lightest $CP$-even state is an
SM-like Higgs boson.

The strong SI suppression of this point arises from the complete NMSSM
$CP$-even amplitude rather than from the simple decoupled-MSSM light-Higgs
blind spot alone. Numerically,
$m_{\chi_1^0}+\mu_{\rm eff}\sin2\beta\simeq24~{\rm GeV}$, so the simple
relation $m_\chi+\mu\sin2\beta\simeq0$ is not exactly satisfied. The full
$CP$-even spectrum and its mixing instead yield
$\sigma_{\rm SI}^{p}=7.9\times10^{-54}~{\rm cm}^2$ and
$\sigma_{\rm SI}^{n}=3.5\times10^{-52}~{\rm cm}^2$.


\paragraph{Thermal abundance and the singlet-pseudoscalar resonance.}

The relic abundance of this point is not set primarily by conventional
bino--Higgsino well tempering. The lightest pseudoscalar satisfies
\begin{align}
 m_{a_1} &=1015.39~{\rm GeV},
~
 2m_{\chi_1^0}=1010.61~{\rm GeV}, \notag \\
~
 \Gamma_{a_1}&=1.11~{\rm GeV}.
 \label{eq:nmssm_as_resonance}
\end{align}
The pseudoscalar mixing matrix gives $|P_{13}|^2=0.892$, showing that
$a_1$ is predominantly singlet-like. Its pole lies only
$4.78~{\rm GeV}$ above the neutralino-pair threshold, placing the
benchmark close to the $a_1$ resonance and thereby enhancing neutralino
annihilation during thermal freeze-out. The pole is
therefore close enough to be sampled by the thermal velocity distribution at
freeze-out, while being appreciably farther from the $v\to0$ kinematics
relevant for late-time annihilation.

The benchmark reproduces the observed thermal relic abundance,
$\Omega_\chi h^2\simeq0.123$, with an effective annihilation rate at
freeze-out of
\[
\langle\sigma v\rangle_{\rm fo}
\simeq 1.5\times10^{-26}~{\rm cm^3\,s^{-1}}.
\]
Annihilation of the bino-like LSP into top-quark pairs ($t\bar t$) provides the
dominant contribution,
$
\chi_1\chi_1\to t\bar t \simeq 77\%,
$
driven by the nearby singlet-dominated pseudoscalar resonance.
The remaining important contributions arise from coannihilation with
the Higgsino-like electroweakinos,
$\chi_1\chi_2^0\to t\bar t\simeq9\%$,
$\chi_1\chi_3^0\to t\bar t\simeq7\%$, and
$\chi_1^\pm\chi_1\to t\bar b\simeq3\%$.
Although the electroweakino mass splittings are sizeable,
$m_{\chi_1}+m_{\chi_2^0}\simeq1159~{\rm GeV}$ and
$m_{\chi_1}+m_{\chi_3^0}\simeq1162~{\rm GeV}$ both lie close to the
heavy neutral-Higgs masses,
$m_{a_2}\simeq1165~{\rm GeV}$ and
$m_{h_3}\simeq1171~{\rm GeV}$.
This kinematic proximity can resonantly enhance the neutralino
coannihilation channels and partly compensate their Boltzmann
suppression at freeze-out.

The benchmark consequently realizes the elastic interpretation through
three complementary ingredients.  The bino--Higgsino admixture generates
the required axial $Z$ coupling, the $CP$-even Higgs sector suppresses the
spin-independent scattering amplitude, and the singlet-like pseudoscalar
provides the dominant annihilation channel needed to obtain the observed
thermal relic abundance.

\subsubsection{Indirect-detection properties.}

At present-day halo velocities, the benchmark predicts
\begin{equation}
 \langle\sigma v\rangle_{v\to0}
 =8.8\times10^{-27}~{\rm cm^3\,s^{-1}} .
 \label{eq:nmssm_vzero}
\end{equation}
The annihilation is almost entirely into $t\bar t$, with an effective
fraction of $99.5\%$, while the $b\bar b$ contribution is below the
percent level. The loop-induced line rates are small,
\begin{equation}
 \langle\sigma v\rangle_{\gamma\gamma}
 =3.08\times10^{-31},
 \qquad
 \langle\sigma v\rangle_{Z\gamma}
 =2.07\times10^{-31}~{\rm cm^3\,s^{-1}},
 \label{eq:nmssm_line_rates}
\end{equation}
and are far below the reported line-search sensitivities near this mass
~\cite{Fermi-LAT:2015kyq}. The pseudoscalar itself has
${\rm BR}(a_1\to t\bar t)=98.84\%$. 

The dominant continuum constraint is therefore the $t\bar t$ channel. The combined analysis of dwarf spheroidal galaxies by Fermi-LAT, HAWC,
H.E.S.S., MAGIC, and VERITAS provides the relevant current
constraint~\cite{Fermi-LAT:2025gei}. Its strongest combined limit for the $t\bar t$ channel near
$m_\chi\simeq505~{\rm GeV}$ gives an approximate
95\% C.L. upper limit of
$(1.1$--$1.5)\times10^{-26}~{\rm cm^3\,s^{-1}}$,
depending on the treatment of the dwarf $J$ factors. The benchmark prediction is therefore below, but
not parametrically far below, current sensitivity, with a predicted-to-limit
ratio of roughly $0.6$--$0.8$. The suppression relative to freeze-out is a
direct consequence of the pole being above threshold: thermal velocities can
approach the resonance, whereas present-day halo neutralinos cannot.

For CMB energy injection, a conservative estimate gives
\begin{equation}
 p_{\rm ann}=f_{\rm eff}\frac{\langle\sigma v\rangle_{v\to0}}{m_\chi}
 \simeq1.74\times10^{-29}f_{\rm eff}
 ~{\rm cm^3\,s^{-1}\,GeV^{-1}},
 \label{eq:nmssm_pann}
\end{equation}
well below the Planck $95\%$ C.L. bound
$p_{\rm ann}<3.2\times10^{-28}~{\rm cm^3\,s^{-1}\,GeV^{-1}}$ even for
$f_{\rm eff}=1$~\cite{Planck:2018vyg}.

A qualitatively different indirect-detection regime can arise elsewhere
in the singlet-pseudoscalar parameter space.  In the $Z_3$-NMSSM, the
doublet--singlet mixing of the $CP$-odd sector is controlled in part by the
combination $A_\lambda-2\kappa s$.  Near the corresponding mixing
cancellation,
\begin{equation}
 A_\lambda \simeq 2\kappa s ,
 \label{eq:cpodd_decoupling}
\end{equation}
the light pseudoscalar can become highly singlet dominated.  Its
tree-level couplings to Standard Model fermions, which are inherited
from the doublet component, are then correspondingly suppressed.  By
contrast, the coupling of a singlet-like pseudoscalar to Higgsino-like
charginos originates from the $\lambda \hat S\hat H_u\!\cdot\!\hat H_d$
interaction and remains proportional to $\lambda$.  Chargino loops can
therefore become comparatively important and enhance the loop-induced
$\gamma\gamma$ and $Z\gamma$ decay modes.  In sufficiently
singlet-dominated regions, the diphoton branching fraction can reach the
tens-of-percent level, with
${\rm BR}(a_s\to\gamma\gamma)\sim30\%$ possible for suitable spectra and
couplings~\cite{Guchait:2016pes,Ellwanger:2022jtd}.

If the pseudoscalar pole simultaneously remains sufficiently close to
$2m_\chi$, the same structure can lead to an enhanced monochromatic
$\gamma$-ray component in present-day dark-matter annihilation.  This
defines a qualitatively different indirect-detection regime from the
benchmark considered above, whose late-time annihilation is dominated by
$t\bar t$ and for which the $\gamma\gamma$ and $Z\gamma$ rates are
negligible.  A large radiative branching fraction alone, however, does
not imply an observable line signal: the absolute rates
$\langle\sigma v\rangle_{\gamma\gamma}$ and
$\langle\sigma v\rangle_{Z\gamma}$ also depend on the total present-day
annihilation cross section and on the proximity of the pseudoscalar pole
to the neutralino-pair threshold.  Regions in which the $CP$-odd mixing
cancellation is accompanied by a sufficiently large late-time
annihilation rate therefore provide an interesting complementary target
for monochromatic $\gamma$-ray searches.

\paragraph{Solar-capture neutrino constraints.}
For the elastic axial-$Z$ interaction, solar capture proceeds
predominantly through SD scattering on hydrogen and is therefore
controlled by the same coupling $\kappa_A=|X|$ that enters the
terrestrial recoil calculation.  The explicit benchmark has
$\sigma_{\rm SD}^p\simeq1.9\times10^{-42}~{\rm cm^2}$.
For comparison, assuming capture--annihilation equilibrium~\cite{Jungman:1995df} and
annihilation entirely into $W^+W^-$, IceCube obtains a
$90\%$ C.L. upper limit
$\sigma^p_{\rm SD}\simeq5.7\times10^{-42}~{\rm cm}^2$
at $m_\chi=500~{\rm GeV}$~\cite{IceCube:2025fcu}.
The benchmark therefore lies about a factor of three below
this hard-channel limit, and the solar-neutrino channel provides a relevant
constraint on the elastic realization. 

The present-day halo annihilation of the benchmark is dominated by
$t\bar t$.  Since the corresponding neutrino spectrum is softer than
the standard $W^+W^-$ channel template~\cite{Liu:2020ckq}, the comparison above
should be regarded as conservative rather than as a dedicated
$t\bar t$ IceCube limit.  The reduced capture rate is consequently
the primary reason that the elastic bino-like realization is less
constrained by solar-neutrino searches than the inelastic Higgsino
scenario discussed in Sec.~\ref{sec:realization}.
Furthermore, as discussed above, shifting $A_\lambda$ toward $2\kappa s$ enhances the
$\gamma\gamma$ and $Z\gamma$ annihilation modes of the dark matter.
The enhanced diphoton branching fraction reduces the neutrino yield from
annihilation in the solar core and can therefore significantly weaken the
IceCube sensitivity in this region of parameter space.


\subsubsection{Spin-dependent normalization and recoil expectation.}

The axial coupling $\kappa_A=|X|$ that generates the high-recoil signal also
controls conventional low-energy spin-dependent scattering. For
$m_\chi\simeq500~{\rm GeV}$, the reported neutron-only LZ limit is
~\cite{LZ:2024zvo}
\begin{equation}
 \sigma^n_{\rm SD}\lesssim2.5\times10^{-42}~{\rm cm}^2,
 \label{eq:LZSD}
\end{equation}
which we use only as a face-value guide to the normalization. As emphasized
in Sec.~\ref{benchmarkelastic}, the experimental limit assumes neutron-only
couplings, a specific nuclear response, and the full detector likelihood,
whereas the $Z$ interaction considered here fixes correlated proton and
neutron amplitudes and is evaluated with GCN5082. A like-for-like exclusion
test would require a dedicated recast.

As discussed above, the neutron-only SD bound provides only an
approximate normalization guide for the correlated axial-$Z$
interaction.  The explicit benchmark has
$X=0.0069$ and
$\sigma_{\rm SD}^{n}=1.53\times10^{-42}~{\rm cm^2}$.
For the exact benchmark mass and coupling,
$m_\chi=505~{\rm GeV}$ and $X=0.0069$, rerunning the validated
GCN5082 recoil calculation gives
\begin{align}
 N^{\rm proxy}_{5.4-100} &= 1.47,
 \notag\\
 N^{\rm proxy}_{100-270} &= 0.28,
 \notag\\
 N^{\rm proxy}_{5.4-270} &= 1.75 .
 \label{eq:x007_proxy_yields}
\end{align}
At fixed dark-matter mass, halo model, and nuclear response, the elastic recoil rate
scales as $X^2$.  Using the illustrative interval in Eq.~\eqref{eq:Xillustrative}, the
corresponding recoil yields are
\begin{align}
 N^{\rm proxy}_{5.4-100} &\simeq 0.77\text{--}2.49,
 \notag\\
 N^{\rm proxy}_{100-270} &\simeq 0.144\text{--}0.467,
 \notag\\
 N^{\rm proxy}_{5.4-270} &\simeq 0.91\text{--}2.95 .
 \label{eq:Xrange_proxy_yields}
\end{align}
The benchmark value $X=0.0069$ lies within this range and gives a
total proxy yield of approximately $1.75$ below $270~{\rm keV}$.

As a simple signal-only counting diagnostic, the benchmark high-recoil
mean $\lambda_H=0.277$ gives
\begin{eqnarray}
 P(N_H\ge1)&=&1-e^{-\lambda_H}=0.24,
 \nonumber\\
 P(N_H=1)&=&\lambda_H e^{-\lambda_H}=0.21.
 \label{eq:x007_poisson}
\end{eqnarray}
Across the same illustrative interval, the expected
high-recoil proxy yield varies from approximately $0.144$ to $0.467$,
corresponding to
\begin{equation}
 0.134 \lesssim P(N_H\ge1) \lesssim 0.373 .
\end{equation}
Thus the signal-only probability of at least one high-recoil proxy
event ranges from approximately $13\%$ to $37\%$.  These probabilities
are not LZ local significances and should not be interpreted as
detector-level likelihood probabilities; they use only the recoil-level
proxy mean.  A quantitative statistical assessment requires the full
$\{S1_c,\log_{10}S2_c\}$ likelihood.
For completeness, the exact benchmark itself predicts raw elastic
recoil yields of $0.0237$ in $269.9$--$350~{\rm keV}$ and $0.0175$ in
$350$--$590~{\rm keV}$.  No efficiency or high-sideband acceptance is
applied to these two quantities.

\subsubsection{Searches at the LHC}

The representative benchmark contains two Higgsino-like neutralinos,
$m_{\chi_2^0}\simeq654~{\rm GeV}$ and
$m_{\chi_3^0}\simeq657~{\rm GeV}$, together with a Higgsino-like
chargino, $m_{\chi_1^\pm}\simeq653~{\rm GeV}$, above the bino-like LSP
with $m_{\chi_1^0}\simeq505~{\rm GeV}$. The corresponding mass
splittings allow the two-body decays
\begin{equation}
 \chi_1^\pm\to W^\pm\chi_1^0,\qquad
 \chi_{2,3}^0\to Z\chi_1^0,\;h\chi_1^0 ,
\end{equation}
giving rise to the standard $WW$, $WZ$, and $Wh$ final states
accompanied by missing transverse momentum ($\slashed E_T$).

Direct searches for electroweakinos at the LHC probe both wino- and
Higgsino-like scenarios, including Higgsino--bino spectra relevant to
the present benchmark~\cite{ATLAS:2021moa, ATLAS:2021yqv,ATLAS:2023lfr, CMS:2024gyw, 
CMS:2024gyw}. The sensitivity to Higgsino-like states is nevertheless
weaker than for winos of comparable mass because of their smaller
electroweak production cross sections~\cite{Abdallah:2020yag}.
In addition, the present benchmark has a comparatively modest
Higgsino--LSP mass splitting of about $150~{\rm GeV}$. Although the
decays to on-shell $W$, $Z$, and $h$ bosons are kinematically open,
the resulting Standard-Model particles are less energetic than in the
large-splitting configurations for which these searches attain their
strongest sensitivity. The reduced production rate together with the
lower acceptance therefore considerably weakens the corresponding mass
bounds. The benchmark consequently remains allowed by the existing
electroweakino searches, while the high-luminosity LHC will extend the
sensitivity to this region.

The $Wh+\slashed E_T$ topology provides an additional probe whenever
$\chi_{2,3}^0\to h\chi_1^0$ is appreciable. LHC analyses have exploited
several Higgs decay modes, including the dominant $h\to b\bar b$ channel
as well as the cleaner $h\to\gamma\gamma$ final state
~\cite{ATLAS:2018qmw,ATLAS:2020qlk}. Despite its smaller branching
fraction, the diphoton mode benefits from the well-resolved Higgs
invariant-mass peak and therefore provides a complementary search
strategy. More generally, the sensitivity of the $WZ$ and $Wh$
searches depends on the physical branching fractions of the two
Higgsino-like neutralinos, so that the relevant collider rates are set
by the corresponding production cross sections times branching
fractions rather than by the Higgsino mass alone.

The extended Higgs sector provides a complementary set of collider
tests. Our benchmark contains a predominantly singlet-like pseudoscalar
$a_1$ near $1.02~{\rm TeV}$, while the remaining heavy neutral Higgs
states and the charged Higgs state lie in the $1$--$1.2~{\rm TeV}$ range,
with $m_{H^\pm}\simeq1.15~{\rm TeV}$. At the low value
$\tan\beta\simeq2.2$ realized by the benchmark, searches for heavy
neutral Higgs bosons decaying into $t\bar t$ are particularly relevant
~\cite{ATLAS:2024vxm}. The coupling of the doublet-like heavy Higgs
states to top quarks scales approximately as $\cot\beta$, making the
$t\bar t$ channel an important probe of the low-$\tan\beta$ region.
Experimentally, however, this search is challenging because of the
large Standard-Model $t\bar t$ background and the sizable
signal--background interference. The predominantly singlet nature of
$a_1$ further suppresses its direct production through Standard-Model
initial states.

Complementary sensitivity arises from searches for a heavy charged
Higgs boson produced in association with top and bottom quarks and
decaying through $H^\pm\to tb$~\cite{ATLAS:2021upq, ATLAS:2024itc}. The benchmark
value $m_{H^\pm}\simeq1.15~{\rm TeV}$ lies within the mass range
targeted by these searches. The heavy neutral Higgs masses also lie
close to the $\chi_1^0\chi_{2,3}^0$ thresholds, so decays into
electroweakinos can become kinematically accessible and may redistribute
the conventional heavy-Higgs branching fractions. The benchmark remains
consistent with the existing electroweakino and heavy-Higgs constraints,
while the increased sensitivity of the high-luminosity LHC in the
$WZ/Wh+\slashed E_T$, $t\bar t$, and $tb$ channels will provide
complementary probes of this region of parameter space.

\section{Supersymmetric Realization II : Inelastic pseudo-Dirac Higgsino}
\label{sec:uv}
\label{sec:realization}
\label{sec:uvinel}

For the inelastic branch, the central model-building question is the origin
of the sub-MeV Majorana splitting of an otherwise pseudo-Dirac Higgsino
pair.  We consider splittings generated by electroweak gauginos in the MSSM
and also by the singlino mixing in the NMSSM, and then confront the resulting
Higgsino interpretation with solar-capture constraints.  

\subsection{Off-diagonal $Z$ interaction and the Higgsino mass scale}

The neutral-Higgsino gauge current is
\begin{equation}
 \mathcal L_Z
 =
 \frac{g}{2c_W} Z_\mu
 \left(
 \widetilde H_d^{0\dagger}\bar\sigma^\mu\widetilde H_d^0
 -
 \widetilde H_u^{0\dagger}\bar\sigma^\mu\widetilde H_u^0
 \right).
 \label{eq:higgsinogaugecurrent}
\end{equation}
For positive supersymmetric Higgsino mass
parameter $\mu$, define
\begin{equation}
 \eta_1=
 \frac{\widetilde H_d^0-\widetilde H_u^0}{\sqrt2},
 \qquad
 \eta_2=
 i\frac{\widetilde H_d^0+\widetilde H_u^0}{\sqrt2}.
\end{equation}
Writing the corresponding four-component Majorana fields as
$\chi_i=(\eta_i,\eta_i^\dagger)^T$ gives
\begin{equation}
 \mathcal L_Z
 =
 i\,\frac{g}{2c_W}
 Z_\mu\bar\chi_2\gamma^\mu\chi_1 .
 \label{eq:uvvector}
\end{equation}
A pure Higgsino therefore realizes the off-diagonal interaction of
Eq.~\eqref{eq:inelL} with $\kappa_V=1$.

If the Higgsino constitutes the dark matter through standard thermal
freeze-out, its relic abundance fixes the mass scale independently of the
LZ recoil kinematics.  The two neutral Higgsinos and the charged Higgsino
are nearly degenerate at freeze-out, so coannihilation makes an essential
contribution to the effective annihilation cross section.  A useful
approximation is~\cite{Mizuta:1992qp}
\begin{equation}
 \Omega_{\widetilde H}h^2
 \simeq
 0.10
 \left(
 \frac{|\mu|}{1\,\mathrm{TeV}}
 \right)^2.
 \label{eq:relic}
\end{equation}
Together with the observed dark-matter abundance,
$\Omega_{\rm DM}h^2\simeq0.120$~\cite{Planck:2018vyg},
Eq.~\eqref{eq:relic} gives the familiar thermal-Higgsino scale
$|\mu|\simeq1.1\,\mathrm{TeV}$.  More complete calculations of neutralino
and chargino coannihilation, including Sommerfeld effects, retain a thermal
Higgsino mass close to this value
~\cite{Hryczuk:2010zi,Beneke:2014hja}.

We therefore use $m_\chi=1.1\,\mathrm{TeV}$ as the canonical thermal
benchmark.  In the pure-Higgsino limit, the off-diagonal
$Z\chi_1\chi_2$ coupling fixes the inelastic free-neutron normalization
through Eq.~\eqref{eq:inel_sigma}; it is not adjusted independently to the
observed event.  Once the halo and nuclear inputs are specified, the
neutral-state splitting $\delta$ determines the recoil kinematics.

For a Higgsino at this mass scale, freeze-out occurs at a temperature of
order several tens of GeV.  A splitting of a few hundred keV is therefore
negligible during freeze-out and does not appreciably modify the Boltzmann
populations or coannihilation dynamics.  Our baseline calculation assumes
that the Higgsino constitutes all of the cosmological and local dark matter.
If instead it accounts for a fraction
$f_\chi=\Omega_{\widetilde H}/\Omega_{\rm DM}$, we take
$\rho_\chi=f_\chi\rho_{\rm DM}$.  Under this assumption, the
direct-detection rate scales as $f_\chi$, whereas the annihilation rate in
the Galactic halo scales as $f_\chi^2$.

The $m_\chi=500~\mathrm{GeV}$ Higgsino benchmark used in the recoil
analysis has a different cosmological interpretation.  Standard thermal
freeze-out at this mass gives a relic abundance below the observed
dark-matter density.  If such a state is nevertheless assumed to constitute
essentially all of the present-day dark matter, a nonstandard cosmological
history is required, for example late production of Higgsinos from decays
of heavier fields after freeze-out.

\begin{figure*}[t]
\centering
\includegraphics[width=0.47\linewidth]{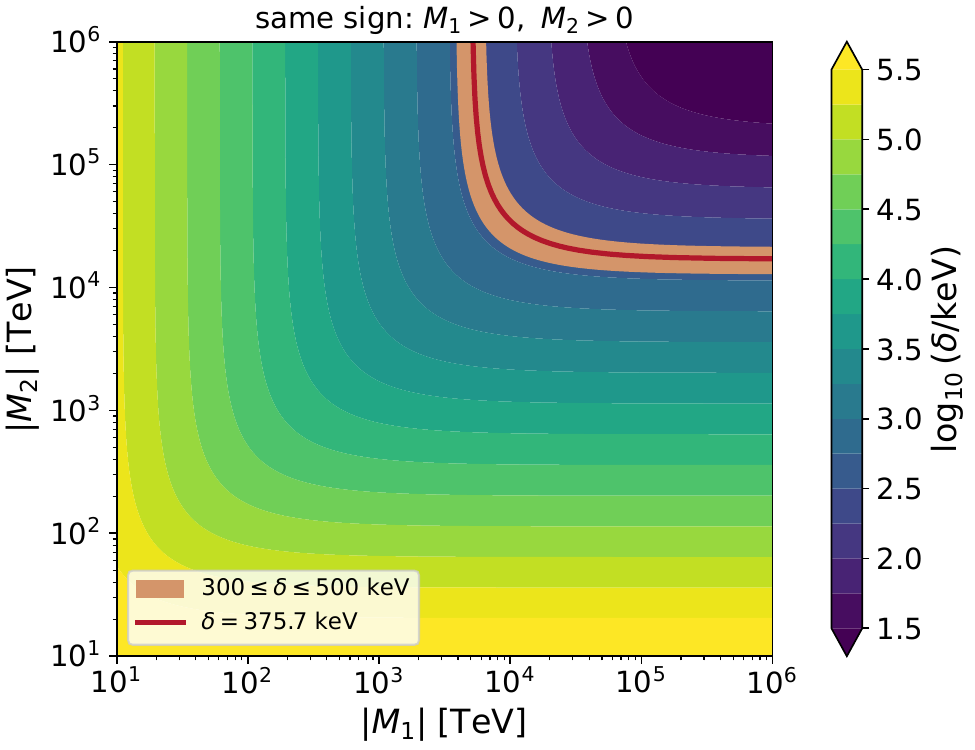} \,\,\,
\includegraphics[width=0.47\linewidth]{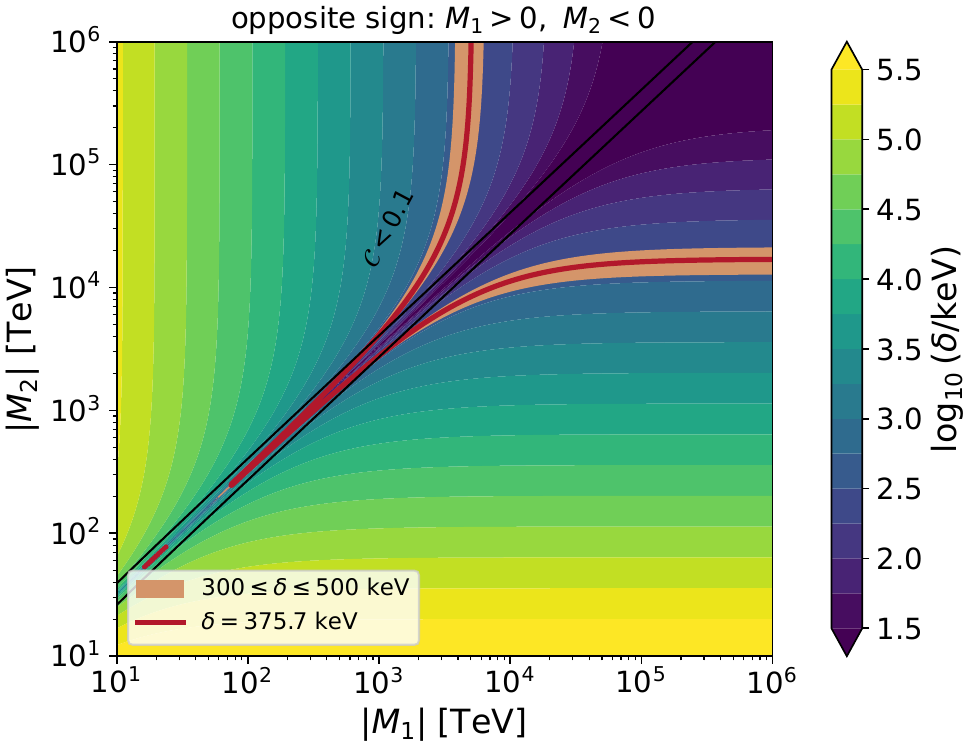} \,\,\,
\caption{Neutral-Higgsino mass splitting in MSSM in the same-sign (left) and opposite-sign (right) gaugino region
for $\mu=1.1\,\mathrm{TeV}$ and $\tan\beta=10$.  The color scale shows
$\log_{10}(\delta /\mathrm{keV})$, the orange band denotes
$300\leq\delta \leq500\,\mathrm{keV}$, and the red curve marks
$\delta =375.7\,\mathrm{keV}$.}
\label{fig:mssm_same_sign}
\end{figure*}
\subsection{MSSM: gaugino-induced Majorana splitting}

Away from Higgsino--gaugino degeneracies, expansion in the electroweak
mixing gives
\begin{equation}
\delta \simeq
m_Z^2\left|
s_W^2\frac{M_1+\mu\sin2\beta}{M_1^2-\mu^2}
+
c_W^2\frac{M_2+\mu\sin2\beta}{M_2^2-\mu^2}
\right|,
\label{eq:mssm_general_split}
\end{equation}
or equivalently
\begin{align}
\delta \simeq\frac{m_Z^2}{2}\bigg|
&s_W^2
\left(
\frac{1+\sin2\beta}{M_1-\mu}
+
\frac{1-\sin2\beta}{M_1+\mu}
\right)
\nonumber\\
+{}&c_W^2
\left(
\frac{1+\sin2\beta}{M_2-\mu}
+
\frac{1-\sin2\beta}{M_2+\mu}
\right)
\bigg|.
\label{eq:mssm_fullsplit}
\end{align}
These expressions retain the relative signs of $M_1$, $M_2$, and $\mu$
and agree with the heavy-gaugino effective description of
Ref.~\cite{Nagata:2014aoa}.  For
$|M_{1,2}|\gg|\mu|$ they reduce to
\begin{equation}
 \delta \simeq
 m_Z^2
 \left|
 \frac{s_W^2}{M_1}
 +
 \frac{c_W^2}{M_2}
 \right|.
 \label{eq:mssm_deep_split}
\end{equation}

For $M_1M_2>0$, the leading bino and wino contributions add.  As an
illustrative example, take
\begin{equation}
 M_2=2M_1,
 \qquad
 \mu=1.1\,\mathrm{TeV}.
\end{equation}
Then, independently of the value of $\tan\beta$, 
$\delta =375.7~{\rm keV}$ corresponds to
\begin{equation}
 M_1\simeq13~{\rm PeV},
 \qquad
 M_2\simeq27~{\rm PeV}.
\end{equation}
A generic same-sign spectrum therefore places the electroweak gauginos at
a very high scale for a splitting of a few hundred keV.

For $M_1\cdot M_2<0$, the leading contributions can instead be canceled.  In the
heavy-gaugino limit the cancellation occurs along
\begin{equation}
 \frac{M_2}{M_1}
 =
 -\frac{c_W^2}{s_W^2}
 =
 -\cot^2\theta_W
 \simeq-3.325.
 \label{eq:gaugino_cancel}
\end{equation}
The individual gaugino masses can then be substantially lower than in the
same-sign case, although the resulting sub-MeV splitting becomes more
sensitive to subleading terms and radiative corrections.  We quantify the
cancellation by
\begin{equation}
 {\cal C}
 =
 \frac{|t_1+t_2|}{|t_1|+|t_2|},
 \label{eq:cancellation_measure}
\end{equation}
with
\begin{equation}
 t_1=
 s_W^2\frac{M_1+\mu\sin2\beta}{M_1^2-\mu^2},
 \qquad
 t_2=
 c_W^2\frac{M_2+\mu\sin2\beta}{M_2^2-\mu^2}.
\end{equation}
Values ${\cal C}\ll1$ identify a cancellation rather than an additional
decoupling limit.

Figure~\ref{fig:mssm_same_sign} shows the same-sign and opposite-sign regions
for $\mu=1.1\,\mathrm{TeV}$ and $\tan\beta=10$.  The splittings are obtained
by diagonalizing the complete neutralino mass matrix in
Eq.~\eqref{eq:uvmatrix}; Eqs.~\eqref{eq:mssm_general_split} and
\eqref{eq:mssm_deep_split} describe the corresponding limiting behavior.

The MSSM therefore realizes a sub-MeV neutral-Higgsino splitting in two
distinct ways: generic gaugino decoupling at the multi-PeV scale, or a
cancellation between bino and wino contributions for opposite signed
masses.  The latter permits substantially lower individual gaugino scales
but is correspondingly more sensitive to subleading and radiative effects.
Because the recoil-preferred value of $\delta$ depends on the extreme
high-speed halo population, the inferred gaugino scale or degree of
cancellation inherits the astrophysical dependence found in
Sec.~\ref{sec:inelastic}.


\subsection{NMSSM: singlino-induced Higgsino splitting}

A singlino provides a second source of the Majorana perturbation.  
The singlino mixes with the neutral Higgsinos through
$-\lambda v_u$ and $-\lambda v_d$.  When the gaugino contribution is small
compared with the target splitting and the singlino is not nearly degenerate
with either Higgsino eigenstate, perturbation theory gives
\begin{equation}
 \delta^{(\widetilde S)}
 \simeq
 \left|
 \frac{
 \lambda^2v^2
 (m_{\widetilde S}-\mu_{\rm eff}\sin2\beta)}
 {m_{\widetilde S}^2-\mu_{\rm eff}^2}
 \right|.
 \label{eq:uvssplit}
\end{equation}
This expansion requires
$|\lambda v|\ll|m_{\widetilde S}\pm\mu_{\rm eff}|$.  For a
parametrically heavy singlino,
\begin{equation}
 \delta^{(\widetilde S)}
 \simeq
 \frac{\lambda^2v^2}{|m_{\widetilde S}|}.
 \label{eq:heavySinglinoSplit}
\end{equation}

The recoil-preferred splittings can therefore be generated with weak
Higgsino--singlino mixing.  For example, with
$\lambda=0.02$ and $\tan\beta=10$, the positive-singlino solutions occur
near $35~{\rm TeV}$ and $32~{\rm TeV}$ for the
$500~{\rm GeV}$ and $1.1~{\rm TeV}$ recoil targets, respectively.  In the
minimal $Z_3$ theory, however, Eq.~\eqref{eq:z3masses} correlates
$m_{\widetilde S}$ with $\kappa/\lambda$, so a heavy singlino at fixed
$\mu_{\rm eff}$ can require sizeable $\kappa$.

This correlation is a property of the $Z_3$ construction rather than of
inelastic Higgsino dark matter itself.  In the general NMSSM,
supersymmetric singlet and doublet mass terms allow
\begin{equation}
 \mu_{\rm tot}=\mu+\lambda s,
 \qquad
 m_{\widetilde S}=\mu_S+2\kappa s ,
 \label{eq:gnmssmMasses}
\end{equation}
so the singlino scale can vary independently of the ratio
$2\kappa/\lambda$.

For moderate or large $\lambda$, the same singlet sector can also open
annihilation channels such as
\begin{equation}
 \chi\chi\rightarrow h_s a_s,\qquad
 h_sh_s,\qquad
 a_sa_s .
\end{equation}
These channels supplement the standard electroweak annihilation modes, such as \(W^+W^-\) and \(ZZ\). In a standard thermal history, they can allow a Higgsino heavier than the canonical \(1.1~{\rm TeV}\) value to reproduce the observed relic abundance. Conversely, for a Higgsino mass near \(1.1~{\rm TeV}\), the additional annihilation channels increase the total annihilation rate and therefore reduce the thermal relic abundance below the observed dark-matter density. They can also modify the annihilation branching fractions relevant for solar-neutrino searches. These possibilities are not included in the recoil analysis presented in this work and we reserve this for a future study.


\subsection{Solar-capture neutrino constraints}
\label{sec:solar}

Solar capture provides an important complementary probe of endothermic
Higgsino dark matter.  Dark-matter particles falling into the solar
gravitational potential can reach speeds of order
$1400~{\rm km\,s^{-1}}$ in the solar interior, allowing inelastic
up-scattering on heavy elements for splittings beyond the terrestrial
kinematic reach~\cite{Nussinov:2009ft, Menon:2009qj}.  After capture, subsequent scattering can reduce the
orbital energy of the dark-matter population and enhance its annihilation
rate in the solar interior.  For a nearly pure Higgsino, annihilation
predominantly into $W^+W^-$ and $ZZ$ produces energetic neutrinos, making
IceCube and Super-Kamiokande sensitive to this scenario
~\cite{Pospelov:2026ewn,Bose:2026ndd,DiMauro:2026dqp,Nguyen:2026lui}.

For a full-density thermal Higgsino with
$m_\chi\simeq1.08~{\rm TeV}$,
Ref.~\cite{Pospelov:2026ewn} finds that tree-level inelastic capture
constrains splittings below approximately
$\delta\simeq506~{\rm keV}$.  Once the inelastic channel becomes
inefficient, loop-induced elastic scattering can continue to reduce the
orbital size of the captured population, including through SD scattering
on solar hydrogen.  Under the post-capture evolution assumed in that
analysis, this yields the stronger approximate requirement
$\delta\gtrsim566~{\rm keV}$.  Independent analyses obtain constraints at
a similar scale: Ref.~\cite{Bose:2026ndd} finds strong sensitivity below
approximately $557~{\rm keV}$, while
Refs.~\cite{DiMauro:2026dqp,Nguyen:2026lui} likewise find strong
constraints in the few-hundred-keV region relevant for the LZ recoil.

The precise numerical values of these constraints are not determined by
inelastic kinematics alone.  They also depend on the capture rate and on
the subsequent evolution of the captured population, including orbital
damping, thermalization, and the approach to capture--annihilation
equilibrium.  Once inelastic up-scattering becomes kinematically closed,
further energy loss proceeds through much weaker elastic interactions.
If these interactions are sufficiently suppressed, the captured
population can remain on extended orbits for longer times, making the
thermalization history and possible planetary perturbations relevant.
Uncertainties in the solar heavy-element abundances and in neutrino
propagation through the Sun provide additional systematics.  The quoted
$\delta\gtrsim566~{\rm keV}$ condition should therefore be interpreted as
a strong constraint on the canonical full-density thermal Higgsino under
the assumptions of the corresponding solar-capture calculation, rather
than as a model-independent lower bound on the neutral-state splitting.
Recent discussions of these assumptions and possible nonminimal
realizations are given in Ref.~\cite{Bisal:2026khf}.

For the fiducial SHM defined in Sec.~\ref{sec:conventions}, our
Helm-fiducial $1.1~{\rm TeV}$ recoil solution has
$\delta\simeq375.7~{\rm keV}$.  Under the canonical full-density
thermal-Higgsino assumptions, the solar-capture analyses instead
require splittings in the approximate $0.51$--$0.57~{\rm MeV}$ range
or larger.  The tension arises from the combination of the full local
Higgsino density, the electroweak-strength off-diagonal $Z$ coupling,
and the hard annihilation final states, rather than from an
off-diagonal $Z$ interaction by itself.  The canonical full-density
thermal-Higgsino interpretation of the fiducial-SHM recoil solution is
therefore strongly disfavored by the solar-neutrino constraints.  This
conclusion applies specifically to the minimal thermal-Higgsino
realization and should not be interpreted as a model-independent
exclusion of all realizations with an off-diagonal $Z$ interaction.

The Helm-fiducial $500~{\rm GeV}$ benchmark in
Table~\ref{tab:inelastic} has $\delta=347.7~{\rm keV}$.  Standard thermal
freeze-out at this mass underproduces the observed dark-matter abundance,
so a full-density interpretation requires nonthermal production.  A
nonthermal production history, however, does not by itself reduce solar
capture if the Higgsino still constitutes essentially all of the local
dark matter and retains its electroweak transition coupling.  The
numerical $\delta\gtrsim566~{\rm keV}$ requirement obtained near
$m_\chi\simeq1.08~{\rm TeV}$ should not be transferred directly to the
$m_\chi=500~{\rm GeV}$ case.  Nevertheless, the dedicated full-density
nonthermal analysis of Ref.~\cite{Bose:2026ndd} places this benchmark in
the solar-neutrino-constrained region under the assumptions adopted
there. 

More general realizations can modify both the capture and annihilation
stages of the solar signal.  Mixing with additional neutral states can
modify, and in particular reduce, the off-diagonal $Z$ coupling and hence
the inelastic capture rate, while changing the dark-matter mass modifies
the reduced mass and therefore the region of the Sun in which
up-scattering is kinematically allowed.  Suppressed elastic scattering
can also slow the post-capture thermalization once the inelastic channel
has closed, while additional annihilation channels can alter the
high-energy neutrino yield.  These effects are inherently model
dependent and are not included in our recoil benchmarks.  A quantitative
test of a specific ultraviolet realization therefore requires the solar
capture rate, post-capture orbital evolution, annihilation branching
fractions, and neutrino propagation to be evaluated consistently.

The terrestrial high-velocity-tail dependence discussed in
Sec.~\ref{sec:inelastic} does not by itself eliminate the solar-neutrino
tension.  Modifications of the extreme halo tail can substantially shift
the value of $\delta$ preferred by terrestrial scattering, whereas inside
the Sun the large gravitational acceleration makes the capture kinematics
less sensitive to modest changes in the asymptotic halo velocity.  For
the LMC-influenced simulated halo considered in
Ref.~\cite{Ghosh:2026txe}, the largest splitting capable of producing one
expected event in the $200$--$300~{\rm keV}$ LZ recoil interval is
$\delta\simeq514~{\rm keV}$.  This remains below the
$\sim566~{\rm keV}$ value obtained near
$m_\chi\simeq1.08~{\rm TeV}$ under the canonical thermal-Higgsino
solar-capture assumptions of Ref.~\cite{Pospelov:2026ewn}.  Thus, the
astrophysical variations considered so far do not by themselves
reconcile the minimal thermal-Higgsino interpretation with the quoted
solar-neutrino constraints.  A quantitative Milky-Way--LMC assessment of
our benchmarks would require a specified simulation-derived
Galactic-frame phase-space distribution, as emphasized in
Sec.~\ref{sec:inelastic}.

\section{Conclusions}
\label{sec:conclusions}

We have investigated two supersymmetric dark-matter interpretations of the
recently reported $248~\mathrm{keV}$ LZ nuclear-recoil candidate, arising
from distinct couplings to the $Z$ boson. In the elastic interpretation, a
bino- or singlino-like Majorana neutralino scatters through a diagonal axial
$Z$ coupling. In the inelastic interpretation, a pseudo-Dirac Higgsino
undergoes endothermic scattering through an off-diagonal vector $Z$ current.
Although both mechanisms can populate the candidate region, they predict
qualitatively different recoil spectra away from it.

For the elastic interpretation, we first defined the recoil-level
reference normalization $N^{\rm proxy}_{100-270}=1$.  At
$m_\chi=500~{\rm GeV}$ with the GCN5082 response this corresponds to
$\kappa_A=|X|\simeq0.013$ and $R_{\rm low}\simeq5.3$.
The operator-specific comparison with the public LZ
$L_{15}^{v}$ limits gives an approximate observed upper limit
$|X|\lesssim0.012$ near $m_\chi\simeq505~{\rm GeV}$, so the unit-yield
reference lies slightly above this limit.  We therefore use
$|X|\simeq0.007$ as a representative benchmark normalization below
the observed endpoint; the public release does not determine a
statistically preferred value of $X$.
For the explicit point $m_\chi=505~{\rm GeV}$ and $X=0.0069$, the
GCN5082 calculation gives $1.47$ proxy recoils in
$5.4$--$100~{\rm keV}$, $0.277$ in $100$--$270~{\rm keV}$, and
$1.75$ in total below $270~{\rm keV}$.  The corresponding simple
signal-only probability for at least one high-recoil proxy event is
$24.2\%$.  These proxy yields and Poisson probabilities are recoil-level
diagnostics, not a reconstruction of the LZ detector likelihood.

For the inelastic interpretation, an electroweak-strength transition requires
a neutral-state splitting of a few hundred keV. In the baseline annually
averaged SHM, fixing one raw recoil in $215$--$269.9~\mathrm{keV}$ gives
$\delta=347.7~\mathrm{keV}$ for $m_\chi=500~\mathrm{GeV}$ and
$\delta=375.7~\mathrm{keV}$ for $m_\chi=1.1~\mathrm{TeV}$ with the fiducial
Helm response. The $500~\mathrm{GeV}$ solution predicts about $94\%$ fewer
raw recoils in $350$--$590~\mathrm{keV}$ than the $1.1~\mathrm{TeV}$
solution, while the independently matched Vietze response gives a reduction
of about $90\%$. This ordering is not halo independent: changes in the
extreme high-speed tail can substantially modify, and within the deformations
considered here even reverse, the relative high-recoil populations. Under
the canonical full-density thermal-Higgsino assumptions, the $1.1~\mathrm{TeV}$
realization is also strongly constrained by solar-neutrino searches.

The elastic recoil requirement fixes the Higgsino asymmetry rather than the
total Higgsino fraction. In the bino--Higgsino MSSM, the axial requirement
alone permits a broad range of bino--Higgsino separations, whereas imposing
SI suppression selects a more restricted low-$\tan\beta$ branch. In the
$Z_3$ NMSSM, the additional tree-level Higgs quartic and singlet sector make
this region easier to realize. Our representative benchmark has
$m_{\chi_1^0}=505.30~{\rm GeV}$ and $X=0.0069$, with $\Omega_\chi h^2\simeq0.123$ and a strongly suppressed
tree-level SI amplitude, lying very close to an NMSSM blind spot. The thermal relic
abundance is controlled primarily by annihilation through a
singlet-dominated pseudoscalar whose pole lies slightly above the
neutralino-pair threshold. Thermal velocities at freeze-out allow the
neutralinos to access the resonance efficiently, whereas the much smaller
velocities at late times move the annihilation away from the pole. The present-day annihilation rate is therefore suppressed relative to its
freeze-out value, allowing the benchmark to remain consistent with current
indirect-detection constraints.

For the inelastic realization, the off-diagonal $Z$ current arises naturally
for a pseudo-Dirac Higgsino pair, while the required sub-MeV Majorana splitting
can be generated by electroweak-gaugino mixing in the MSSM or by singlino
mixing in the NMSSM. The supersymmetric origin of the splitting therefore
connects the recoil phenomenology to otherwise much higher mass scales and,
in the gaugino-induced case, to the relative signs and possible cancellations
among the electroweak gaugino contributions.

Taken together, the recoil correlations away from the $248~\mathrm{keV}$
candidate provide the most direct distinction between the two mechanisms in
our recoil-level analysis. Elastic axial scattering predicts a correlated
lower-energy population, whereas endothermic scattering suppresses this
continuation and shifts the discriminating information toward a mass- and
halo-dependent high-recoil tail. A quantitative confrontation with the LZ
data requires the full $(S1c,S2c)$ detector response, the low-energy
background likelihood, and the nuclear-recoil acceptance in the high-energy
sideband. Additional LZ exposure and improved sensitivity to spin-dependent
scattering can therefore test these complementary spectral predictions.

~\\
{\bf Acknowledgments.} 
SR would like to acknowledge the many useful discussions that took place at Fermilab. PS would like to thank F.~Elahi for discussions and collaboration on a related topic. 
The work of SR at the University of Cincinnati has been supported by the DOE grants DE-SC0011784, DE-SC0026301, and the NSF grant  OAC-2411215.
 SR was also supported by the U.S.~Department of Energy under contracts No.\ DEAC02-06CH11357 at the Argonne National Laboratory. SR would like
to thank the University of Chicago and Fermilab where a significant
part of this work was carried out.  The work of CW\ at the University of Chicago has been supported by the DOE grant DE-SC0013642,
 and by a Distinguished Visiting Research Chair  at the Perimeter Institute for Theoretical Physics. Work in Mainz is supported by the Cluster of Excellence ``PRISMA$^{++}$'' funded by the German Research Foundation (DFG) within the German Excellence Strategy (Project No. 390831469).

\appendix
\section{Nuclear-response conventions}
\label{app:nuclear}

\subsection{Coherent inelastic transition}

For the endothermic process $\chi_1 A\to\chi_2 A$, the xenon nucleus
remains in its ground state and the dominant vector--vector contribution is
coherent.  For isotope $A$, with weak charge
$ Q_W^{(A)}
 =
 N_A-\left(1-4s_W^2\right)Z_A$,
we define
\begin{align}
 \sigma_A^0
 &=
 \sigma_n^{\rm inel}
 \frac{\mu_A^2}{\mu_n^2}
 \left[Q_W^{(A)}\right]^2 ,
 \notag\\
 \frac{d\sigma_A^{\rm coh}}{dE_R}
 &=
 \frac{m_A\sigma_n^{\rm inel}}
      {2\mu_n^2\beta^2}
 \left[Q_W^{(A)}\right]^2
 F_A^2(q),
 \qquad
 F_A(0)=1 ,
 \label{eq:coherentFF}
\end{align}
where $\beta=v/c$ and
$q=\sqrt{2m_AE_R}$.  The quantity $\sigma_A^0$ fixes the
zero-momentum normalization of the coherent matrix element and should not
be interpreted as a physical zero-threshold inelastic cross section; the
endothermic threshold is imposed separately through
$v_{\min}(E_R)$.

For the fiducial coherent response, we use the Helm nuclear form factor in the Lewin–Smith parametrization~\cite{Lewin:1995rx}. The Helm model approximates the nuclear density as a uniform sphere convolved with a Gaussian surface profile, providing a simple analytic description of both the finite nuclear radius and surface diffuseness. The resulting form factor is
\begin{align}
 F_{A,\rm Helm}^2(q)
 &=
 \left[
 \frac{3j_1(qr_{n,A}/\hbar c)}
      {qr_{n,A}/\hbar c}
 \right]^2
 e^{-(qs/\hbar c)^2},
 \notag\\
 r_{n,A}^2
 &=
 c_A^2+\frac{7\pi^2a^2}{3}-5s^2,
 \notag\\
 c_A
 &=
 (1.23A^{1/3}-0.60)\,{\rm fm},
 \notag\\
 a&=0.52\,{\rm fm},~~~
 s=0.9\,{\rm fm}.
 \label{eq:helmFF}
\end{align}
Here, $c_A$ characterizes the effective nuclear radius, $a$ controls the diffuseness of the underlying density distribution, and $s$ parametrizes the Gaussian surface thickness.
Here
\begin{equation}
 j_1(x)=\frac{\sin x}{x^2}-\frac{\cos x}{x},
\end{equation}
with the analytic $q\to0$ limit understood.

To assess the dependence of the high-recoil spectrum on the coherent
finite-momentum response, we also use the elastic ground-state
spin-independent structure factors of Ref.~\cite{Vietze:2014vsa}, normalized to
their zero-momentum values,
\begin{equation}
 F_{A,\rm V}^2(q)
 =
 \frac{S_S^{(A)}(q)}
      {S_S^{(A)}(0)} .
 \label{eq:vietzeFF}
\end{equation}
The fits and oscillator lengths are taken from Table~II of
Ref.~\cite{Vietze:2014vsa}.  For $^{131}$Xe we retain the complete ground-state
fit, including the $L=2$ contribution.  The isotopes $^{124}$Xe and
$^{126}$Xe are not included in the Vietze fits; for these isotopes we use
the Helm form factor.  Their combined natural abundance is approximately
$0.184\%$.  Nuclear-excitation structure factors are not included.

The Helm and Vietze prescriptions provide two independent descriptions of the coherent finite-momentum nuclear response: the former is an analytic density-based form factor, while the latter is obtained from shell-model nuclear structure factors. They are not identified
with the nuclear implementation used in the LZ analysis.  LZ evaluates its
recoil spectra with \texttt{WimPyDD}~\cite{Jeong:2021bpl}, using one-body
density matrices developed for \texttt{DMFormFactor-v6} together with the
modifications described in the extended-energy analysis
~\cite{LZ:2026axp}.  Following Ref.~\cite{Rodd:2026tyn}, we use the Vietze
response as an independent shell-model shape comparison.  The Helm and
Vietze benchmarks are normalized separately to the same recoil condition;
their difference therefore quantifies the dependence of our recoil-level
results on the adopted coherent form factor and is not treated as a
statistical uncertainty.

\subsection{Natural xenon and elastic axial responses}

Natural isotope atom fractions $x_A$ and atomic masses
$m_A^{\rm atom}$ are taken from NIST~\cite{NISTXe}.  The corresponding
mass fractions used for target counting are
\begin{equation}
 \xi_A
 =
 \frac{x_A m_A^{\rm atom}}
      {\sum_B x_B m_B^{\rm atom}} .
\end{equation}
Nuclear masses are obtained by subtracting the electron rest masses
from the atomic masses, neglecting electronic binding energies.  Each
isotope retains its own nuclear mass and kinematics in the recoil
calculation.

The elastic axial calculation uses the Berkeley GCN5082 one-body
density matrices~\cite{Hoferichter:2020osn, Berkeley}, evaluated in the NREFT framework
~\cite{Fitzpatrick:2012ix,Anand:2013yka} with \texttt{dmscatter}~\cite{Gorton:2022eed}.
With $q=\sqrt{2m_AE_R}$ in GeV and the oscillator length $b$ in fm,
we use
\begin{equation}
 y=\left(\frac{qb}{2\hbar c}\right)^2,
~~
 u=2y,~~
 b=
 \sqrt{\frac{41.467}
 {45A^{-1/3}-25A^{-2/3}}}\ {\rm fm}.
\end{equation}
The odd-mass isotopes $^{129}$Xe and $^{131}$Xe generate the spin
responses relevant to $\mathcal O_4$, while even-even isotopes can
contribute to the coherent $M$ response generated by
$\mathcal O_8$.  The calculation retains the absolute nuclear
responses, proton--neutron interference, and all operators generated
by the adopted leading one-body $Z$ current; Helm and Vietze
spin-independent form factors are not used for the elastic axial
response.

We adopt GCN5082 as the fiducial elastic nuclear response for xenon.
GCN5082 is a shell-model interaction defined in the valence space between
the $50$ and $82$ shell closures and has been used extensively in
calculations of xenon nuclear responses.  As an alternative
nuclear-structure input, we also evaluate the AFH response
functions~\cite{Anand:2013yka}.  For $m_\chi=500~{\rm GeV}$, imposing
$N^{\rm proxy}_{100-270}=1$~gives
\begin{align}
 &\text{GCN5082:}
 \notag\\[-2pt]
 &\kappa_A \simeq 0.013,\quad
 \sigma_{\rm SD}^n \simeq 5.21\times10^{-42}\,{\rm cm}^2,\quad
 R_{\rm low} \simeq 5.3,
 \notag\\[5pt]
 &\text{AFH:}
 \notag\\[-2pt]
 &\kappa_A \simeq 0.021,\quad
 \sigma_{\rm SD}^n \simeq 1.36\times10^{-41}\,{\rm cm}^2,\quad
 R_{\rm low} \simeq 7.2 \nonumber.
 \label{eq:elastic_nuclear_comparison}
\end{align}
The nuclear-response choice therefore affects both the coupling required
to reproduce the high-recoil normalization and its translation into
$\sigma_{\rm SD}^n$; for these two calculations, the inferred
cross sections differ by a factor of approximately $2.6$.  The spectral
conclusion is less sensitive to this choice: both responses predict
several correlated lower-energy recoils when normalized to the
high-recoil population.  We therefore use GCN5082 as our fiducial
response and AFH to illustrate the dependence on the nuclear-structure
input.  Their difference is not interpreted as a statistical uncertainty.

\begin{table}
\begin{ruledtabular}
\begin{tabular}{rlrrr}
\multicolumn{5}{c}{Coherent inelastic transition} \\
$m_\chi$ [GeV] & Form factor & $\delta$ [keV] & $N^{\rm raw}_{269.9-350}$ & $N^{\rm raw}_{350-590}$ \\
500 & Helm & 347.7 & 0.94 & 0.3 \\
500 & Vietze & 342.0 & 1.53 & 0.5 \\
1100 & Helm & 375.7 & 3.3 & 5.4 \\
1100 & Vietze & 368.2 & 4.1 & 4.8 \\
\hline
\multicolumn{5}{c}{Elastic axial interaction} \\
$m_\chi$ [GeV] & Response & $\kappa_A$ & $\sigma_{\rm SD}^n$ & $R_{\rm low}$ \\
500 & AFH & 0.021 & 13.6 & 7.2 \\
500 & GCN5082 & 0.013 & 5.21 & 5.3 \\
\end{tabular}
\end{ruledtabular}
\caption{Coherent-form-factor cross-check and separate elastic-response comparison in the fiducial SHM. All inelastic rows use $\sigma_n^{\rm ref}=7.3\times10^{-39}$ cm$^2$ and are separately matched to one raw recoil in $215$--$269.9$ keV. The elastic rows use one efficiency-proxy recoil in $100$--$270$ keV, with $\sigma_{\rm SD}^n$ in $10^{-42}$ cm$^2$.}
\label{tab:responses_comp}
\end{table}

\section{Relativistic-to-nonrelativistic matching}
\label{app:operators}

\subsection{Quark and nucleon current conventions}

The tree-level Standard Model neutral current is
\begin{equation}
 \mathcal L_Z^q
 =
 \frac{g}{2c_W}Z_\mu
 \sum_q
 \bar q\gamma^\mu
 \left[
 \left(T_3^q-2Q_qs_W^2\right)-T_3^q\gamma^5
 \right]q,
 \label{eq:Zquarkcurrent_app}
\end{equation}
where $T_3^q$ and $Q_q$ denote the weak isospin and electric charge of
quark $q$, respectively.  For momentum transfers satisfying
$q^2\ll m_Z^2$, the $Z$ boson can be integrated out using
$ \frac{G_F}{\sqrt2}
 =
 \frac{g^2}{8m_W^2}$.

With the weak-charge convention of Eq.~\eqref{eq:wN}, the vector
couplings satisfy
$ \frac{f_p}{f_n}
 =
 -(1-4s_W^2)$,
while the nucleon axial charges are defined by
$\langle N|\bar q\gamma^\mu\gamma^5q|N\rangle
 =
 2\,\Delta q^N s_N^\mu$.
These conventions fix the relative signs and normalizations entering
Eq.~\eqref{eq:matching}.

For the axial current we use
$\Delta u^p=0.781$,
 $\Delta d^p=-0.440$ and $\Delta s^p=-0.055$,
in the $\overline{\rm MS}$ scheme at $2~{\rm GeV}$
~\cite{Park:2025rxi}, with the neutron light-quark charges obtained by
$u\leftrightarrow d$.  Their uncertainties are not propagated.  As in
Sec.~\ref{sec:effective}, we define $D_N=\Delta u^N-\Delta d^N-\Delta s^N$.


\subsection{Nonrelativistic matching}

We define $\mathbf q$ as the momentum transferred to the dark-matter
particle, use the metric $(+---)$, and express velocities in units of
$c$.  For scattering from a free nucleon $N=p,n$,
\begin{align}
 \mu_{\chi N}
 &=
 \frac{m_\chi m_N}{m_\chi+m_N},~~~~
 \mathbf v_{\chi N}^{\perp,\rm inel}
 =
 \mathbf v_{\chi N}
 +\frac{\mathbf q}{2\mu_{\chi N}}
 +\frac{\delta\,\mathbf q}{q^2},
 \label{eq:nucleonvelocity}
\end{align}
with energy conservation implying
$ \mathbf v_{\chi N}\cdot\mathbf q = -\frac{q^2}{2\mu_{\chi N}}-\delta$,
consequently, $\mathbf v_{\chi N}^{\perp,\rm inel}\cdot\mathbf q=0$~\cite{Barello:2014uda}.

To distinguish the physical nucleon mass $m_N$ entering
$\mu_{\chi N}$ from the fixed mass used in the NREFT operator
normalization, we define
$m_{\rm ref}\equiv m_n$.
The operators generated by the interactions in
Sec.~\ref{sec:effective} are
\begin{align}
 \mathcal O_1&=1,
 &
 \mathcal O_4&=\mathbf S_\chi\cdot\mathbf S_N,
 \notag\\
 \mathcal O_7&=\mathbf S_N\cdot\mathbf v_{\chi N}^{\perp},
 &
 \mathcal O_8&=\mathbf S_\chi\cdot\mathbf v_{\chi N}^{\perp},
 \notag\\
 \mathcal O_9
 &=
 i\mathbf S_\chi\cdot
 \left(
 \mathbf S_N\times\frac{\mathbf q}{m_{\rm ref}}
 \right),
 &
 \mathcal O_{10}
 &=
 i\mathbf S_N\cdot\frac{\mathbf q}{m_{\rm ref}} .
 \label{eq:nroperators}
\end{align}
Here $\mathbf v_{\chi N}^{\perp}$ denotes
Eq.~\eqref{eq:nucleonvelocity} for inelastic scattering and its
$\delta=0$ limit for elastic scattering.

After dividing the relativistic amplitude by $4m_\chi m_N$, the
point-nucleon coefficients are
\begin{align}
 c_1^N&=f_N,
 &
 c_7^N&=-2B_N,
 \notag\\
 c_9^N&=
 2\frac{m_{\rm ref}}{m_\chi}B_N,
 &
 c_{10}^N&=
 -2i\frac{m_{\rm ref}\delta}{q^2}B_N
 \qquad(\text{inelastic}),
 \label{eq:coeffinel}\\
 c_4^N&=-4C_N,
 &
 c_8^N&=c_9^N=2b_N
 \qquad(\text{elastic}).
 \label{eq:coeffel}
\end{align}
The coefficient $c_{10}^N$ is imaginary because the inelastic
longitudinal contribution has been expressed in the transverse operator
basis; it vanishes in the limit $\delta\to0$.  The quantities
$f_N$, $B_N$, $b_N$, and $C_N$ are defined in
Sec.~\ref{sec:effective}, so the coefficients generated within each
interaction are fixed by a single underlying $Z$ coupling.

For the central inelastic calculation we retain only the dominant
vector--vector contribution, corresponding to $\mathcal O_1$ and the
coherent nuclear response.  The coefficients
$c_7^N$, $c_9^N$, and $c_{10}^N$ specify the remaining leading
point-nucleon terms but are not added to the Helm or Vietze spectra.
For elastic scattering, all nonzero contributions generated by
Eq.~\eqref{eq:coeffel} are retained.


\subsection{Nuclear embedding and response functions}

Embedding the one-nucleon operators in a nucleus separates the target
center-of-mass motion from the internal nucleon velocities
~\cite{Anand:2013yka,Barello:2014uda}.  The transverse
dark-matter--nucleus velocity entering the nuclear response functions is
\begin{align}
 \mathbf v_T^{\perp,\rm inel}
 &=
 \mathbf v_{\chi A}
 +\frac{\mathbf q}{2\mu_A}
 +\frac{\delta\,\mathbf q}{q^2},
 \notag\\
 \mathbf v_{\chi A}\cdot\mathbf q
 &=
 -\frac{q^2}{2\mu_A}-\delta,
 \qquad
 \mathbf v_T^{\perp,\rm inel}\cdot\mathbf q=0 ,
 \label{eq:targetvelocity}
\end{align}
where $\mu_A=\frac{m_\chi m_A}{m_\chi+m_A}$.
It follows that
$ v_T^{\perp2} = v_{\chi A}^2-v_{\min,A}^2$,
with $v_{\min,A}$ fixed by the dark-matter--nucleus kinematics.
This target-level velocity is distinct from the one-nucleon quantity
defined in Eq.~\eqref{eq:nucleonvelocity}.

For each operator we use the proton--neutron coefficient vector
$ \mathbf c_i
 =
 \left(
 c_i^p+c_i^n,\,
 c_i^p-c_i^n
 \right).$
This convention is twice the usual half-sum vector
$ \left(
 \frac{c_i^p+c_i^n}{2},
 \frac{c_i^p-c_i^n}{2}
 \right)$,
so the corresponding response matrices are one quarter of those written
in that convention.

Defining
$ [i,j]_K
 \equiv
 \operatorname{Re}
 \left(
 \mathbf c_i^\dagger W_K\mathbf c_j
 \right)$,
the contributions to the reduced squared nuclear amplitude are
\begin{align}
 P_M
 &=
 [1,1]_M
 +\frac{v_T^{\perp2}}{4}[8,8]_M,
 \notag\\
 P_{\Sigma'}
 &=
 \frac{[4,4]_{\Sigma'}}{16}
 +\frac{q^2[9,9]_{\Sigma'}}{16m_{\rm ref}^2}
 +\frac{v_T^{\perp2}[7,7]_{\Sigma'}}{8},
 \notag\\
 P_{\Sigma''}
 &=
 \frac{[4,4]_{\Sigma''}}{16}
 +\frac{q^2[10,10]_{\Sigma''}}{4m_{\rm ref}^2},
 \notag\\
 P_\Delta
 &=
 \frac{q^2}{4m_{\rm ref}^2}[8,8]_\Delta,
 \notag\\
 P_{\Delta\Sigma'}
 &=
 -\frac{q^2}{4m_{\rm ref}^2}
 [8,9]_{\Delta\Sigma'} .
 \label{eq:nuclearresponses}
\end{align}
The sign of the $\Delta\Sigma'$ interference term is retained explicitly;
it is not treated as an independent positive contribution.

The elastic rate therefore contains both the usual inverse-speed moment
and the target transverse-speed moment associated with the
velocity-dependent operators.  In all cases
$v_{\min,A}$ is evaluated with the dark-matter--nucleus reduced mass.

The reduced spin-averaged nuclear amplitude is
\begin{equation}
 P_A
 =
 \frac{4\pi}{2J_A+1}
 \left(
 P_M+P_{\Sigma'}+P_{\Sigma''}
 +P_\Delta+P_{\Delta\Sigma'}
 \right),
\end{equation}
and the corresponding differential nuclear cross section is
\begin{equation}
 \frac{d\sigma_A}{dE_R}
 =
 \frac{m_A P_A}{2\pi\beta^2}
 (\hbar c)^2,
 \qquad
 \beta=\frac{v_{\chi A}}{c}.
 \label{eq:responsecross}
\end{equation}
Here $P_A$ is the spin-averaged squared amplitude divided by
$(4m_\chi m_A)^2$.  With masses and recoil energies expressed in GeV,
Eq.~\eqref{eq:responsecross} gives $d\sigma_A/dE_R$ per GeV of recoil
energy; the conversion to a rate per keV is made when evaluating
Eq.~\eqref{eq:targetrate}.  These conventions fix the target
normalization without any additional Majorana or isospin factor.
Even-even xenon isotopes have vanishing spin responses but can contribute
to the coherent $M$ response generated by $\mathcal O_8$.

As a free-nucleon normalization check, the axial coefficients satisfy
$ a_N
 =
 \frac{C_N}{2\sqrt{2}G_F}
 =
 -\frac{\kappa_A D_N}{4}$.
For spin-$1/2$ dark matter and nucleons,
$ \left\langle
 (\mathbf S_\chi\cdot\mathbf S_N)^2
 \right\rangle
 =
 \frac{3}{16}$,
which reproduces Eq.~\eqref{eq:sigma}.  No additional Majorana factor is
required.

Finally, for the reference-normalized inelastic calculation we define
\begin{equation}
 \widehat\kappa_V^2
 \equiv
 \frac{\sigma_n^{\rm inel}}{\sigma_n^{\rm ref}},
 \qquad
 \kappa_V
 =
 \widehat\kappa_V
 \sqrt{
 \frac{\sigma_n^{\rm ref}}
 {G_F^2\mu_n^2(\hbar c)^2/(2\pi)}
 } .
 \label{eq:kappaconversion}
\end{equation}
Thus $\widehat\kappa_V$ is a rate-normalization parameter, whereas
$\kappa_V$ is the physical $Z$-boson transition coupling defined in
Sec.~\ref{sec:effective}.  The reference spectra in
Table~\ref{tab:inelastic} correspond to
$\widehat\kappa_V=1$.

\section{Numerical and halo treatment}
\label{app:halo}

The baseline Standard Halo Model (SHM) uses a truncated Maxwellian
Galactic-frame velocity distribution,
\begin{equation}
 f_{\rm gal}(\mathbf v)\propto
 e^{-v^2/v_0^2}\,
 \Theta(v_{\rm esc}-v),
\end{equation}
normalized to unity before multiplication by the local dark-matter
density.  In Galactic coordinates, the laboratory velocity is
\begin{align}
 \mathbf v_{\rm lab}(\phi)
 &=
 (11.1,\,238+12.2,\,7.3)
 \notag\\
 &\quad
 +29.8
 \left(
 \mathbf e_1\cos\phi+\mathbf e_2\sin\phi
 \right)
 \quad [{\rm km/s}],
 \notag\\
 \mathbf e_1&=(0.9941,\,0.1088,\,0.0042),
 \notag\\
 \mathbf e_2&=(-0.0504,\,0.4946,\,-0.8677),
 \label{eq:vlab}
\end{align}
following the circular-orbit prescription of
Ref.~\cite{Baxter:2021pqo}.  The annual rate is evaluated using
$730$ equally spaced orbital phases.  This corresponds to a uniform
annual average rather than the actual time distribution of the LZ live
exposure.  The density and exposure are those defined in
Sec.~\ref{sec:conventions}, and we use
$\sin^2\theta_W=0.23122$ in the weak charge.

The GCN5082 response used for the auxiliary coherent-response check is
tabulated over $0$--$600~{\rm keV}$ with $0.25~{\rm keV}$ spacing and is
refined to $0.05~{\rm keV}$ over $180$--$380~{\rm keV}$.  Its normalized
quadratic form is interpolated with a non-extrapolating PCHIP, while the
GCN spin-response matrices are linearly interpolated on the
$0.25~{\rm keV}$ grid.  The AFH response polynomials are evaluated
directly, and the Helm and Vietze form factors are evaluated analytically.
Recoil-energy integration with $0.1~{\rm keV}$ spacing was checked against
a $0.05~{\rm keV}$ grid.  The GCN5082 density-matrix inputs are taken from
the Berkeley data set~\cite{Hoferichter:2020osn,Berkeley}; detailed input
provenance is given in the reproducibility supplement.

The normalization and units were cross-checked against the corresponding
\texttt{WimPyDD}~\cite{Jeong:2021bpl} and AFH conventions.  The analytic
truncated-Maxwellian velocity integrals were independently checked against
direct numerical integration.  The annual average, energy integration,
matching roots, isotope sums, and response interpolation were also tested
under numerical refinement.  These checks establish numerical convergence
but do not include nuclear-response or detector-model systematics.

The high-speed-tail study of Sec.~\ref{sec:inelastic} uses normalized
deformations of the detector-frame speed distribution evaluated within the
same annual-averaging prescription.  These deformations are used only to
diagnose sensitivity to the extreme high-speed population; they are neither
specific Galactic models nor statistical uncertainty bands.  A quantitative
Milky-Way--LMC prediction would instead require a specified
simulation-derived Galactic-frame phase-space distribution, including its
normalization, directional structure, and time dependence.

\section{Efficiency proxy and detector response}
\label{app:detector}

The recoil-efficiency proxy is obtained by digitizing the total
nuclear-recoil efficiency shown in Fig.~S2 of
Ref.~\cite{LZ:2026axp}.  The digitized efficiency is evaluated using a shape-preserving
piecewise-cubic interpolation over the energy range shown in
Fig.~S2, extending to approximately $300~{\rm keV}$, without
extrapolation beyond this range.  The corresponding digitized values
are provided in the reproducibility supplement.  The value
$269.9~{\rm keV}$ is the high-energy $50\%$ efficiency crossing and is not
the endpoint of the digitized curve.

This efficiency is used only in the proxy quantities
$N^{\rm proxy}$ and $R_{\rm low}$.  The recoil populations in
$269.9$--$350~{\rm keV}$ and $350$--$590~{\rm keV}$ are quoted as raw
populations and are not efficiency weighted.  In particular, the reported
one-dimensional efficiency does not determine the nuclear-recoil acceptance
in the detector-space high-energy sideband and is not extrapolated into that
region.

The efficiency proxy also does not specify the full detector response
$P(S1c,S2c\mid E_R)$, the background probability densities, or the nuisance
parameters entering the LZ likelihood.  It therefore cannot reconstruct the
multidimensional LZ likelihood or its best-fit $\mathcal O_4$ normalization.
Accordingly, $N^{\rm proxy}$ and $R_{\rm low}$ are used as
efficiency-weighted recoil-level diagnostics rather than detector-level
event predictions.

As a numerical cross-check, we approximately reproduced the
recoil-to-sideband prescription of Ref.~\cite{Rodd:2026tyn}.  Assuming unit
sideband acceptance gives a comparable matched splitting and sideband-count
scale.  This check is not used to define any benchmark in the present
analysis.

\section{Finite-momentum current limitations}
\label{app:finiteq}

The recoil calculation includes finite nuclear size through the coherent
form factors in the inelastic analysis and the finite-momentum NREFT nuclear
responses in the elastic analysis.  The underlying nucleon weak currents,
however, are treated at leading point-nucleon order.

At the momentum transfers relevant to the high-recoil region, a more
complete treatment would include the momentum dependence of the vector and
axial nucleon form factors, weak-magnetism and induced-pseudoscalar terms,
and consistent two-body axial currents
~\cite{Bishara:2017pfq,Klos:2013rwa}.  In particular, longitudinal
pion-pole contributions need not follow a naive momentum-suppression
estimate~\cite{Bishara:2017pfq}.  Incorporating these effects consistently
requires matching the finite-momentum hadronic currents onto the
corresponding nuclear responses and lies beyond the scope of the present
analysis.

We therefore do not apply an ad hoc correction or assign a theory
uncertainty to the omitted current effects.  The percent-level
non-$\mathcal O_4$ contribution quoted in the main text refers only to the
additional operators generated within the adopted leading one-body
point-nucleon $Z$ current.  It should not be interpreted as an estimate of
the full hadronic or nuclear-response uncertainty.

\section{Additional NMSSM aspects of the elastic interpretation}
\label{app:elastic_nmssm}

In this appendix we collect additional aspects of the singlino-like elastic
realization and the scalar-sector correlations that make this branch more
restrictive in the $Z_3$-invariant NMSSM. The benchmark-specific thermal and
indirect-detection properties are discussed directly in the main text.

\subsection{Singlino-like dark matter and scalar-sector correlations}
\label{app:singlino}

The explicit $Z_3$-NMSSM benchmark presented in the main text follows the
bino-like branch.  A singlino-like LSP nevertheless provides a second
realization of elastic axial-$Z$ scattering.  With both electroweak
gauginos decoupled, the Higgsino components of a singlino-like LSP
satisfy~\cite{Badziak:2015exr}
\begin{align}
 \frac{N_{13}}{N_{15}}
 &=
 \frac{\lambda v
 \left(
 m_\chi\sin\beta-\mu_{\rm eff}\cos\beta
 \right)}
 {\mu_{\rm eff}^2-m_\chi^2},
 \notag\\
 \frac{N_{14}}{N_{15}}
 &=
 \frac{\lambda v
 \left(
 m_\chi\cos\beta-\mu_{\rm eff}\sin\beta
 \right)}
 {\mu_{\rm eff}^2-m_\chi^2}.
 \label{eq:singlinoHiggsinoComponents}
\end{align}
The resulting Higgsino asymmetry is
\begin{equation}
 X=
 |N_{15}|^2
 \frac{\lambda^2v^2\cos2\beta}
 {\mu_{\rm eff}^2-m_\chi^2}.
 \label{eq:uvsinglino}
\end{equation}
As in the bino--Higgsino case, the leading axial $Z$ coupling vanishes at
$\tan\beta=1$.  Away from this limit, a controlled Higgsino admixture can
generate the $|X|\sim10^{-2}$ required by the recoil interpretation.

If the SM-like Higgs boson dominates the SI amplitude and the
nonstandard $CP$-even states are effectively decoupled, the
singlino--Higgsino blind spot occurs approximately at
\begin{equation}
 \frac{m_\chi}{\mu_{\rm eff}}
 \simeq
 \sin2\beta .
 \label{eq:singlinoHiggsinoBlind}
\end{equation}
This relation differs in both origin and sign structure from the
bino--Higgsino blind spot~\cite{Badziak:2015exr,Badziak:2015nrb}.
A nearby bino opens additional cancellation possibilities.  When
higher-order mixing terms are subleading, the corresponding SI
blind-spot condition becomes approximately~\cite{Roy:2024yoh}
\begin{equation}
 \left(
 m_\chi+
 \frac{g_1^2v^2}{M_1-m_\chi}
 \right)
 \frac{1}{\mu_{\rm eff}\sin2\beta}
 \simeq1 .
 \label{eq:singlinoBinoBlind}
\end{equation}
The bino contribution can therefore compensate the
singlino--Higgsino contribution and suppress the SI amplitude while
retaining a nonzero Higgsino asymmetry.  The NMSSM also admits
additional SD blind spots in which cancellations among the neutralino
components drive $X\to0$; such regions are not relevant to the present
recoil interpretation.

The relic abundance need not be controlled by the same Higgsino
admixture.  Nearby bino- or Higgsino-like neutralinos can participate in
coannihilation, while singlet-like scalar or pseudoscalar resonances can
enhance the annihilation rate
~\cite{Badziak:2015exr,Roy:2024yoh}.  Thus the recoil requirement by
itself does not determine the thermal abundance.

A more restrictive feature of the singlino-like realization arises from
the scalar sector of the $Z_3$-invariant NMSSM.  Neglecting
doublet-dependent contributions, singlet--doublet mixing, and loop
corrections, the singlet-sector diagonal entries satisfy
\begin{align}
 m_{\widetilde S}
 &=
 2\kappa s,
 \notag\\
 m_{h_s}^2
 &\simeq
 \kappa s\left(A_\kappa+4\kappa s\right),
 \notag\\
 m_{a_s}^2
 &\simeq
 -3\kappa s A_\kappa ,
 \label{eq:singletSectorMasses}
\end{align}
which implies
\begin{equation}
 m_{\widetilde S}^2
 \simeq
 m_{h_s}^2+\frac13m_{a_s}^2 .
 \label{eq:singletMassCorrelation}
\end{equation}
Interpreting these diagonal entries directly as physical masses requires
small scalar mixing, but Eq.~\eqref{eq:singletMassCorrelation} makes
explicit that the singlino mass is correlated with the singlet-scalar
spectrum.  Bringing the singlino to the LSP mass scale therefore tends
to bring at least part of the singlet scalar sector into a
phenomenologically relevant mass range.

This correlation is particularly important for SI direct detection.
Even when the approximate singlino--Higgsino blind-spot condition in
Eq.~\eqref{eq:singlinoHiggsinoBlind} is satisfied, a relatively light
singlet-like $CP$-even scalar with a small doublet admixture can regenerate
a sizeable scalar amplitude.  Conversely, destructive interference
between the SM-like Higgs boson and a singlet-like $CP$-even scalar can
itself produce an SI blind spot
~\cite{Badziak:2015exr,Badziak:2015nrb}.  A singlino-like interpretation
therefore requires the complete $CP$-even scalar amplitude to be evaluated,
rather than relying only on the simple blind-spot relation.

The simultaneous realization of the required Higgsino asymmetry,
suppressed SI scattering, and the observed thermal abundance is therefore
more correlated in the singlino-like branch.  The bino-like realization
used for our explicit benchmark offers greater parametric freedom:
$M_1$ can set the LSP mass largely independently of the singlet-scalar
spectrum, while the singlet sector can separately participate in the SI
cancellation and provide a near-resonant annihilation channel.  This is
the reason we use the bino-like branch to provide an explicit numerical
realization of the elastic recoil interpretation.

The additional supersymmetric mass parameters of the general NMSSM relax
these $Z_3$ correlations, allowing the singlino mass, scalar spectrum,
and annihilation channels to vary more independently.  A systematic
study of this enlarged parameter space subject to the recoil requirement
is beyond the scope of the present analysis.

\bibliographystyle{apsrev4-2}
\bibliography{references}

@article{Slavich:2020zjv,
    author = "Slavich, P. and others",
    editor = "Slavich, P. and Heinemeyer, S.",
    title = "{Higgs-mass predictions in the MSSM and beyond}",
    eprint = "2012.15629",
    archivePrefix = "arXiv",
    primaryClass = "hep-ph",
    reportNumber = "DESY 20-229, DESY-20-229, IFT-UAM/CSIC-20-184, FR-PHENO-2020-021, KA-TP-23-2020, MPP-2020-235, KA-TP-23-2020,
  MPP-2020-235, P3H-20-086, TTK-20-53, FERMILAB-PUB-21-575-T",
    doi = "10.1140/epjc/s10052-021-09198-2",
    journal = "Eur. Phys. J. C",
    volume = "81",
    number = "5",
    pages = "450",
    year = "2021"
}

@article{Draper:2013oza,
    author = "Draper, Patrick and Lee, Gabriel and Wagner, Carlos E. M.",
    title = "{Precise estimates of the Higgs mass in heavy supersymmetry}",
    eprint = "1312.5743",
    archivePrefix = "arXiv",
    primaryClass = "hep-ph",
    reportNumber = "SCIPP-13-15, EFI-PREPRINT-13-29",
    doi = "10.1103/PhysRevD.89.055023",
    journal = "Phys. Rev. D",
    volume = "89",
    number = "5",
    pages = "055023",
    year = "2014"
}

@article{Giudice:2004tc,
    author = "Giudice, G. F. and Romanino, A.",
    title = "{Split supersymmetry}",
    eprint = "hep-ph/0406088",
    archivePrefix = "arXiv",
    reportNumber = "CERN-PH-TH-2004-100",
    doi = "10.1016/j.nuclphysb.2004.08.001",
    journal = "Nucl. Phys. B",
    volume = "699",
    pages = "65--89",
    year = "2004",
    note = "[Erratum: Nucl.Phys.B 706, 487--487 (2005)]"
}

@article{PardoVega:2015eno,
    author = "Pardo Vega, Javier and Villadoro, Giovanni",
    title = "{SusyHD: Higgs mass Determination in Supersymmetry}",
    eprint = "1504.05200",
    archivePrefix = "arXiv",
    primaryClass = "hep-ph",
    doi = "10.1007/JHEP07(2015)159",
    journal = "JHEP",
    volume = "07",
    pages = "159",
    year = "2015"
}

@article{ATLAS:2024itc,
    author = "Aad, Georges and others",
    collaboration = "ATLAS",
    title = "{ATLAS searches for additional scalars and exotic Higgs boson decays with the LHC Run~2 dataset}",
    eprint = "2405.04914",
    archivePrefix = "arXiv",
    primaryClass = "hep-ex",
    reportNumber = "CERN-EP-2024-094",
    doi = "10.1016/j.physrep.2024.09.002",
    journal = "Phys. Rept.",
    volume = "1116",
    pages = "184--260",
    year = "2025"
}

@article{ATLAS:2021moa,
    author = "Aad, Georges and others",
    collaboration = "ATLAS",
    title = "{Search for chargino{\textendash}neutralino pair production in final states with three leptons and missing transverse momentum in $\sqrt{s} = 13$~TeV pp collisions with the ATLAS detector}",
    eprint = "2106.01676",
    archivePrefix = "arXiv",
    primaryClass = "hep-ex",
    reportNumber = "CERN-EP-2021-059",
    doi = "10.1140/epjc/s10052-021-09749-7",
    journal = "Eur. Phys. J. C",
    volume = "81",
    number = "12",
    pages = "1118",
    year = "2021"
}

@article{ATLAS:2020zms,
    author = "Aad, Georges and others",
    collaboration = "ATLAS",
    title = "{Search for heavy Higgs bosons decaying into two tau leptons with the ATLAS detector using $pp$ collisions at $\sqrt{s}=13$ TeV}",
    eprint = "2002.12223",
    archivePrefix = "arXiv",
    primaryClass = "hep-ex",
    reportNumber = "CERN-EP-2020-014",
    doi = "10.1103/PhysRevLett.125.051801",
    journal = "Phys. Rev. Lett.",
    volume = "125",
    number = "5",
    pages = "051801",
    year = "2020"
}

@article{CMS:2022goy,
    author = "Tumasyan, Armen and others",
    collaboration = "CMS",
    title = "{Searches for additional Higgs bosons and for vector leptoquarks in $\tau\tau$ final states in proton-proton collisions at $\sqrt{s}$ = 13 TeV}",
    eprint = "2208.02717",
    archivePrefix = "arXiv",
    primaryClass = "hep-ex",
    reportNumber = "CMS-HIG-21-001, CERN-EP-2022-137",
    doi = "10.1007/JHEP07(2023)073",
    journal = "JHEP",
    volume = "07",
    pages = "073",
    year = "2023"
}

@article{Nagata:2014aoa,
    author = "Nagata, Natsumi and Shirai, Satoshi",
    title = "{Electroweakly-Interacting Dirac Dark Matter}",
    eprint = "1411.0752",
    archivePrefix = "arXiv",
    primaryClass = "hep-ph",
    reportNumber = "DESY-14-201, FTPI-MINN-14-38, IPMU14-0332",
    doi = "10.1103/PhysRevD.91.055035",
    journal = "Phys. Rev. D",
    volume = "91",
    number = "5",
    pages = "055035",
    year = "2015"
}

@article{LZ:2024zvo,
    author = "Aalbers, J. and others",
    collaboration = "LZ",
    title = "{Dark Matter Search Results from 4.2{\,}{\,}Tonne-Years of Exposure of the LUX-ZEPLIN (LZ) Experiment}",
    eprint = "2410.17036",
    archivePrefix = "arXiv",
    primaryClass = "hep-ex",
    reportNumber = "FERMILAB-PUB-24-0796-V",
    doi = "10.1103/4dyc-z8zf",
    journal = "Phys. Rev. Lett.",
    volume = "135",
    number = "1",
    pages = "011802",
    year = "2025"
}

@article{Baum:2017enm,
    author = "Baum, Sebastian and Carena, Marcela and Shah, Nausheen R. and Wagner, Carlos E. M.",
    title = "{Higgs portals for thermal Dark Matter. EFT perspectives and the NMSSM}",
    eprint = "1712.09873",
    archivePrefix = "arXiv",
    primaryClass = "hep-ph",
    reportNumber = "NORDITA-2017-130, FERMILAB-PUB-17-611-T, EFI-17-25, WSU-HEP-1715",
    doi = "10.1007/JHEP04(2018)069",
    journal = "JHEP",
    volume = "04",
    pages = "069",
    year = "2018"
}

@misc{LZ:2026axp,
    author = "Akerib, D. S. and others",
    collaboration = "LZ",
    title = "{Search for dark matter particle interactions in an extended nuclear recoil energy window with the LUX-ZEPLIN (LZ) experiment}",
    eprint = "2609.02823",
    archivePrefix = "arXiv",
    primaryClass = "hep-ex",
    doi = "10.17182/hepdata.182472.v1",
    month = "9",
    year = "2026"
}

@misc{Fan:2026kxx,
    author = "Fan, JiJi and Reece, Matthew",
    title = "{Higgsino Above the Sea of Fog}",
    eprint = "2609.01504",
    archivePrefix = "arXiv",
    primaryClass = "hep-ph",
    month = "9",
    year = "2026"
}

@misc{Freese:2026sga,
    author = "Freese, Katherine and Theodosopoulos, Dionysios P.",
    title = "{Higgsino Dark Matter Interpretation of the LUX-ZEPLIN 248 keV Nuclear-Recoil Event}",
    eprint = "2609.01583",
    archivePrefix = "arXiv",
    primaryClass = "hep-ph",
    month = "9",
    year = "2026"
}

@misc{Rodd:2026tyn,
    author = "Rodd, Nicholas L. and Safdi, Benjamin R. and Slatyer, Tracy R. and Xu, Weishuang Linda",
    title = "{Confronting the Higgsino Interpretation of the LZ Event with the High-Energy Sideband}",
    eprint = "2609.04175",
    archivePrefix = "arXiv",
    primaryClass = "hep-ph",
    month = "9",
    year = "2026"
}

@misc{Wu:2026nhi,
    author = "Wu, Lei and Zhang, Yang and Zhu, Bin",
    title = "{TeV Higgsino Dark Matter from LZ Nuclear Recoil to Fermi-LAT Gamma Rays}",
    eprint = "2609.01590",
    archivePrefix = "arXiv",
    primaryClass = "hep-ph",
    month = "9",
    year = "2026"
}

@misc{Du:2026guj,
    author = "Du, Xiaokang and Wang, Fei",
    title = "{TeV Higgsino Interpretation of the LZ High-Recoil Event with Intermediate-Scale Electroweak Gauginos}",
    eprint = "2609.04163",
    archivePrefix = "arXiv",
    primaryClass = "hep-ph",
    month = "9",
    year = "2026"
}

@misc{Bisal:2026khf,
    author = "Bisal, Subhadip and Cao, Junjie and Li, Fei",
    title = "{Higgsino Dark Matter Interpretation of the LZ High-Recoil Event in the GNMSSM with TeV-Scale Gauginos}",
    eprint = "2609.07811",
    archivePrefix = "arXiv",
    primaryClass = "hep-ph",
    month = "9",
    year = "2026"
}

@misc{Langhoff:2026ujr,
    author = "Langhoff, Kevin",
    title = "{Heavy Higgsino Interpretation of the LZ Event}",
    eprint = "2609.09385",
    archivePrefix = "arXiv",
    primaryClass = "hep-ph",
    month = "9",
    year = "2026"
}

@misc{Cheung:2026byg,
    author = "Cheung, Kingman and Kang, Sin Kyu and Kumar, Ranjeet",
    title = "{From LUX-ZEPLIN to Colliders: Probing Higgsino Dark Matter}",
    eprint = "2609.08712",
    archivePrefix = "arXiv",
    primaryClass = "hep-ph",
    month = "9",
    year = "2026"
}

@misc{Chatterjee:2026scv,
    author = "Chatterjee, Arindam and Das, Debottam and Pasha, Syed Adil and Pukhov, Alexander and Puri, Rahul",
    title = "{Radiative Corrections to the Direct Detection of Inelastic Scattering of Higgsino-like Neutralino Dark Matter}",
    eprint = "2609.09830",
    archivePrefix = "arXiv",
    primaryClass = "hep-ph",
    month = "9",
    year = "2026"
}

@misc{Frolovsky:2026tvq,
    author = "Frolovsky, Daniel and Ketov, Sergei V.",
    title = "{Higgsino dark matter in the Starobinsky supergravity with the MSSM in light of the LUX-ZEPLIN event}",
    eprint = "2609.11241",
    archivePrefix = "arXiv",
    primaryClass = "hep-ph",
    reportNumber = "IPMU26-0034",
    month = "9",
    year = "2026"
}

@article{Fitzpatrick:2012ix,
    author = "Fitzpatrick, A. Liam and Haxton, Wick and Katz, Emanuel and Lubbers, Nicholas and Xu, Yiming",
    title = "{The Effective Field Theory of Dark Matter Direct Detection}",
    eprint = "1203.3542",
    archivePrefix = "arXiv",
    primaryClass = "hep-ph",
    doi = "10.1088/1475-7516/2013/02/004",
    journal = "JCAP",
    volume = "02",
    pages = "004",
    year = "2013"
}

@article{Anand:2013yka,
    author = "Anand, Nikhil and Fitzpatrick, A. Liam and Haxton, W. C.",
    title = "{Weakly interacting massive particle-nucleus elastic scattering response}",
    eprint = "1308.6288",
    archivePrefix = "arXiv",
    primaryClass = "hep-ph",
    doi = "10.1103/PhysRevC.89.065501",
    journal = "Phys. Rev. C",
    volume = "89",
    number = "6",
    pages = "065501",
    year = "2014"
}

@misc{Elahi:2026vlm,
    author = "Elahi, Fatemeh and Schwaller, Pedro",
    title = "{A Vector-Like Lepton Interpretation of the High-Energy Nuclear Recoil Candidate in LUX-ZEPLIN}",
    eprint = "2609.08993",
    archivePrefix = "arXiv",
    primaryClass = "hep-ph",
    reportNumber = "MITP-26-043",
    month = "9",
    year = "2026"
}

@article{Nagata:2014wma,
    author = "Nagata, Natsumi and Shirai, Satoshi",
    title = "{Higgsino Dark Matter in High-Scale Supersymmetry}",
    eprint = "1410.4549",
    archivePrefix = "arXiv",
    primaryClass = "hep-ph",
    reportNumber = "DESY-14-180, FTPI-MINN-14-37, IPMU14-0320",
    doi = "10.1007/JHEP01(2015)029",
    journal = "JHEP",
    volume = "01",
    pages = "029",
    year = "2015"
}

@article{Arkani-Hamed:2006wnf,
    author = "Arkani-Hamed, N. and Delgado, A. and Giudice, G. F.",
    title = "{The Well-tempered neutralino}",
    eprint = "hep-ph/0601041",
    archivePrefix = "arXiv",
    reportNumber = "CERN-PH-TH-2005-260",
    doi = "10.1016/j.nuclphysb.2006.02.010",
    journal = "Nucl. Phys. B",
    volume = "741",
    pages = "108--130",
    year = "2006"
}

@article{Cheung:2012qy,
    author = "Cheung, Clifford and Hall, Lawrence J. and Pinner, David and Ruderman, Joshua T.",
    title = "{Prospects and Blind Spots for Neutralino Dark Matter}",
    eprint = "1211.4873",
    archivePrefix = "arXiv",
    primaryClass = "hep-ph",
    doi = "10.1007/JHEP05(2013)100",
    journal = "JHEP",
    volume = "05",
    pages = "100",
    year = "2013"
}

@article{Huang:2014xua,
    author = "Huang, Peisi and Wagner, Carlos E. M.",
    title = "{Blind Spots for neutralino Dark Matter in the MSSM with an intermediate $m_A$}",
    eprint = "1404.0392",
    archivePrefix = "arXiv",
    primaryClass = "hep-ph",
    doi = "10.1103/PhysRevD.90.015018",
    journal = "Phys. Rev. D",
    volume = "90",
    number = "1",
    pages = "015018",
    year = "2014"
}

@article{Badziak:2015exr,
    author = "Badziak, Marcin and Olechowski, Marek and Szczerbiak, Pawe{\l}",
    title = "{Blind spots for neutralino dark matter in the NMSSM}",
    eprint = "1512.02472",
    archivePrefix = "arXiv",
    primaryClass = "hep-ph",
    doi = "10.1007/JHEP03(2016)179",
    journal = "JHEP",
    volume = "03",
    pages = "179",
    year = "2016"
}

@article{Abdallah:2020yag, 
    author = "Abdallah, Waleed and Datta, AseshKrishna and Roy, Subhojit",
    title = "{A relatively light, highly bino-like dark matter in the Z$_{3}$-symmetric NMSSM and recent LHC searches}",
    eprint = "2012.04026",
    archivePrefix = "arXiv",
    primaryClass = "hep-ph",
    reportNumber = "HRI-RECAPP-2020-10",
    doi = "10.1007/JHEP04(2021)122",
    journal = "JHEP",
    volume = "04",
    pages = "122",
    year = "2021"
}

@article{Chatterjee:2022pxf, 
    author = "Chatterjee, Arindam and Datta, AseshKrishna and Roy, Subhojit",
    title = "{Electroweak phase transition in the Z$_{3}$-invariant NMSSM: Implications of LHC and Dark matter searches and prospects of detecting the gravitational waves}",
    eprint = "2202.12476",
    archivePrefix = "arXiv",
    primaryClass = "hep-ph",
    reportNumber = "HRI-RECAPP-2022-001",
    doi = "10.1007/JHEP06(2022)108",
    journal = "JHEP",
    volume = "06",
    pages = "108",
    year = "2022"
}

@article{Datta:2022bvg,
    author = "Datta, AseshKrishna and Guchait, Monoranjan and Roy, Arnab and Roy, Subhojit",
    title = "{Hunting ewinos and a light scalar of Z$_{3}$-NMSSM with a bino-like dark matter in top squark decays at the LHC}",
    eprint = "2211.05905",
    archivePrefix = "arXiv",
    primaryClass = "hep-ph",
    reportNumber = "HRI-RECAPP-2022-014",
    doi = "10.1007/JHEP11(2023)081",
    journal = "JHEP",
    volume = "11",
    pages = "081",
    year = "2023"
}

@article{Roy:2024yoh,
    author = "Roy, Subhojit and Wagner, Carlos E. M.",
    title = "{Dark Matter searches with photons at the LHC}",
    eprint = "2401.08917",
    archivePrefix = "arXiv",
    primaryClass = "hep-ph",
    reportNumber = "EFI 24-1",
    doi = "10.1007/JHEP04(2024)106",
    journal = "JHEP",
    volume = "04",
    pages = "106",
    year = "2024"
}

@article{Arganda:2025fhx,
    author = "Arganda, Ernesto and de los Rios, Mart{\'\i}n and Perez, Andres D. and Roy, Subhojit and Sand{\'a} Seoane, Rosa M. and Wagner, Carlos E. M.",
    title = "{Shedding light on dark matter at the LHC with machine learning}",
    eprint = "2509.15121",
    archivePrefix = "arXiv",
    primaryClass = "hep-ph",
    reportNumber = "EFI-25-12, IFT-UAM/CSIC-25-93",
    doi = "10.1103/m1cd-1sfb",
    journal = "Phys. Rev. D",
    volume = "113",
    number = "9",
    pages = "095013",
    year = "2026"
}

@article{Ellwanger:2009dp,
    author = "Ellwanger, Ulrich and Hugonie, Cyril and Teixeira, Ana M.",
    title = "{The Next-to-Minimal Supersymmetric Standard Model}",
    eprint = "0910.1785",
    archivePrefix = "arXiv",
    primaryClass = "hep-ph",
    reportNumber = "LPT-ORSAY-09-76, CFTP-09-032, LPTA-09-066",
    doi = "10.1016/j.physrep.2010.07.001",
    journal = "Phys. Rept.",
    volume = "496",
    pages = "1--77",
    year = "2010"
}

@article{Baxter:2021pqo,
    author = "Baxter, D. and others",
    title = "{Recommended conventions for reporting results from direct dark matter searches}",
    eprint = "2105.00599",
    archivePrefix = "arXiv",
    primaryClass = "hep-ex",
    doi = "10.1140/epjc/s10052-021-09655-y",
    journal = "Eur. Phys. J. C",
    volume = "81",
    number = "10",
    pages = "907",
    year = "2021"
}

@article{Smith-Orlik:2023kyl,
    author = "Smith-Orlik, Adam and others",
    title = "{The impact of the Large Magellanic Cloud on dark matter direct detection signals}",
    eprint = "2302.04281",
    archivePrefix = "arXiv",
    primaryClass = "astro-ph.GA",
    doi = "10.1088/1475-7516/2023/10/070",
    journal = "JCAP",
    volume = "10",
    pages = "070",
    year = "2023"
}

@article{Graham:2024syw,
    author = "Graham, Peter W. and Ramani, Harikrishnan and Wong, Samuel S. Y.",
    title = "{Enhancing direct detection of Higgsino dark matter}",
    eprint = "2409.07768",
    archivePrefix = "arXiv",
    primaryClass = "hep-ph",
    doi = "10.1103/PhysRevD.111.055030",
    journal = "Phys. Rev. D",
    volume = "111",
    number = "5",
    pages = "055030",
    year = "2025"
}

@article{Barello:2014uda,
    author = "Barello, G. and Chang, Spencer and Newby, Christopher A.",
    title = "{A Model Independent Approach to Inelastic Dark Matter Scattering}",
    eprint = "1409.0536",
    archivePrefix = "arXiv",
    primaryClass = "hep-ph",
    doi = "10.1103/PhysRevD.90.094027",
    journal = "Phys. Rev. D",
    volume = "90",
    number = "9",
    pages = "094027",
    year = "2014"
}

@article{Bishara:2017pfq,
    author = "Bishara, Fady and Brod, Joachim and Grinstein, Benjamin and Zupan, Jure",
    title = "{From quarks to nucleons in dark matter direct detection}",
    eprint = "1707.06998",
    archivePrefix = "arXiv",
    primaryClass = "hep-ph",
    reportNumber = "DO-TH-17-10, OUTP-17-07P, CERN-TH-2017-157",
    doi = "10.1007/JHEP11(2017)059",
    journal = "JHEP",
    volume = "11",
    pages = "059",
    year = "2017"
}

@article{Vietze:2014vsa,
    author = "Vietze, L. and Klos, P. and Men{\'e}ndez, J. and Haxton, W. C. and Schwenk, A.",
    title = "{Nuclear structure aspects of spin-independent WIMP scattering off xenon}",
    eprint = "1412.6091",
    archivePrefix = "arXiv",
    primaryClass = "nucl-th",
    doi = "10.1103/PhysRevD.91.043520",
    journal = "Phys. Rev. D",
    volume = "91",
    number = "4",
    pages = "043520",
    year = "2015"
}

@article{Mizuta:1992qp,
    author = "Mizuta, Satoshi and Yamaguchi, Masahiro",
    title = "{Coannihilation effects and relic abundance of Higgsino dominant LSP(s)}",
    eprint = "hep-ph/9208251",
    archivePrefix = "arXiv",
    reportNumber = "TU-409",
    doi = "10.1016/0370-2693(93)91717-2",
    journal = "Phys. Lett. B",
    volume = "298",
    pages = "120--126",
    year = "1993"
}

@article{Planck:2018vyg,
    author = "Aghanim, N. and others",
    collaboration = "Planck",
    title = "{Planck 2018 results. VI. Cosmological parameters}",
    eprint = "1807.06209",
    archivePrefix = "arXiv",
    primaryClass = "astro-ph.CO",
    doi = "10.1051/0004-6361/201833910",
    journal = "Astron. Astrophys.",
    volume = "641",
    pages = "A6",
    year = "2020",
    note = "[Erratum: Astron.Astrophys. 652, C4 (2021)]"
}

@article{Hryczuk:2010zi,
    author = "Hryczuk, Andrzej and Iengo, Roberto and Ullio, Piero",
    title = "{Relic densities including Sommerfeld enhancements in the MSSM}",
    eprint = "1010.2172",
    archivePrefix = "arXiv",
    primaryClass = "hep-ph",
    doi = "10.1007/JHEP03(2011)069",
    journal = "JHEP",
    volume = "03",
    pages = "069",
    year = "2011"
}

@article{Beneke:2014hja,
    author = "Beneke, M. and Hellmann, Charlotte and Ruiz-Femenia, P.",
    title = "{Heavy neutralino relic abundance with Sommerfeld enhancements - a study of pMSSM scenarios}",
    eprint = "1411.6930",
    archivePrefix = "arXiv",
    primaryClass = "hep-ph",
    reportNumber = "TUM-HEP-955-14, TTK-14-22, SFB-CPP-14-70, IFIC-14-59",
    doi = "10.1007/JHEP03(2015)162",
    journal = "JHEP",
    volume = "03",
    pages = "162",
    year = "2015"
}

@article{Martin:1997ns,
    author = "Martin, Stephen P.",
    editor = "Kane, Gordon L.",
    title = "{A Supersymmetry primer}",
    eprint = "hep-ph/9709356",
    archivePrefix = "arXiv",
    reportNumber = "FERMILAB-PUB-97-425-T",
    doi = "10.1142/9789812839657_0001",
    journal = "Adv. Ser. Direct. High Energy Phys.",
    volume = "18",
    pages = "1--98",
    year = "1998"
}

@article{Djouadi:2001kba,
    author = "Djouadi, A. and Drees, Manuel and Fileviez Perez, P. and Muhlleitner, M.",
    title = "{Loop induced Higgs and Z boson couplings to neutralinos and implications for collider and dark matter searches}",
    eprint = "hep-ph/0109283",
    archivePrefix = "arXiv",
    reportNumber = "PM-01-28, TUM-HEP-434-01, MPI-TH-32-01",
    doi = "10.1103/PhysRevD.65.075016",
    journal = "Phys. Rev. D",
    volume = "65",
    pages = "075016",
    year = "2002"
}

@misc{Pierce:2013rda,
    author = "Pierce, Aaron and Shah, Nausheen R. and Freese, Katherine",
    title = "{Neutralino Dark Matter with Light Staus}",
    eprint = "1309.7351",
    archivePrefix = "arXiv",
    primaryClass = "hep-ph",
    reportNumber = "MCTP-13-30",
    month = "9",
    year = "2013"
}

@misc{Pospelov:2026ewn,
    author = "Pospelov, Maxim and Ramani, Harikrishnan",
    title = "{Strong Constraints on Higgsino Dark Matter from Solar Capture}",
    eprint = "2609.02775",
    archivePrefix = "arXiv",
    primaryClass = "hep-ph",
    month = "9",
    year = "2026"
}

@misc{Bose:2026ndd,
    author = "Bose, Debajit and others",
    title = "{Not so good $\nu$s for Higgsino dark matter as LZ excess: stringent limits from Super-Kamiokande and IceCube}",
    eprint = "2609.07807",
    archivePrefix = "arXiv",
    primaryClass = "hep-ph",
    month = "9",
    year = "2026"
}

@misc{Nguyen:2026lui,
    author = "Nguyen, Thong T. Q. and Linden, Tim and Hooper, Dan",
    title = "{Solar Neutrino Constraints on Inelastic Dark Matter Scattering in Light of Recent LUX-ZEPLIN Observations}",
    eprint = "2609.11833",
    archivePrefix = "arXiv",
    primaryClass = "hep-ph",
    month = "9",
    year = "2026"
}

@misc{Ghosh:2026txe,
    author = "Ghosh, Aditya and Chavez, Ilumi and Kelso, Chris",
    title = "{Confronting the Higgsino Interpretation of the LZ Event with Astrophysical Uncertainties and the Solar Capture Constraints}",
    eprint = "2609.15321",
    archivePrefix = "arXiv",
    primaryClass = "hep-ph",
    month = "9",
    year = "2026"
}

@article{Badziak:2015nrb,
    author = "Badziak, M. and Olechowski, M. and Szczerbiak, P.",
    editor = "Antoniadis, Ignatios and Leontaris, George K. and Tamvakis, Kyriakos",
    title = "{Blind spots for neutralinos in NMSSM with light singlet scalar}",
    eprint = "1601.00768",
    archivePrefix = "arXiv",
    primaryClass = "hep-ph",
    journal = "PoS",
    volume = "PLANCK2015",
    pages = "130",
    year = "2015"
}

@article{Klos:2013rwa,
    author = "Klos, P. and Men{\'e}ndez, J. and Gazit, D. and Schwenk, A.",
    title = "{Large-scale nuclear structure calculations for spin-dependent WIMP scattering with chiral effective field theory currents}",
    eprint = "1304.7684",
    archivePrefix = "arXiv",
    primaryClass = "nucl-th",
    doi = "10.1103/PhysRevD.88.083516",
    journal = "Phys. Rev. D",
    volume = "88",
    number = "8",
    pages = "083516",
    year = "2013",
    note = "[Erratum: Phys.Rev.D 89, 029901 (2014)]"
}

@article{Jeong:2021bpl,
    author = "Jeong, Injun and Kang, Sunghyun and Scopel, Stefano and Tomar, Gaurav",
    title = "{WimPyDD: An object{\textendash}oriented Python code for the calculation of WIMP direct detection signals}",
    eprint = "2106.06207",
    archivePrefix = "arXiv",
    primaryClass = "hep-ph",
    reportNumber = "CQUeST-2021-0663, TUM-HEP 1343/21",
    doi = "10.1016/j.cpc.2022.108342",
    journal = "Comput. Phys. Commun.",
    volume = "276",
    pages = "108342",
    year = "2022"
}

@article{Park:2025rxi,
    author = "Park, Sungwoo and Gupta, Rajan and Bhattacharya, Tanmoy and He, Fangcheng and Mondal, Santanu and Lin, Huey-Wen and Yoon, Boram",
    title = "{Flavor diagonal nucleon charges using clover fermions on MILC HISQ ensembles}",
    eprint = "2503.07100",
    archivePrefix = "arXiv",
    primaryClass = "hep-lat",
    reportNumber = "LLNL-JRNL-858665, LA-UR-22053",
    doi = "10.1103/hn5v-9zgm",
    journal = "Phys. Rev. D",
    volume = "112",
    number = "5",
    pages = "054508",
    year = "2025"
}

@article{Gorton:2022eed,
    author = "Gorton, Oliver C. and Johnson, Calvin W. and Jiao, Changfeng and Nikoleyczik, Jonathan",
    title = "{dmscatter: A fast program for WIMP-nucleus scattering}",
    eprint = "2209.09187",
    archivePrefix = "arXiv",
    primaryClass = "nucl-th",
    doi = "10.1016/j.cpc.2022.108597",
    journal = "Comput. Phys. Commun.",
    volume = "284",
    pages = "108597",
    year = "2023"
}

@article{Lewin:1995rx,
    author = "Lewin, J. D. and Smith, P. F.",
    title = "{Review of mathematics, numerical factors, and corrections for dark matter experiments based on elastic nuclear recoil}",
    reportNumber = "RAL-TR-95-024",
    doi = "10.1016/S0927-6505(96)00047-3",
    journal = "Astropart. Phys.",
    volume = "6",
    pages = "87--112",
    year = "1996"
}

@article{Hoferichter:2020osn,
    author = "Hoferichter, Martin and Men{\'e}ndez, Javier and Schwenk, Achim",
    title = "{Coherent elastic neutrino-nucleus scattering: EFT analysis and nuclear responses}",
    eprint = "2007.08529",
    archivePrefix = "arXiv",
    primaryClass = "hep-ph",
    reportNumber = "INT-PUB-20-026",
    doi = "10.1103/PhysRevD.102.074018",
    journal = "Phys. Rev. D",
    volume = "102",
    number = "7",
    pages = "074018",
    year = "2020"
}

@misc{Berkeley,
  author       = {Haxton, W. C. and McElvain, K. S.},
  title        = {Elastic: One-Body Density Matrices for Common Experimental Targets},
  howpublished = {\url{https://github.com/Berkeley-Electroweak-Physics/Elastic}},
  note         = {GCN5082 xenon files, commit
                  475d69bebcecffaf1fce1d5a4f0a24eb554bf16c;
                  accessed September 14, 2026},
  year         = {2022}
}

@misc{NISTXe,
  author       = {Coursey, J. S. and Schwab, D. J. and Dragoset, R. A.},
  title        = {Atomic Weights and Isotopic Compositions},
  howpublished = {\url{https://physics.nist.gov/Comp}},
  note         = {National Institute of Standards and Technology,
                  Gaithersburg, MD; accessed September 14, 2026}
}

@article{Guchait:2016pes,
    author = "Guchait, Monoranjan and Kumar, Jacky",
    title = "{Diphoton Signal of light pseudoscalar in NMSSM at the LHC}",
    eprint = "1608.05693",
    archivePrefix = "arXiv",
    primaryClass = "hep-ph",
    doi = "10.1103/PhysRevD.95.035036",
    journal = "Phys. Rev. D",
    volume = "95",
    number = "3",
    pages = "035036",
    year = "2017"
}

@article{Ellwanger:2022jtd,
    author = "Ellwanger, Ulrich and Hugonie, Cyril",
    title = "{Benchmark planes for Higgs-to-Higgs decays in the NMSSM}",
    eprint = "2203.05049",
    archivePrefix = "arXiv",
    primaryClass = "hep-ph",
    reportNumber = "LUPM 22-004",
    doi = "10.1140/epjc/s10052-022-10364-3",
    journal = "Eur. Phys. J. C",
    volume = "82",
    number = "5",
    pages = "406",
    year = "2022"
}

@article{Fermi-LAT:2025gei,
    author = "Abdollahi, S. and others",
    collaboration = "Fermi-LAT, HAWC, H.E.S.S., MAGIC, VERITAS",
    title = "{Combined dark matter search towards dwarf spheroidal galaxies with Fermi-LAT, HAWC, H.E.S.S., MAGIC, and VERITAS}",
    eprint = "2508.20229",
    archivePrefix = "arXiv",
    primaryClass = "astro-ph.HE",
    doi = "10.1088/1475-7516/2026/03/035",
    journal = "JCAP",
    volume = "03",
    pages = "035",
    year = "2026"
}

@article{Fermi-LAT:2015kyq,
    author = "Ackermann, M. and others",
    collaboration = "Fermi-LAT",
    title = "{Updated search for spectral lines from Galactic dark matter interactions with pass 8 data from the Fermi Large Area Telescope}",
    eprint = "1506.00013",
    archivePrefix = "arXiv",
    primaryClass = "astro-ph.HE",
    reportNumber = "FERMILAB-PUB-15-673-AE",
    doi = "10.1103/PhysRevD.91.122002",
    journal = "Phys. Rev. D",
    volume = "91",
    number = "12",
    pages = "122002",
    year = "2015"
}

@article{Essig:2007az,
    author = "Essig, Rouven",
    title = "{Direct Detection of Non-Chiral Dark Matter}",
    eprint = "0710.1668",
    archivePrefix = "arXiv",
    primaryClass = "hep-ph",
    reportNumber = "RUNHETC-2007-20",
    doi = "10.1103/PhysRevD.78.015004",
    journal = "Phys. Rev. D",
    volume = "78",
    pages = "015004",
    year = "2008"
}

@article{Kim:1983dt,
    author = "Kim, Jihn E. and Nilles, Hans Peter",
    title = "{The mu Problem and the Strong CP Problem}",
    reportNumber = "UGVA-DPT 1983/10-410",
    doi = "10.1016/0370-2693(84)91890-2",
    journal = "Phys. Lett. B",
    volume = "138",
    pages = "150--154",
    year = "1984"
}

@article{Ellwanger:2004xm,
    author = "Ellwanger, Ulrich and Gunion, John F. and Hugonie, Cyril",
    title = "{NMHDECAY: A Fortran code for the Higgs masses, couplings and decay widths in the NMSSM}",
    eprint = "hep-ph/0406215",
    archivePrefix = "arXiv",
    reportNumber = "LPT-ORSAY-04-32, UCD-04-22, IFIC-04-33",
    doi = "10.1088/1126-6708/2005/02/066",
    journal = "JHEP",
    volume = "02",
    pages = "066",
    year = "2005"
}

@article{Ellwanger:2005dv,
    author = "Ellwanger, Ulrich and Hugonie, Cyril",
    title = "{NMHDECAY 2.0: An Updated program for sparticle masses, Higgs masses, couplings and decay widths in the NMSSM}",
    eprint = "hep-ph/0508022",
    archivePrefix = "arXiv",
    reportNumber = "LPT-ORSAY-05-52",
    doi = "10.1016/j.cpc.2006.04.004",
    journal = "Comput. Phys. Commun.",
    volume = "175",
    pages = "290--303",
    year = "2006"
}

@article{Das:2011dg,
    author = "Das, Debottam and Ellwanger, Ulrich and Teixeira, Ana M.",
    title = "{NMSDECAY: A Fortran Code for Supersymmetric Particle Decays in the Next-to-Minimal Supersymmetric Standard Model}",
    eprint = "1106.5633",
    archivePrefix = "arXiv",
    primaryClass = "hep-ph",
    reportNumber = "LPT-ORSAY-11-56, PCCF-RI-1104",
    doi = "10.1016/j.cpc.2011.11.021",
    journal = "Comput. Phys. Commun.",
    volume = "183",
    pages = "774--779",
    year = "2012"
}

@article{Belanger:2006is,
    author = "Belanger, G. and Boudjema, F. and Pukhov, A. and Semenov, A.",
    title = "{MicrOMEGAs 2.0: A Program to calculate the relic density of dark matter in a generic model}",
    eprint = "hep-ph/0607059",
    archivePrefix = "arXiv",
    reportNumber = "LAPTH-1152-06",
    doi = "10.1016/j.cpc.2006.11.008",
    journal = "Comput. Phys. Commun.",
    volume = "176",
    pages = "367--382",
    year = "2007"
}

@article{Belanger:2008sj,
    author = "Belanger, G. and Boudjema, F. and Pukhov, A. and Semenov, A.",
    title = "{Dark matter direct detection rate in a generic model with micrOMEGAs 2.2}",
    eprint = "0803.2360",
    archivePrefix = "arXiv",
    primaryClass = "hep-ph",
    reportNumber = "LAPTH-1237-08",
    doi = "10.1016/j.cpc.2008.11.019",
    journal = "Comput. Phys. Commun.",
    volume = "180",
    pages = "747--767",
    year = "2009"
}

@misc{IceCube:2025fcu,
    author = "Abbasi, R. and others",
    collaboration = "IceCube",
    title = "{Search for High-Energy Neutrinos From the Sun Using Ten Years of IceCube Data}",
    eprint = "2507.08457",
    archivePrefix = "arXiv",
    primaryClass = "hep-ex",
    month = "7",
    year = "2025"
}

@misc{Yin:2026jnn,
    author = "Yin, Wen",
    title = "{A PQ-Symmetric High-Scale SUSY Interpretation of the LZ High-Energy Recoil}",
    eprint = "2609.01892",
    archivePrefix = "arXiv",
    primaryClass = "hep-ph",
    month = "9",
    year = "2026"
}

@article{Tucker-Smith:2001myb,
    author = "Tucker-Smith, David and Weiner, Neal",
    title = "{Inelastic dark matter}",
    eprint = "hep-ph/0101138",
    archivePrefix = "arXiv",
    reportNumber = "UCB-PTH-00-43, LBNL-47234, UW-PT-00-17",
    doi = "10.1103/PhysRevD.64.043502",
    journal = "Phys. Rev. D",
    volume = "64",
    pages = "043502",
    year = "2001"
}

@misc{DiMauro:2026dqp,
    author = "Di Mauro, Mattia and Shaikh, Halim",
    title = "{Solar Capture Tests of Inelastic Dark Matter after the LZ High-Recoil Event}",
    eprint = "2609.06760",
    archivePrefix = "arXiv",
    primaryClass = "hep-ph",
    month = "9",
    year = "2026"
}

@article{Menendez:2012tm,
    author = "Menendez, J. and Gazit, D. and Schwenk, A.",
    title = "{Spin-dependent WIMP scattering off nuclei}",
    eprint = "1208.1094",
    archivePrefix = "arXiv",
    primaryClass = "astro-ph.CO",
    doi = "10.1103/PhysRevD.86.103511",
    journal = "Phys. Rev. D",
    volume = "86",
    pages = "103511",
    year = "2012"
}

@article{Jungman:1995df,
    author = "Jungman, Gerard and Kamionkowski, Marc and Griest, Kim",
    title = "{Supersymmetric dark matter}",
    eprint = "hep-ph/9506380",
    archivePrefix = "arXiv",
    reportNumber = "SU-4240-605, UCSD-PTH-95-02, IASSNS-HEP-95-14, CU-TP-677",
    doi = "10.1016/0370-1573(95)00058-5",
    journal = "Phys. Rept.",
    volume = "267",
    pages = "195--373",
    year = "1996"
}

@article{Liu:2020ckq,
    author = {Liu, Qinrui and Lazar, Jeffrey and Arg{\"u}elles, Carlos A. and Kheirandish, Ali},
    title = "{$\chi$aro$\nu$: a tool for neutrino flux generation from WIMPs}",
    eprint = "2007.15010",
    archivePrefix = "arXiv",
    primaryClass = "hep-ph",
    doi = "10.1088/1475-7516/2020/10/043",
    journal = "JCAP",
    volume = "10",
    pages = "043",
    year = "2020"
}

@article{Nussinov:2009ft,
    author = "Nussinov, Shmuel and Wang, Lian-Tao and Yavin, Itay",
    title = "{Capture of Inelastic Dark Matter in the Sun}",
    eprint = "0905.1333",
    archivePrefix = "arXiv",
    primaryClass = "hep-ph",
    doi = "10.1088/1475-7516/2009/08/037",
    journal = "JCAP",
    volume = "08",
    pages = "037",
    year = "2009"
}

@article{Menon:2009qj,
    author = "Menon, Arjun and Morris, Rob and Pierce, Aaron and Weiner, Neal",
    title = "{Capture and Indirect Detection of Inelastic Dark Matter}",
    eprint = "0905.1847",
    archivePrefix = "arXiv",
    primaryClass = "hep-ph",
    reportNumber = "MCTP-09-14",
    doi = "10.1103/PhysRevD.82.015011",
    journal = "Phys. Rev. D",
    volume = "82",
    pages = "015011",
    year = "2010"
}

@article{Alguero:2023zol,
    author = "Alguero, G. and Belanger, G. and Boudjema, F. and Chakraborti, S. and Goudelis, A. and Kraml, S. and Mjallal, A. and Pukhov, A.",
    title = "{micrOMEGAs 6.0: N-component dark matter}",
    eprint = "2312.14894",
    archivePrefix = "arXiv",
    primaryClass = "hep-ph",
    doi = "10.1016/j.cpc.2024.109133",
    journal = "Comput. Phys. Commun.",
    volume = "299",
    pages = "109133",
    year = "2024"
}

@article{ATLAS:2021yqv,
    author = "Aad, Georges and others",
    collaboration = "ATLAS",
    title = "{Search for charginos and neutralinos in final states with two boosted hadronically decaying bosons and missing transverse momentum in $pp$ collisions at $\sqrt {s}$ = 13{\,}{\,}TeV with the ATLAS detector}",
    eprint = "2108.07586",
    archivePrefix = "arXiv",
    primaryClass = "hep-ex",
    reportNumber = "CERN-EP-2021-127",
    doi = "10.1103/PhysRevD.104.112010",
    journal = "Phys. Rev. D",
    volume = "104",
    number = "11",
    pages = "112010",
    year = "2021"
}

@article{ATLAS:2023lfr,
    author = "Aad, Georges and others",
    collaboration = "ATLAS",
    title = "{Search for direct production of winos and higgsinos in events with two same-charge leptons or three leptons in pp collision data at $ \sqrt{s} $ = 13 TeV with the ATLAS detector}",
    eprint = "2305.09322",
    archivePrefix = "arXiv",
    primaryClass = "hep-ex",
    reportNumber = "CERN-EP-2023-063",
    doi = "10.1007/JHEP11(2023)150",
    journal = "JHEP",
    volume = "11",
    pages = "150",
    year = "2023"
}

@article{CMS:2024gyw,
    author = "Hayrapetyan, Aram and others",
    collaboration = "CMS",
    title = "{Combined search for electroweak production of winos, binos, higgsinos, and sleptons in proton-proton collisions at s=13{\,}{\,}TeV}",
    eprint = "2402.01888",
    archivePrefix = "arXiv",
    primaryClass = "hep-ex",
    reportNumber = "CMS-SUS-21-008, CERN-EP-2023-238",
    doi = "10.1103/PhysRevD.109.112001",
    journal = "Phys. Rev. D",
    volume = "109",
    number = "11",
    pages = "112001",
    year = "2024"
}

@article{ATLAS:2018qmw,
    author = "Aaboud, Morad and others",
    collaboration = "ATLAS",
    title = "{Search for chargino and neutralino production in final states with a Higgs boson and missing transverse momentum at $\sqrt{s} = 13$ TeV with the ATLAS detector}",
    eprint = "1812.09432",
    archivePrefix = "arXiv",
    primaryClass = "hep-ex",
    reportNumber = "CERN-EP-2018-306",
    doi = "10.1103/PhysRevD.100.012006",
    journal = "Phys. Rev. D",
    volume = "100",
    number = "1",
    pages = "012006",
    year = "2019"
}

@article{ATLAS:2020qlk,
    author = "Aad, Georges and others",
    collaboration = "ATLAS",
    title = "{Search for direct production of electroweakinos in final states with missing transverse momentum and a Higgs boson decaying into photons in pp collisions at $ \sqrt{s} $ = 13 TeV with the ATLAS detector}",
    eprint = "2004.10894",
    archivePrefix = "arXiv",
    primaryClass = "hep-ex",
    reportNumber = "CERN-EP-2019-204",
    doi = "10.1007/JHEP10(2020)005",
    journal = "JHEP",
    volume = "10",
    pages = "005",
    year = "2020"
}

@article{ATLAS:2024vxm,
    author = "Aad, Georges and others",
    collaboration = "ATLAS",
    title = "{Search for heavy neutral Higgs bosons decaying into a top quark pair in 140 fb$^{−1}$ of proton-proton collision data at $ \sqrt{s} $ = 13 TeV with the ATLAS detector}",
    eprint = "2404.18986",
    archivePrefix = "arXiv",
    primaryClass = "hep-ex",
    reportNumber = "CERN-EP-2024-090",
    doi = "10.1007/JHEP08(2024)013",
    journal = "JHEP",
    volume = "08",
    pages = "013",
    year = "2024"
}

@article{ATLAS:2021upq,
    author = "Aad, Georges and others",
    collaboration = "ATLAS",
    title = "{Search for charged Higgs bosons decaying into a top quark and a bottom quark at $ \sqrt{\mathrm{s}} $ = 13 TeV with the ATLAS detector}",
    eprint = "2102.10076",
    archivePrefix = "arXiv",
    primaryClass = "hep-ex",
    reportNumber = "CERN-EP-2021-004",
    doi = "10.1007/JHEP06(2021)145",
    journal = "JHEP",
    volume = "06",
    pages = "145",
    year = "2021"
}

@misc{DiMauro:2026ldr,
    author = "Di Mauro, Mattia",
    title = "{Dark Matter at the Kinematic Edge: Interpreting the 248 keV LZ Nuclear-Recoil Candidate}",
    eprint = "2609.02608",
    archivePrefix = "arXiv",
    primaryClass = "hep-ph",
    month = "9",
    year = "2026"
}
\end{document}